\documentstyle[aps,pre,epsfig,psfig,eqsecnum]{revtex}

\def\eps{\varepsilon}
\def\k{{\bf k}}
\def\n{{\bf n}}
\def\p{{\bf p}}
\def\q{{\bf q}}
\def\l{{\bf l}}
\def\x{{\bf x}}
\def\y{{\bf y}}
\def\z{{\bf z}}
\def\B{{\cal B}}
\def\C{{\cal C}}
\def\Dm{\widetilde{\cal D}_{\mu}}
\def\nueff{\nu_{\it eff}}
\def\balpha{\mbox{\boldmath $\alpha$}}
\def\bbeta{\mbox{\boldmath $\beta$}}

\begin{document}

\title{Calculation of the anomalous exponents in the rapid-change model
of passive scalar advection to order $\varepsilon^{3}$.}

\author{L.\,Ts.\,Adzhemyan,  N.\,V.\,Antonov, V.\,A.\,Barinov,
Yu.\,S.\,Kabrits,  and A.\,N.\,Vasil'ev }

\address{Department of Theoretical Physics, St Petersburg University,
Uljanovskaja 1, St Petersburg---Petrodvorez, 198504, Russia}

\draft

\date{24 May 2001}

\maketitle

\begin{abstract}
The field theoretic renormalization group and operator product expansion are
applied to the model of a passive scalar advected by the Gaussian velocity
field with zero mean and correlation function
$\propto\delta(t-t')/k^{d+\eps}$. Inertial-range anomalous exponents,
identified with the critical dimensions of various scalar and tensor
composite operators constructed of the scalar gradients, are calculated
within the $\varepsilon$ expansion to order $\varepsilon^{3}$ (three-loop
approximation), including the exponents in anisotropic sectors. The main goal
of the paper is to give the complete derivation of this third-order result,
and to present and explain in detail the corresponding calculational
techniques. The character and convergence properties of the $\varepsilon$
expansion are discussed; the improved ``inverse'' $\varepsilon$
expansion is proposed and the comparison with the existing nonperturbative
results is given.
\end{abstract}
\pacs{PACS number(s): 47.27.$-$i, 47.10.+g, 05.10.Cc}

\section{Introduction} \label{sec:Intro}

The investigation of intermittency and anomalous scaling in fully developed
turbulence remains essentially an open theoretical problem. Both the natural
and numerical experiments suggest that the deviation from the classical
Kolmogorov theory \cite{Legacy} is even more strongly pronounced for a
passively advected scalar field than for the velocity field itself; see,
e.g., Ref.~\cite{Sree} and literature cited therein. At the same time, the
problem of passive advection appears easier tractable theoretically: even
simplified models describing the advection by a ``synthetic'' velocity field
with a given Gaussian statistics reproduce many of the anomalous features of
genuine turbulent heat or mass transport observed in experiments. Therefore,
the problem of passive scalar advection, being of practical importance in
itself, may also be viewed as a starting point in studying intermittency and
anomalous scaling in the turbulence as a whole. Detailed review of the recent
theoretical research on the passive scalar problem and the bibliography can
be found in Ref.~\cite{FGV}.

Most progress has been achieved for Kraichnan's rapid-change model
\cite{Kraich1}: for the first time, the anomalous exponents have been
derived on the basis of a microscopic model and within controlled
approximations \cite{GK,Falk1,Pumir,Siggia}.

In that model, the advection of a passive scalar field $\theta(x)\equiv
\theta(t,{\bf x})$ is described by the stochastic equation
\begin{equation}
\nabla_{t}\theta =\nu _0\partial^{2} \theta+f, \qquad
\nabla_{t} \equiv \partial _t + v_{i}\partial_{i},
\label{1}
\end{equation}
where $\partial _t \equiv \partial /\partial t$,
$\partial _i \equiv \partial /\partial x_{i}$, $\nu _0$
is the molecular diffusivity coefficient, $\partial^{2}$ is the Laplace
operator, ${\bf v}(x)\equiv \{v_{i}(x)\}$ is the transverse (owing
to the incompressibility) velocity field, and $f\equiv f(x)$ is an
artificial Gaussian scalar noise with zero mean and correlation function
\begin{equation}
\langle  f(x)  f(x')\rangle = \delta(t-t')\, C({\bf r}/L), \qquad
{\bf r}={\bf x}-{\bf x'}.
\label{2}
\end{equation}
The parameter $L$ is an integral scale related to the noise, and
$C({\bf r}/L)$ is some function finite as $L\to\infty$.

In the real problem, the field  ${\bf v}(x)$ satisfies the Navier--Stokes
equation. In the rapid-change model it obeys a
Gaussian distribution with zero mean and correlation function
\begin{eqnarray}
\langle v_{i}(x) v_{j}(x')\rangle = D_{0}\, \delta(t-t')\,
(2\pi)^{-d} \int d\k \, P_{ij}({\bf k})\, N_{k} \, \exp [{\rm
i}{\bf k}\cdot({\bf x}-{\bf x'})] , \qquad N_{k} = \Theta(k-m)\,
k^{-d-\eps} , \label{3}
\end{eqnarray}
where $P_{ij}({\bf k}) = \delta _{ij} - k_i k_j / k^2$ is the transverse
projector, $k\equiv |{\bf k}|$, $D_{0}>0$ is an amplitude factor, $d$ is the
dimensionality of the ${\bf x}$ space, $\Theta\,(\cdots)$ is the step
function and $0<\eps<2$ is a parameter with the real (``Kolmogorov'')
value $\eps=4/3$.

The infrared (IR) regularization is provided by the cut-off in the integral
(\ref{3}) from below at $k=m$, where $m\equiv 1/\ell$ is the reciprocal of
another integral scale $\ell$. The anomalous exponents are independent on
the precise form of the IR regularization; the sharp cut-off is the most
convenient choice from the calculational viewpoints. In what follows, we
shall not distinguish the two IR scales, setting $L\sim\ell$. The relations
\begin{equation}
D_{0}/\nu_0 = g_{0} = \Lambda^{\eps}
\label{Lambda}
\end{equation}
define the coupling constant $g_{0}$ (i.e., the formal expansion parameter
in the ordinary perturbation theory) and the characteristic ultraviolet
(UV) momentum scale $\Lambda$.

The issue of interest is, in particular, the behavior of the
equal-time structure functions
\begin{equation}
S_{n}({\bf r}) =\Big\langle[\theta(t,{\bf x})-\theta(t,{\bf x'})]^{n}
\Big\rangle
\label{struc}
\end{equation}
in the inertial-convective range $\Lambda \gg 1/r \gg m$.

In the isotropic model (\ref{1})--(\ref{3}), the odd multipoint correlation
functions of the scalar field vanish, while the even equal-time functions
satisfy linear partial differential equations \cite{Kraich1,GK,Falk1}. The
solution for the pair correlation function is obtained
explicitly; it shows that the structure function $S_{2} \propto r^{2-\eps}$
is finite at $m=0$ \cite{Kraich1}. The higher-order
correlators are not found explicitly, but their inertial-range
behavior can be extracted from the analysis of
the nontrivial zero modes of the corresponding differential
operators in the limits $\eps\to0$ \cite{GK,Pumir},
$1/d\to0$ \cite{Falk1}, or $\eps\to2$ \cite{Pumir,Siggia}.
It was shown that the even structure functions in the
inertial-convective range exhibit anomalous scaling behavior:
\begin{equation}
S_{n}(r)\propto D_{0}^{-n/2}\, r^{n(1-\eps/2)}\, (mr)^{\Delta_{n}},
\qquad  r =|{\bf x}-{\bf x'}|
\label{HZ1}
\end{equation}
with negative anomalous exponents $\Delta_{n}$, whose first terms of the
expansion in $\eps$ \cite{GK} and $1/d$ \cite{Falk1} have the forms
\begin{eqnarray}
\Delta_{n}= -n(n-2)\,\eps/2(d+2)+O(\eps^{2})= -n(n-2)\,\eps/2d +O(1/d^{2}).
\label{HZ3}
\end{eqnarray}

Another quantity of interest is the local dissipation rate of scalar
fluctuations, $E(x)=\nu_0 \partial_{i}\theta(x)\partial_{i}\theta(x)$.
The equal-time correlation functions of its powers in the
inertial range have the forms \cite{GK,Falk1}:
\begin{equation}
\langle E^{n}(x)\,E^{p}(x')\rangle
\propto (\Lambda r)^{-\Delta_{2n}-\Delta_{2p}} (mr)^{\Delta_{2n+2p}}
\label{HZ2}
\end{equation}
with $\Lambda$ from Eq.~(\ref{Lambda}) and $\Delta_{n}$ from~(\ref{HZ1}).
Relations of the form (\ref{HZ2}) are characteristic of the models
with multifractal behavior \cite{DL}.

In Ref. \cite{RG} and subsequent papers \cite{RG,RG1,RG2,RG3,Juha,cube},
the field theoretic renormalization group (RG) and operator product
expansion (OPE) were applied to model (\ref{1})--(\ref{3}).
In the RG approach, the anomalous scaling for the structure functions and
various pair correlators is established as a consequence of the existence
in the corresponding operator product expansions of ``dangerous'' composite
operators, whose {\it negative} critical dimensions are identified with
the anomalous exponents $\Delta_{n}$. This allows one to construct a
systematic perturbation expansion for the anomalous exponents, analogous
to the well-known $\eps$ expansion in the RG theory of critical behavior.

The key role in the RG and OPE approach to model (\ref{1})--(\ref{3}) is
played by the critical dimensions $\Delta_{nl}$, associated with the
irreducible tensor composite operators
\begin{equation}
F_{nl}= {\cal IRP}\, \bigr[ \partial_{i_{1}}
\theta\cdots\partial_{i_{l}}\theta\,
(\partial_{i}\theta\partial_{i}\theta)^{p} \bigl],
\label{Fnp}
\end{equation}
where $l$ is the number of the free vector indices and $n=l+2p$
is the total number of the fields $\theta$ entering the operator;
the vector indices of the symbol $F_{nl}$ are omitted. The symbol
${\cal IRP}$ denotes the irreducible part, obtained by subtracting the
appropriate expression involving the delta symbols, such that the
resulting tensor is traceless with respect to any pair of indices.
In particular, $ {\cal IRP} \bigr[ \partial_{i}\theta \partial_{j}\theta
\bigl] = \partial_{i}\theta \partial_{j}\theta - \delta_{ij}
(\partial_{k}\theta \partial_{k}\theta) /d$ and so on.

The dimension $\Delta_{n}\equiv\Delta_{n0}$ of the scalar operator is
nothing other than the anomalous exponent in Eq. (\ref{HZ1});
see Ref.~\cite{RG}.
The dimensions with $l\ne0$ come into play if the forcing (\ref{2})
becomes anisotropic: $\Delta_{nl}$ corresponds to the leading zero-mode
contribution to the $l$-th term of the Legendre decomposition
for the function $S_{n}$; see Ref.~\cite{RG3}.
They can be systematically calculated as series in $\eps$:
\begin{equation}
\Delta_{nl} = \sum_{k=1}^{\infty}  \, \Delta^{(k)}_{nl} \, \eps^{k}
\label{epsilon}
\end{equation}
with the first-order coefficient \cite{RG3}
\begin{equation}
\Delta^{(1)}_{nl} = -\frac{n\,(n-2)}{2(d+2)} +
\frac{(d+1)\,l\, (d+l-2)} {2(d-1)(d+2)}\, .
\label{Qnp}
\end{equation}
For $l=0$ this gives the result of \cite{GK}, while for $n=3$ and $l=1,3$
the results of Refs.~\cite{Pumir} are recovered. The result (\ref{Qnp})
was rederived later in Refs.~\cite{ReG,Wiese}.

The coefficients $\Delta^{(2)}_{n0}$ and $\Delta^{(2)}_{n2}$ were obtained
in Ref.~\cite{RG} for any $n$ and $d$; the result for general $l$ is
presented in \cite{Juha}. In particular, one has
\begin{mathletters}
\label{Qnp2}
\begin{eqnarray}
\Delta^{(2)}_{nl} = n(n-2)(0.000203n -0.02976) - l^{2} (0.01732n + 0.01223)
\label{Qnp22}
\end{eqnarray}
for $d=2$ and
\begin{eqnarray}
\Delta^{(2)}_{nl}= n(n-2)(0.00203n-0.00384 ) - l(l+1) (0.00710n - 0.00619)
\label{Qnp23}
\end{eqnarray}
\end{mathletters}
for $d=3$ (analytical results are too cumbersome and will not be given here;
see Refs. \cite{RG} for $l=0,2$ and \cite{Juha} for general $l$.

The $O(\eps^{3})$ contribution to $\Delta_{nl}$ was presented in
Ref.~\cite{cube}:
\begin{mathletters}
\label{Qnp3}
\begin{eqnarray}
\Delta^{(3)}_{nl} = n(n-2)(0.00454n^{2}+0.06486n+0.06505) +l^{2}
\left(-0.01974n^{2}-0.10423n+0.24094+0.01748\, l^{2}\, \right)
\label{Qnp32}
\end{eqnarray}
for $d=2$ and
\[ \Delta^{(3)}_{nl} = n(n-2)(0.00140n^{2}+0.01992n+0.03437)
+l(l+1) \bigl[-0.00420n^{2}-0.02421n+0.00280\, l(l+1) +
0.05065\, \bigr] \]
\begin{eqnarray}
{}
\label{Qnp33}
\end{eqnarray}
\end{mathletters}
for $d=3$. Here, we have presented the $O(\eps^{3})$ results with improved
accuracy in numerical coefficients and corrected a misprint in the expression
for $d=3$ in Ref.~\cite{cube}. No analytical formula for $\Delta^{(3)}_{nl}$
is available for general $d$, but the numerical result of the form
(\ref{Qnp3}) can be obtained for any given $d$. The large $d$ limit is
discussed in Sec.~\ref{sec:larged}.

Besides the calculational efficiency, an important advantage of the RG
approach is its universality: it can also be applied to the case of finite
correlation time or non-Gaussian advecting field; see Refs.~\cite{RG3}.
For passively advected vector fields, any calculation of the exponents
for higher-order correlations calls for the RG techniques already in the
$O(\eps)$ approximation \cite{Lanotte2,amodel,vektor}. Detailed introduction
to the RG approach in the statistical theory of turbulence and the
bibliography can be found in Refs.~\cite{UFN,turbo}.

The main goal of this paper is to give the complete and detailed derivation
of the third-order result (\ref{Qnp3}) announced in the Rapid Communication
\cite{cube}, and to present and explain in detail the corresponding
calculational technique. It might be useful not only for the
rapid-change model (\ref{1})--(\ref{3}) and its descendants, but also in a
wider context of the statistical models of fully developed turbulence and
critical dynamics.

Another scope of the paper is to discuss the nature and convergence
properties of the $\eps$ expansion. The knowledge of the three terms
allows one to obtain reasonable predictions for finite values of
$\eps\sim1$ and to compare them with the existing nonperturbative
results: analytical and numerical solutions of the zero-mode equations
\cite{Falk1,Pumir} and numerical simulations \cite{VMF1,VMF2,MM}.

The plan of the paper is as follows. In Sec.~\ref{sec:RG} we recall
the field theoretic formulation of the model, diagrammatic technique,
renormalization and RG equations. In Sec.~\ref{sec:OPE}, we briefly
discuss the OPE, renormalization of composite operators (\ref{Fnp})
and their relationship to the issue of anomalous scaling. Since the
``ideology''  of the RG and OPE approach to the model (\ref{1})--(\ref{3})
is explained in Refs. \cite{RG,RG1,RG2,RG3} in detail, here we confine
ourselves to only the necessary information. In Sec.~\ref{sec:Operators},
we present the general scheme of the calculation of the critical dimensions
of the operators $F_{nl}$. In Sec.~\ref{sec:Oneloop}, the calculation in
the one-loop and two-loop approximations is presented in great detail.
Section~\ref{sec:Three} is devoted to the three-loop calculation;
some results for the relevant quantities are given in the Appendix.
In Sec.~\ref{sec:Inverse} we discuss the convergence of the $\eps$
expansion, the improved inverse $\eps$ expansion and comparison with
existing nonperturbative results.
The main ideas of the paper are briefly reviewed in the Conclusion.

\section{Field theoretic formulation of the model, diagrammatic
technique, renormalization, and RG equations}   \label{sec:RG}

The stochastic problem (\ref{1})--(\ref{3}) is equivalent
to the field theoretic model of the set of three fields
$\Phi\equiv\{\theta, \theta',{\bf v}\}$ with action functional
\begin{equation}
{\cal S}(\Phi)=\theta' D_{\theta}\theta' /2+
\theta' \left[ - \partial_{t} + \nu _0\partial^{2} -(v\partial) \right]
\theta -{\bf v} D_{v}^{-1} {\bf v}/2.
\label{action}
\end{equation}
The first four terms in (\ref{action}) represent the
Martin--Siggia--Rose-type action for the stochastic
problem (\ref{1}), (\ref{2}) at fixed ${\bf v}$, and the last term
represents the Gaussian averaging over ${\bf v}$. Here $D_{\theta}$
and $D_{v}$ are the correlators (\ref{2}) and (\ref{3}), respectively,
the required integrations over $x=(t,{\bf x})$ and summations over
the vector indices are understood.

The model (\ref{action}) corresponds to a standard Feynman diagrammatic
technique with the bare propagators $vv$, $\theta\theta$, $\theta\theta'$
(the line $\theta'\theta'$ is absent).
In the diagrams, these propagators are represented by the lines:
\begin{equation}
\langle vv \rangle_{0}
=\raisebox{-0.20cm}{\psfig{file=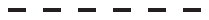,height=0.5cm,width=1.5cm}},
\quad \langle \theta\theta \rangle_{0}
=\raisebox{-0.20cm}{\psfig{file=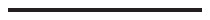,height=0.5cm,width=1.5cm}},
\quad \langle \theta\theta' \rangle_{0}
=\raisebox{-0.20cm}{\psfig{file=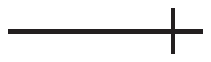,height=0.5cm,width=1.5cm}},
\label{linesD}
\end{equation}
the slashed end of a solid line corresponds to the field $\theta'$, the end
without a slash corresponds to $\theta$. The triple vertex
$V(\Phi)= -\theta'v_{j}\partial_{j}\theta = \partial_{j}\theta'\cdot v_{j}
\theta$ (the equality holds due to the integration over ${\bf x}$
in Eq. (\ref{action})) is represented as
\begin{equation}
V(\Phi) = -\,\,
\raisebox{-0.68cm}{\psfig{file=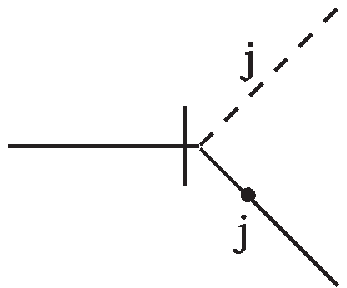,height=1.5cm}} =
\raisebox{-0.70cm}{\psfig{file=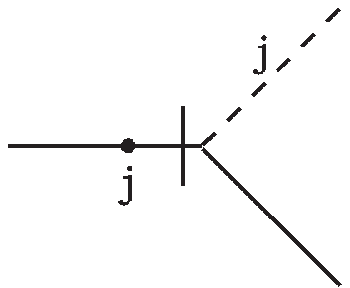,height=1.5cm}}.
\label{vertex}
\end{equation}
The dot with the index $j$ on the solid line denotes the differentiation
$\partial/\partial x_{j}$ with respect to the argument ${\bf x}$ of the
end of the line attached to the vertex; the index $j$ of the derivative
is contracted with the index of the end of the line $vv$ attached to the
vertex. Owing to the transversality of the $vv$ line, the dot can be moved
onto another line, as shown in Eq.~(\ref{vertex}). In the momentum
representation the vertex (\ref{vertex}) corresponds to the factor
$- {\rm i}k_{j} $, or, equivalently, $+{\rm i}q_{j}$, where ${\bf k}$
and ${\bf q}$ are the momenta flowing into the vertex via
the fields $\theta'$ and $\theta$,
respectively. The sum of the three momenta flowing into the vertex via
the fields $\theta,\theta',v$ is equal to zero.

The line $vv$ in the diagrams corresponds to correlation function (\ref{3}),
and the lines $\theta\theta'$ and $\theta\theta$ in
model (\ref{action}) in the $\omega,\k$ representation correspond to
the bare propagators
\begin{eqnarray}
\bigl\langle \theta\theta' \bigr\rangle _0 =
(-{\rm i}\omega + \epsilon_{\k})^{-1}, \qquad
\bigl\langle \theta\theta \bigr\rangle _0 = C(\k)\,
(\omega^{2}+ \epsilon_{\k}^{2})^{-1}, \qquad \epsilon_{\k}\equiv \nu_0
k^{2},
\label{lines}
\end{eqnarray}
where $C(\k)$ is the Fourier transform of the function $C$ from
Eq.~(\ref{2}) and $\omega$ and $\k$ ``flow via the line from the left to
the right''  if the standard form of the Fourier transform with respect to
the coordinate and time differences is used.

In what follows, we shall work in the $t,\k$ representation, where
the propagators (\ref{lines}) have the forms
\begin{eqnarray}
\bigl\langle \theta \theta' \bigr\rangle _0 =
\Theta(t-t') \exp \bigl\{ - (t-t')\epsilon_{\k} \bigr\},
\qquad
\bigl\langle \theta \theta\bigr\rangle _0
= \bigl\{C(\k)/2\epsilon_{\k}\bigr\}\, \exp \bigl\{ - |t-t'| \,
\epsilon_{\k} \bigr\};
\label{Lines}
\end{eqnarray}
in $\langle \theta \theta' \rangle _0$, $t$ is the time argument of
$\theta$ and $t'$ is the argument of $\theta'$.

The model (\ref{action}) is logarithmic (the coupling constant $g_{0}$
in Eq. (\ref{Lambda}) is dimensionless) at $\eps=0$, and
the UV divergences have the form of the poles in $\eps$ in the correlation
functions of the fields $\theta$, $\theta'$.
Superficial UV divergences, whose removal requires counterterms,
are present only in the 1-irreducible function
$\bigl\langle \theta'\theta\bigr\rangle_{\rm 1-ir}$,
and the corresponding counterterm
reduces to the form $\theta'\partial^{2}\theta$; see Ref. \cite{RG}.
Thus for the complete elimination of the UV divergences it is
sufficient to perform the multiplicative renormalization of the
parameters $\nu_0$ and $g_{0}=D_{0}/\nu_0$
with the only independent renormalization constant $Z_{\nu}$:
\begin{equation}
\nu_0=\nu Z_{\nu}, \qquad g_{0}=g\mu^{\eps}Z_{g},
\qquad Z_{g}=Z_{\nu}^{-1} \qquad (D_{0} = g_{0}\nu_0 = g\mu^{\eps} \nu).
\label{18}
\end{equation}
Here $\mu$ is the renormalization mass in the minimal subtraction (MS)
scheme, which we always use in what follows, $g$ and $\nu$
are renormalized analogs of the bare parameters $g_{0}$ and $\nu_0$,
and $Z=Z(g,\eps,d)$ are the renormalization constants.
In the MS scheme they have the form ``1 + only poles in $\eps$.''
The last relation in (\ref{18}) results from the absence of renormalization
of the contribution with $D_{v}$ in (\ref{action}).

The renormalized action is obtained from the functional (\ref{action})
by the substitution (\ref{18}) and has the form
\begin{equation}
{\cal S}_{R}(\Phi) =\theta' D_{\theta}\theta' /2+ \theta' \left[
- \partial_{t} + \nu Z_{\nu}\partial^{2} -(v\partial) \right] \theta
-{\bf v} D_{v}^{-1} {\bf v}/2,
\label{renact}
\end{equation}
where the amplitude $D_{0}= g\mu^{\eps} \nu$ from Eq. (\ref{3})
is expressed in renormalized parameters using Eqs.~(\ref{18}).

The exact response function $G\equiv\bigl\langle \theta
\theta'\bigr\rangle$ satisfies the standard Dyson equation, which
in the $\omega,\k$ representation has the form
\begin{equation}
G^{-1}(\omega, \p)= -{\rm i}\omega +\nu_0 p^{2} -
\Sigma_{\theta'\theta} (\omega, \p),
\label{Dyson}
\end{equation}
where the self-energy operator $\Sigma_{\theta'\theta}$ in the
diagrammatic notation (\ref{linesD}), (\ref{vertex}) is represented as
follows:
\begin{equation}
\Sigma_{\theta'\theta}  =
\raisebox{-0.34cm}{\psfig{file=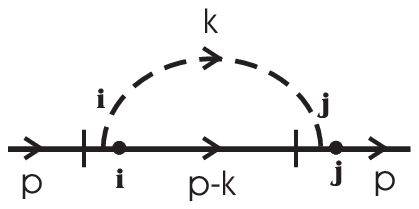,height=1.6cm,width=3cm}}\,\,.
\label{Sigma}
\end{equation}

The multiloop diagrams, which could be added on the right-hand side of
Eq. (\ref{Sigma}), contain effectively closed circuits of retarded
propagators $\langle\theta\theta'\rangle_{0}$ and therefore vanish;
it is crucial here that the correlation function $\langle vv\rangle$
in Eq. (\ref{3}) is proportional to the $\delta$ function in time.
Therefore, the self-energy operator is given by the one-loop approximation
exactly; it is independent of $\omega$ and has the form
\begin{equation}
\Sigma_{\theta'\theta}(\p) =
- \frac {D_{0}\, p_{i} p_{j}}{2(2\pi)^{d}}
\int d\k \, P_{ij}({\bf k})\, N_{k} =
- \frac{D_{0}\,(d-1)\,p^{2}}{2d(2\pi)^{d}}  \,  \int d\k  \,N_{k}
\equiv
- \frac{D_{0}\,(d-1)\,p^{2}}{2d} \,U,
\label{Sigma1}
\end{equation}
where
\begin{equation}
U \equiv (2\pi)^{-d} \int d\k N_{k} = C_{d} \int_{m}^{\infty} dk
k^{-1-\eps} = C_{d} \, m^{-\eps} /\eps,
\qquad  C_{d} \equiv S_{d} / (2\pi)^{d}
\label{V38}
\end{equation}
with $N_{k}$ from Eq. (\ref{3}).
The quantity $S_d = 2\pi ^{d/2}/\Gamma (d/2)$ is the surface area of
the unit sphere in $d$-dimensional space. In Eq. (\ref{Sigma1})
we have used the standard convention $\Theta(t-t')=1/2$ at $t=t'$ for the
step function in the $\theta\theta'$ line and the relation
$ \bigl\langle P_{ij}(\k) \bigr\rangle =\delta_{ij}\, (d-1)/d$ for the
angular averaging of the transverse projector in $d$ dimensions.

From the Dyson equation and Eqs.~(\ref{Sigma1}) and (\ref{V38}) it
follows that the exact response function $\langle\theta\theta'\rangle$
is obtained from its bare counterpart $\langle\theta\theta'\rangle_{0}$
in Eqs. (\ref{lines}), (\ref{Lines}) simply by the replacement
\begin{equation}
\nu_{0} \to \nueff \equiv \nu_{0} +  C_{d}\, (d-1)\, D_{0} \, m^{-\eps}
/ 2d\eps.
\label{nueffective}
\end{equation}
The renormalization constant $Z_{\nu}$ in Eq. (\ref{18}) can be found
exactly from the requirement that the ``effective diffusivity'' $\nueff$
be UV finite in renormalized theory (\ref{renact}), i.e., have no poles
in $\eps$ when expressed in renormalized variables (\ref{18}).
In the MS scheme this gives
\begin{equation}
Z_{\nu} = 1- u\,(d-1)/2d\eps, \qquad
u \equiv g\, C_{d}, \qquad  C_{d} = S_{d} / (2\pi)^{d},
\label{Z}
\end{equation}
where we have changed to the more convenient coupling constant $u$.
In the renormalized variables, $\nueff$ is given by the expression
\begin{equation}
\nueff = \nu \left[1+ \frac{u(d-1)}{2d} \cdot \frac{(\mu/m)^{\eps}-1}
{\eps} \right],
\label{nueffren}
\end{equation}
which is obviously finite at $\eps\to0$.

The basic RG equation for a multiplicatively renormalizable quantity
$F=Z_{F}F_{R}$ (correlation function, composite operator {\it etc})
has the form
\begin{equation}
\bigl[{\cal D}_{RG}+ \gamma_{F} \bigr] F_{R}=0, \qquad
{\cal D}_{RG} = {\cal D}_{\mu} + \beta \partial_{u}
- \gamma_{\nu}{\cal D}_{\nu}.
\label{RGE}
\end{equation}
Here and below, ${\cal D}_{x}\equiv x\partial_{x}$ for any variable $x$,
and the RG functions (the $\beta$ function and the anomalous dimensions
$\gamma$) are defined as
\begin{equation}
\beta \equiv \Dm u, \qquad
\gamma_{F} \equiv \Dm \ln Z_{F} = \beta \partial_{u} \ln Z_{F}
\quad  {\rm for\ any\ } Z_{F},
\label{RGF1}
\end{equation}
where $\Dm$ denotes the operation $\mu\partial_{\mu}$ at fixed bare
parameters $g_{0}$, $\nu_{0}$. From the definitions, the last relation
in (\ref{18}) and exact expression (\ref{Z}), for the basic RG functions
one obtains:
\begin{equation}
\beta(u) = u\, [-\eps + \gamma_{\nu}], \qquad \gamma_{\nu} = u\,(d-1)/2d.
\label{RGF2}
\end{equation}
From Eq. (\ref{RGF2}) it follows that the RG equations (\ref{RGE})
possess an IR stable positive fixed point:
\begin{equation}
u_{*}= 2d\eps/(d-1), \qquad \beta(u_{*})=0, \qquad \beta'(u_{*})= \eps >0.
\label{FP}
\end{equation}
This fact implies that correlation functions of model (\ref{1})--(\ref{3})
in the IR region ($\Lambda r \gg 1$, $mr \sim 1$) exhibit scaling behavior;
the corresponding critical dimensions $\Delta[F]\equiv\Delta_{F}$ can be
calculated as series in $\eps$. For the basic fields and quantities,
including the composite operators $\theta^{n}$, the dimensions are found
exactly \cite{RG}:
\begin{equation}
\Delta_{\omega}=2-\eps,\qquad \Delta_{m} = 1, \qquad
\Delta _{\theta} = (-1+\eps/2), \qquad
\Delta[\theta^{n}] = n \Delta _{\theta}, \qquad
\Delta_{\theta'} = d+1-\eps/2
\label{critical}
\end{equation}
(no corrections of order $\eps^{2}$ and higher). This is a consequence of the
exact equality $\gamma_{\nu}(u_*)= \eps$, which follows from Eqs.
(\ref{RGF2}) and (\ref{FP}).

In particular, for the structure functions (\ref{struc}) relations
(\ref{critical}) along with dimensionality considerations give:
\begin{equation}
S_{n}({\bf r})= D_{0}^{-n/2}\,  r^{n(1-\eps/2)}\, \xi_{n}(mr),
\label{100}
\end{equation}
with some nontrivial dependence on the IR scale $m$ contained
in the scaling functions $\xi_{n}(mr)$.

\section{Composite operators, operator product expansion, and anomalous
scaling} \label{sec:OPE}

Representations of the form (\ref{100}) for any scaling functions $\xi(mr)$
describe the behavior of the correlation functions for $\Lambda r \gg 1$ and
any fixed value of $mr$. The inertial range corresponds to the additional
condition $mr \ll 1$. The form of the functions $\xi(mr)$ at $mr\to0$ is
studied using the operator product expansion (OPE).

According to the OPE, the behavior of the quantities entering into
the right-hand side of Eq. (\ref{struc}) for
${\bf r} = {\bf x} - {\bf x'} \to 0$  and fixed
$ {\bf x} + {\bf x'} $ is given by the infinite sum
\begin{equation}
\Bigl[ \theta_{r} (t,{\bf x}) - \theta_{r} (t,{\bf x'})\Bigr]^{n}
= \sum_{F} C_{F} ({\bf r})\,
F\left(t,\, \frac{{\bf x}+{\bf x'}}{2} \right),
\label{OPE}
\end{equation}
where $C_{F}$ are coefficients regular in $m^{2}$ and $F$ are all possible
renormalized local composite operators allowed by the symmetry (more
precisely, see below).

In what follows, it is assumed that the expansion (\ref{OPE}) is made in
irreducible tensors (scalars, vectors and traceless tensors); the possible
tensor indices of the operators $F$ are contracted with the corresponding
indices of the coefficients $C_{F}$. With no loss of generality, it can also
be assumed that the expansion is made in ``scaling'' operators, i.e., those
having definite critical dimensions $\Delta_{F}$.

The structure functions (\ref{struc}) are obtained by averaging Eq.
(\ref{OPE}) with the weight $\exp {\cal S}_{R}$, the mean values
$\langle F\rangle$ appear on the right-hand side. Their asymptotic
behavior for $m\to0$ is found from the corresponding RG equations
(\ref{RGE}) and has the form $\langle F\rangle \propto  m^{\Delta_{F}}$.

From the RG representation (\ref{100}) and the operator product expansion
(\ref{OPE}) we therefore find the following expression for the structure
functions in the inertial range ($\Lambda r\gg1$, $mr\ll1$):
\begin{equation}
S_{n}({\bf r})= D_{0}^{-n/2}\, r^{n(1-\eps/2)} \,
\sum_{F} A_{F}(m{\bf r})\, (mr)^{\Delta_{F}} ,
\label{OR}
\end{equation}
with coefficients $A_{F}$ regular in $m^2$.

Some general remarks are now in order.

Owing to translational invariance, the operators having the form of total
derivatives give no contribution to Eq. (\ref{OR}): $\langle \partial F(x)
\rangle = \partial\langle  F(x) \rangle =\partial\times\, {\rm const}=0$.

In model (\ref{1})--(\ref{3}), the operators with an odd number of
fields $\theta$ also have vanishing mean values; their contributions
vanish along with the odd structure functions themselves (they will
be ``activated'' in the presence of a nonzero mixed correlation function
$\langle{\bf v}f\rangle$ or an imposed gradient of the scalar field).

If the function $C$ in Eq. (\ref{2}) depends only on $r=|{\bf r}|$,
the model becomes $SO(d)$ covariant and only the contributions of
scalar operators enter into Eq. (\ref{OR}). Indeed, in the isotropic
case the tensor indices of the
mean values $\langle F\rangle$ of the operators $F$ in Eq. (\ref{OPE}) can
only be those of Kronecker delta symbols. It is impossible, however, to
construct nonzero irreducible (traceless) tensor solely of the delta symbols.

In the presence of anisotropy, irreducible tensor operators acquire
nonzero mean values and their contributions appear on the right-hand
side of Eq. (\ref{OR}). In the simplest case of the uniaxial anisotropy,
specified by an unit vector ${\bf n}$ in the correlation function (\ref{2}),
the mean value of a $l\,$th rank traceless operator is necessarily
proportional to the $l\,$th rank symmetrical traceless
tensor built of the vector ${\bf n}$ along with the delta symbols;
its contraction with the corresponding coefficient $A_{F}$ gives rise
to the $l\,$th order Legendre polynomial $P_{l}(z)$ with
$z= ({\bf r}{\bf n})/r$. In general, the expansion in irreducible tensors
in (\ref{OPE}) after the averaging leads to the decomposition in the
irreducible representations of $SO(d)$, the $l\,$th sector corresponds
to the contribution of the $l\,$th rank composite operators.

The leading term in the $l\,$th anisotropic sector is given by the
$l\,$th rank tensor operator with minimal dimension $\Delta[F]$.
The feature typical to the models describing turbulence is the
existence of composite operators with {\it negative} critical dimensions;
their contributions in the OPE lead to singular behavior of
the scaling functions at $mr\to0$, that is, to the anomalous scaling.
The operators with minimal $\Delta_F$ are those involving the maximal
possible number of fields $\theta$ and the minimal possible number of
derivatives (at least for small $\eps$).
Both the problem (\ref{1})--(\ref{3}) and the quantities
(\ref{struc}) possess the symmetry $\theta\to\theta+{\rm const}$.
It then follows that the expansion (\ref{OPE}) involves only operators
invariant with respect to this shift and therefore built of the
{\it gradients} of $\theta$.

In general, the operators entering into the right-hand side of Eq.
(\ref{OPE}) are those which appear in the Taylor expansion, and those
that admix to them in renormalization. The leading term of the
Taylor expansion for $S_{n}$ is the $n\,$th rank operator which can
symbolically be written as $(\partial\theta)^{n}$; its decomposition in
irreducible tensors gives rise to operators of lower ranks.
In the presence of the noise (\ref{2}), operators of the form
$(\partial\theta)^{k}$ with $k <n$ admix to them in renormalization
and also appear in the OPE. Owing to the linearity of problem (\ref{1}),
operators with $k >n$ (whose contributions would be more important)
do not admix to the terms of the Taylor expansion
for $S_{n}$ and do not appear in the corresponding OPE.

We thus conclude that the leading terms of the inertial-range behavior
are related to the critical dimensions $\Delta_{nl}$ of the infinite
family of irreducible tensor composite operators $F_{nl}$ introduced in
Eq.~(\ref{Fnp}).

In general, operators (\ref{Fnp}) mix in renormalization. One can show that
the corresponding infinite renormalization matrix
\begin{equation}
F_{nl} = \sum_{n'l'} \, Z_{nl,n'l'} \,F^{R}_{n'l'}
\label{Matrix}
\end{equation}
is in fact block-triangular, i.e., $Z_{nl,n'l'} =0$ for $n'> n$,
and so are the matrix of anomalous dimensions $\gamma(u)= Z^{-1}\Dm Z$
and the matrix of critical dimensions $\Delta=n+n \Delta_{\theta}
+ \gamma(u_{*})$. It is then obvious that the dimensions $\Delta_{nl}$,
given by the eigenvalues of the latter matrix, are completely determined
by the finite subblocks with $n'= n$. Therefore, we can neglect all the
elements of the matrix (\ref{Matrix}) other than $Z_{nl,nl'}$. The latter
are determined by the 1-irreducible correlation functions with one
operator $F_{nl}$ and $n$ fields $\theta$. The diagrams for such functions
do not involve the propagator $\langle\theta\theta\rangle_{0}$ from
Eq. (\ref{lines}) and can therefore be calculated directly in the
``unforced'' model without the noise (\ref{2}), that is, without
the first terms in the action functionals (\ref{action}), (\ref{renact}).
In the absence of the forcing, the model becomes $SO(d)$ covariant, the
irreducible operators with different values of $l$ cannot mix in
renormalization, so that the blocks $Z_{nl,nl'}$ appear diagonal.

We thus conclude that the critical dimensions $\Delta_{nl}$ are determined
by the diagonal elements $Z_{nl} \equiv Z_{nl,nl}$ of the matrix
(\ref{Matrix}):
\begin{equation}
\Delta_{nl}=n+n \Delta_{\theta}+ \gamma_{nl} (u_{*})=
n\eps/2+ \gamma_{nl} (u_{*}) ,  \qquad
\gamma_{nl} (u) = \Dm \ln Z_{nl}= \beta \partial_{u} \ln Z_{nl}
\label{Matrix2}
\end{equation}
with $\beta$ from (\ref{RGF1}), $u_{*}$ from (\ref{FP}) and $\Delta_{\theta}$
from (\ref{critical}). Owing to the renormalization, the critical dimension
$\Delta_{nl}$ is not equal to the simple sum of critical dimensions
$\Delta_{\theta}=(-1+\eps/2)$, $\Delta_{\partial}=1$ of the fields and
derivatives constituting $F_{nl}$. The elements $Z_{nl}$ can be calculated
in the model without forcing, in which the operators (\ref{Fnp}) are
renormalized multiplicatively: $F_{nl}=Z_{nl}F_{nl}^{R}$.

\section{Calculation of the critical dimensions of operators
$F_{\lowercase{nl}}$: General scheme} \label{sec:Operators}

From now on, we shall consider composite operators (\ref{Fnp}) in the model
without the noise, that is, with $D_{\theta}=0$ in the action functional
(\ref{action}). They are renormalized multiplicatively,
$F_{nl}=Z_{nl}F_{nl}^{R}$, and the renormalization constants
$Z_{nl}=Z_{nl}(g,\eps,d)$ are determined by the
requirement that the 1-irreducible correlation function
\begin{equation}
\bigl\langle F_{nl}^{R} (x) \theta(x_{1})\dots\theta(x_{n})
\bigr\rangle_{\rm 1-ir}= Z_{nl}^{-1}\bigl\langle
F_{nl}(x) \theta(x_{1})\dots\theta(x_{n})\bigr\rangle
_{\rm 1-ir} \equiv Z_{nl}^{-1}\Gamma_{nl} (x;x_{1},\dots, x_{n})
\label{req}
\end{equation}
be UV finite in renormalized theory (\ref{renact}), i.e., have no poles
in $\eps$ when expressed in renormalized variables
(\ref{18}). This is equivalent to the UV finiteness of the product
$Z_{nl}^{-1}\Gamma_{nl}(x;\theta)$, in which
\begin{equation}
\Gamma_{nl} (x;\theta) = \frac{1}{n!}\, \int d x_{1}\dots \int d x_{n}\,
\Gamma_{nl} (x;x_{1},\dots, x_{n})\, \theta(x_{1})\dots\theta(x_{n})
\label{gena}
\end{equation}
is a functional of the field $\theta(x)$. In the zeroth approximation, the
functional (\ref{gena}) coincides with the operator $F_{nl}(x)$, and in
higher orders the kernel $\Gamma_{nl} (x;x_{1},\dots, x_{n})$ is given by
the sum of diagrams shown in Fig.~1. The analysis of the diagrams shows
that for any argument $x_{s}$, the corresponding spatial derivative
is isolated as an external factor from each diagram. Using the integration
by parts, these derivatives can be moved onto the corresponding fields
$\theta(x_{s})$ in Eq. (\ref{gena}), so that the quantity (\ref{gena}) can
be represented as a functional of the vector field
$w_{i} \equiv \partial _{i} \theta $:
\begin{equation}
\Gamma_{nl} (x;\theta) = \frac{1}{n!}\, \int d x_{1}\dots \int d x_{n}\,
\widetilde \Gamma_{nl}^{i_{1}\dots i_{n}} (x;x_{1},\dots, x_{n})\,
w_{i_{1}}(x_{1}) \dots w_{i_{n}}(x_{n}).
\label{gena2}
\end{equation}

The diagrams that determine the kernel $\widetilde \Gamma$ in Eq.
(\ref{gena2}) contain only logarithmic UV divergencies. Therefore, in order
to find the constant $Z_{nl}^{-1}$ it is sufficient to calculate the
functional $\widetilde \Gamma$ with some special choice of its functional
argument $w_{i}$, namely, one can replace it by its value at the fixed
point $x$, the argument of the operator $F_{nl}$ in Eqs. (\ref{req}). Now
the product $w_{i_{1}}(x)\dots w_{i_{n}}(x)$ can be taken outside
the integrals over $x_{1},\dots, x_{n}$ in (\ref{gena2}), so that the
functional $\Gamma_{nl} (x;\theta)$ turns to a local composite operator. The
integration of the remaining function $\widetilde \Gamma_{nl}$ over
$x_{1},\dots, x_{n}$ gives a quantity independent of any coordinate variables,
and its vector indices can only be those of Kronecker delta symbols. Their
contraction with the indices of the product $w_{i_{1}}(x)\dots w_{i_{n}}(x)$
gives rise to the original operator $F_{nl}(x)$ with some scalar coefficient
$\overline\Gamma$. The integration over $x_{1},\dots, x_{n}$ means that in
the Fourier representation, the corresponding correlation function is
calculated with all its momenta set equal to zero, which is always implied
in what follows.

Now we turn to the derivation of practical formulas for the calculation
of the constants $Z_{nl}^{-1}$ from the diagrams. For the sake of brevity,
we introduce the notation:
\begin{equation}
Z_{nl} \equiv Z_{F}, \qquad F_{nl}(x) \equiv F, \qquad
\Gamma_{nl}(x;\theta) \equiv \Gamma = F \overline\Gamma.
\label{brev}
\end{equation}
Then the UV finiteness of the quantity $Z_{nl}^{-1}\Gamma_{nl}(x;\theta)$
is expressed by the relation
\begin{equation}
{\rm Div.\, p.}\, \left[ Z_{F}^{-1}\, \Gamma \right] =0,
\label{1*}
\end{equation}
where ``Div.\,p.'' is the operation which selects the UV divergent part;
in the MS scheme, it selects only poles in $\eps$. Classifying the
contributions in model (\ref{renact}) according to the powers of the
renormalized coupling constant $u$ from Eq. (\ref{Z}) gives:
\begin{equation}
Z_{F}^{-1}= 1+ \sum _{k=1}^{\infty} \left[ Z^{-1}_{F} \right]_{k} ,
\qquad \Gamma = \sum _{k=0}^{\infty} \Gamma_{k}
=F+F\, \sum _{k=1}^{\infty} \overline\Gamma_{k},
\label{useries}
\end{equation}
where $\left[ Z^{-1}_{F} \right]_{k}$, $\Gamma_{k}$ and
$\overline\Gamma_{k}$ are the contributions of order $u^{k}$ in
the respective quantities.
We substitute expressions (\ref{useries}) into Eq. (\ref{1*}) and omit the
overall factor $F$; this gives:
\begin{equation}
{\rm Div.\, p.}\, \left[ Z^{-1}_{F}  + Z^{-1}_{F}
\, \sum _{k=1}^{\infty} \overline\Gamma_{k} \right]=0.
\label{3*}
\end{equation}
We recall that in the MS scheme one has ``$Z^{-1}_{F} = 1+$ only poles in
$\eps$,'' so that $ {\rm Div.\, p.}\,\left\{ Z^{-1}_{F}\right\}
 = Z^{-1}_{F}-1$.
Substituting this equality into Eq. (\ref{3*}) gives the relation
\begin{equation}
Z^{-1}_{F} = 1- {\rm Div.\, p.}\, \left[ Z^{-1}_{F}
\, \sum _{k=1}^{\infty} \overline\Gamma_{k} \right],
\label{4*}
\end{equation}
which allows for the recurrent calculation of the contributions
$\left[ Z^{-1}_{F} \right]_{k}$ in the expansion (\ref{useries}) from the
quantities $\overline\Gamma_{k}$. Indeed, selecting in Eq. (\ref{4*}) terms
of the same order in $u$ gives:
\begin{mathletters}
\label{deltaZ}
\begin{eqnarray}
\left[Z^{-1}_{F}\right]_{1} &=&- {\rm Div.\, p.}\,\Bigl\{
\overline\Gamma_{1}
\Bigr\},
\label{deltaZ1} \\
\left[ Z^{-1}_{F} \right]_{2} &=& - {\rm Div.\, p.}\, \Bigl\{
\overline\Gamma_{2}+\left[ Z^{-1}_{F} \right]_{1} \overline\Gamma_{1}
\Bigr\},
\label{deltaZ2}  \\
\left[ Z^{-1}_{F} \right]_{3} &=& - {\rm Div.\, p.}\, \Bigl\{
\overline\Gamma_{3}+\left[ Z^{-1}_{F} \right]_{1} \overline\Gamma_{2}
+\left[ Z^{-1}_{F} \right]_{2} \overline\Gamma_{1}
 \Bigr\},
\label{deltaZ3}
\end{eqnarray}
\end{mathletters}
and so on. The relations given in Eq. (\ref{deltaZ}) are sufficient for the
three-loop calculation.

\begin{figure}
\centerline{ {\LARGE $\Gamma^{\scriptscriptstyle(1)}=$}
\raisebox{-8.5ex}{\psfig{file=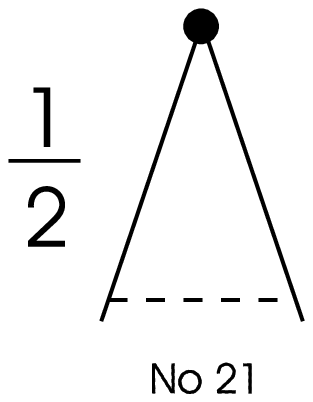,height=15ex}}\quad
{\LARGE ,} \hspace{4em} {\LARGE$\Gamma^{\scriptscriptstyle(2)}=$}
\raisebox{-8.5ex}{\psfig{file=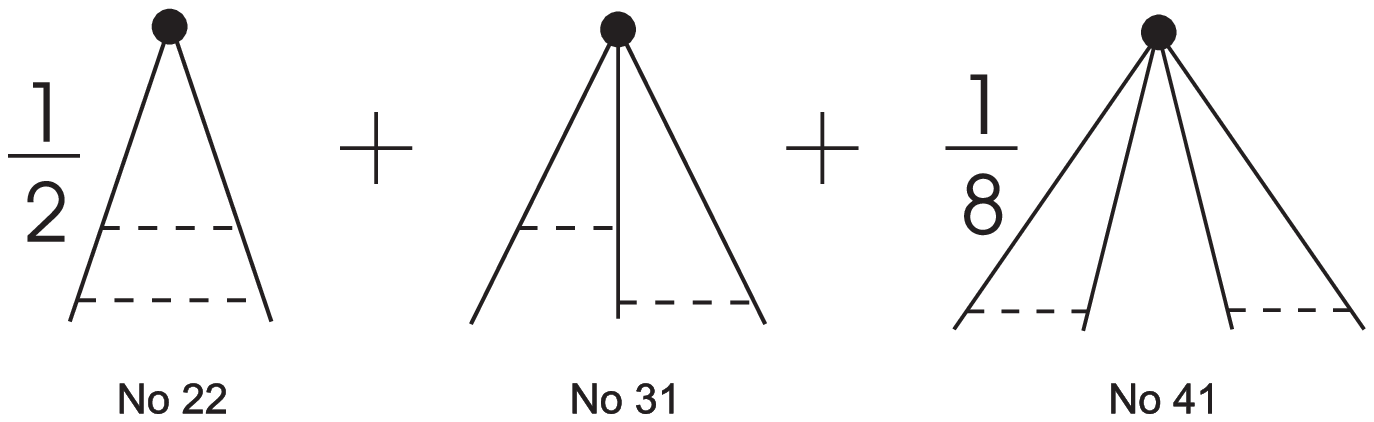,height=15ex}}\quad
{\LARGE ,} } \vspace{2ex} \centerline{
{\LARGE$\Gamma^{\scriptscriptstyle(3)}=$}
\raisebox{-43.5ex}{\psfig{file=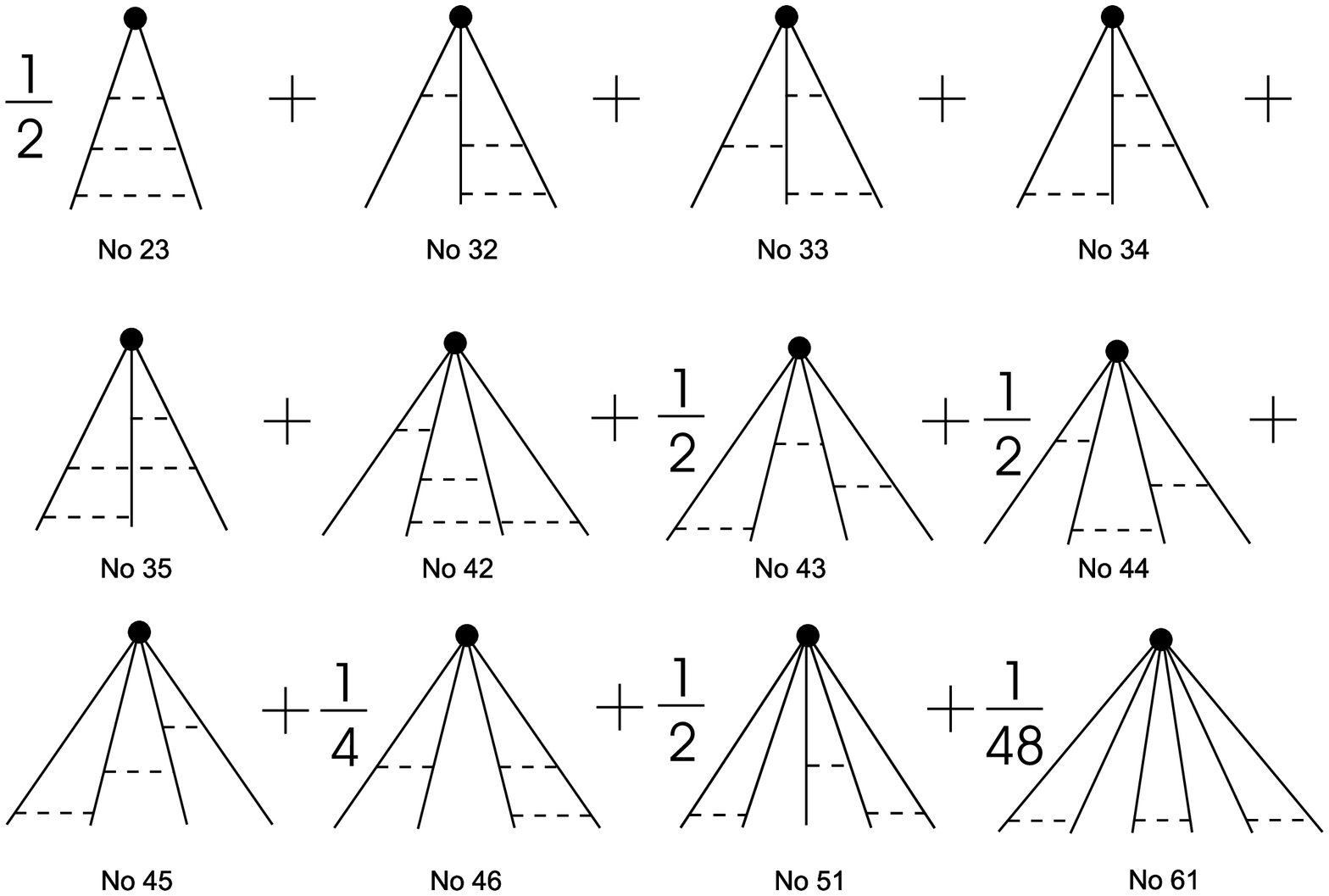,height=50ex}}\quad
\raisebox{-36ex}{\LARGE .}}
\bigskip
\caption{Diagrammatic representation of the function $\Gamma$
in the three-loop approximation.
\label{fig1}}
\end{figure}

In Fig.~1, we present, along with respective symmetry coefficients, all the
diagrams needed for the three-loop calculation of the function $\Gamma$,
except for those with the self-energy insertions of the form (\ref{Sigma})
in the $\theta\theta'$ lines. (The symmetry coefficients which are not shown
are equal to unity.) The index $l$ in $\Gamma^{(l)}$ denotes the number of
loops (that is, independent integration momenta): $l=1,2,3$. Below each
diagram we
give its number, which will always be used in the following to refer to a
diagram. The numbers have the forms ``No~XY,'' where X is the number of
rays in the diagram and Y is the order number (from the left to the right
in Fig.~1) of the diagram in the subset with given X.

The thick dots in the diagrams correspond to the vertices of the
composite operator $F$ (more precisely, see below), all the
horizontal dashed lines
correspond to correlation functions $\langle vv\rangle$ from Eq.
(\ref{3}) with $D_{0}=g\nu\mu^{\eps}$, and the ``rays'' correspond to
chains of lines $\langle\theta\theta'\rangle_{0}
\langle\theta\theta'\rangle_{0}\dots$ in the same order on each ray:
the upper end of each line corresponds to the field $\theta$ and the lower
end corresponds to $\theta'$. For this reason, in contrast with Eq.
(\ref{Sigma}), in Fig.~1 we do not add slashes at the ends of the lines.
If desired, they can easily be restored: a slash should be
added in the lower end of each line belonging to a ray. The lowest
``external lines'' of the diagrams in Fig.~1 correspond to the factors
$\theta$ (and not to the propagators). We also note that the diagrams in
Fig.~1 do not involve the $\theta\theta$ lines (even in the presence of the
noise).

In Fig.~1, we omitted the diagrams which are topologically possible
and would be needed for the general correlation function (\ref{3}),
but in our model with the delta function in time in Eq. (\ref{3}) they
vanish due to the presence of the closed circuits of retarded propagators
$\langle\theta\theta'\rangle_{0}$; cf. the remark in Sec.~\ref{sec:RG}
below Eq.~(\ref{Sigma}).

The contributions of the diagrams with the self-energy insertions
(\ref{Sigma}) will automatically be included if the propagators
$\langle\theta\theta'\rangle_{0}$ in Fig.~1 are taken to be exact, that is,
with $\epsilon_{\k} = \nueff k^{2}$ in Eqs. (\ref{lines}), (\ref{Lines}).
In the renormalized variables, $\nueff$ is given by Eq. (\ref{nueffren}),
the zero order approximation being $\nueff=\nu$. It is easy to see that the
parameter $\nu$ from $\epsilon_{\k} = \nu k^{2}$ enters into the final
answers for the diagrams as $\nu^{-l}$, where $l$ is the number of loops
in the diagram. It is thus sufficient to calculate the diagrams without the
self-energy insertions and with $\epsilon_{\k} = \nu k^{2}$ and then
introduce the additional factor
\begin{equation}
(\nueff/\nu)^{-l} = [1+Q]^{-l}, \qquad Q=
\frac{u(d-1)[(\mu/m)^{\eps}-1]}{2d\eps}
\label{6*}
\end{equation}
for any $l$-loop diagram; see Eq. (\ref{nueffren}). We stress that the
replacement $\nu\to\nueff$ is not needed in the amplitude
$D_{0}=g\nu\mu^{\eps}$ of the correlation function (\ref{3}).

Expression (\ref{nueffren}) corresponds to the special choice of the
function $N_{k}$ in Eq. (\ref{3}). If the specific form of the IR
regularization is different (for example, the function
$N_{k}=(k^{2}+m^{2})^{-d/2-\eps/2}$ was used in Refs. \cite{GK,RG}),
the relation (\ref{Sigma1}) for $\Sigma_{\theta'\theta}$ remains valid,
but the explicit form of the integral $U$ in Eq. (\ref{V38}) changes.
Then the quantity $Q$ in Eq. (\ref{6*}) is given by the following general
relation
\begin{equation}
Q= - \bigl[ \Sigma_{\theta'\theta} - {\rm Div.\, p.}\,\,
\Sigma_{\theta'\theta} \bigr]/\nu p^{2},
\label{7*}
\end{equation}
which recovers the expression (\ref{6*}) for $N_{k}$ from Eq. (\ref{3}).

Introduction of the additional factors (\ref{6*}) to the diagrams of
$\Gamma^{(l)}$ in Fig.~1 with $D_{0}=g\nu\mu^{\eps}$ in (\ref{3}) and
$\epsilon_{\k} = \nu k^{2}$ in (\ref{Lines}) for the quantities
$\overline\Gamma_{n}$ in Eq. (\ref{deltaZ}) gives
\begin{equation}
\overline\Gamma_{1}=\overline\Gamma^{(1)}, \qquad
\overline\Gamma_{2}=\overline\Gamma^{(2)}-Q\,\overline\Gamma^{(1)}, \qquad
\overline\Gamma_{3}=\overline\Gamma^{(3)}-2Q\,\overline\Gamma^{(2)} +
Q^{2}\overline\Gamma^{(1)}.
\label{8*}
\end{equation}
Expanding the quantity $Q$ from Eq. (\ref{6*}) in $\eps$ gives rise to
contributions with the logarithms $\ln (\mu/m)$; similar contributions also
come from the factors $(\mu/m)^{l\eps}$, which naturally arise in any
$l$-loop diagram in $\Gamma^{(l)}$. It is well known that the renormalization
constants in the MS scheme are independent of any mass parameters, like
$m$ and $\mu$ in the case at hand. This means that, after the substitution
of relations (\ref{8*}) into Eqs. (\ref{deltaZ}), all contributions with
$\ln (\mu/m)$ in $Z^{-1}_{F}$ will cancel each other. Such cancellation
provides a good possibility to control the absence of calculational errors.
In the two-loop calculation \cite{RG} we have checked the cancellation of
the contributions $\eps^{-1}\ln (\mu/m)$ in $\left[Z^{-1}_{F}\right]_{2}$.
There, the function $N_{k}$ was taken in the form different from (\ref{3})
(see above), the $\eps$ expansion of the corresponding quantity $Q$ in Eq.
(\ref{7*}) contained constant terms along with powers of the logarithms
$\ln (\mu/m)$, so that it was necessary to take into account the contribution
with $Q$ in expression (\ref{8*}) for $\overline\Gamma_{2}$.

Our present choice for $N_{k}$ in Eq. (\ref{3}) is much more convenient,
because the corresponding integral (\ref{V38}) contains only pole in
$\eps$, and the $\eps$ expansion of the corresponding quantity
$Q$ in Eq. (\ref{6*}) contains only powers of $\ln (\mu/m)$ with no
constant contributions. Since we know in advance that all contributions
with $\ln (\mu/m)$ in $Z^{-1}_{F}$ will cancel each other, we may simply
set $\mu=m$ in the calculation of this quantity. Then all the contributions
with $\ln (\mu/m)$ vanish, in Eq. (\ref{6*}) we obtain $Q=0$
(no corrections from the self-energy insertions are needed),
in Eq. (\ref{8*}) we obtain $\overline\Gamma_{l}=\overline\Gamma^{(l)}$,
and the factors $(\mu/m)^{l\eps}$ in the $l$-loop diagrams turn to unity.
In what follows, in the calculations of the diagrams in
$\overline\Gamma^{(l)}$ we shall retain the factors $(\mu/m)^{l\eps}$ and
replace them with unity only in the last step, that is, in the
calculation of $Z^{-1}_{F}$.

Let us turn to the vertex of the composite operator $F$, denoted in Fig.~1
by thick dots on the top of the diagrams. According to the general rules of
the universal diagrammatic technique (see, e.g., Ref.~\cite{book1}), for any
composite operator $F(x)$ built of the fields $\theta$, the vertex with
$k\ge0$ attached lines corresponds to the vertex factor
\begin{equation}
V_{k} (x;\, x_{1}, \dots, x_{k}) \equiv \delta^{k}
F(x) / {\delta\theta(x_{1}) \dots\delta\theta(x_{k})}.
\label{BigV}
\end{equation}
The arguments $x_{1}\dots x_{k}$ of the quantity (\ref{BigV}) are contracted
with the arguments of the upper ends of the lines $\theta\theta'$ attached to
the vertex. For our operators (\ref{Fnp}), built solely of the gradients
$w_{i}(x)=\partial \theta(x)/\partial x_{i}$ at a single spacetime point $x$,
the factors (\ref{BigV}) contain the product
$\partial_{i_1} \delta(x-x_{1}) \dots \partial_{i_k} \delta(x-x_{k})$,
and the integrations over $x_{1}\dots x_{k}$ are easily performed:
the derivatives move onto the upper ends of the corresponding lines
$\theta\theta'$ attached to the vertex (such derivatives we denote by dots
on the lines), and their arguments $x_{1}\dots x_{k}$ are substituted with
$x$. After the derivatives have been moved inside the diagram, the remaining
vertex factor for the operator $F(x)$ can be understood as an usual
derivative:
\begin{equation}
V_{i_{1}\dots i_{k}} (x) = \partial^{k} F(x) /
\partial w_{i_{1}} (x)\dots \partial w_{i_{k}} (x) .
\label{99}
\end{equation}

In what follows, in order to simplify the notation we shall omit the argument
$x$ and use the numerical indices $1,\dots, k$ instead of $i_{1}\dots i_{k}$.
Then the vertex factor (\ref{99}) for a $k$-ray diagram takes on the form
\begin{equation}
V_{12\dots} = \partial_{1} \partial_{2} \dots \, F \qquad  {\rm with} \quad
\partial_{i} \equiv \partial / \partial w _{i}.
\label{V2}
\end{equation}

Diagrams Nos 41,46,51,61 in Fig.~1 are ``factorizable'' in the sense that
they can be reduced to the products of blocks with a lesser number of rays.
All the other diagrams in Fig.~1 will be termed ``normal.''

\subsection{Scalarization of the diagrams} \label{sec:Scal}

The contribution of a specific diagram into the functional $\Gamma$ in
(\ref{gena2}) for any composite operator $F$, built of the gradients
$w_{i}=\partial_{i} \theta$, is represented in the form
\begin{equation}
\Gamma = V_{12\dots} \, I^{ab\dots}_{12\dots} \, w_{a} w_{b} \dots ,
\label{V1}
\end{equation}
where $V_{12\dots}$ is the vertex factor (\ref{V2}), $I^{ab\dots}_{12\dots}$
is the ``internal block''  of the diagram with free indices, the product
$w_{a} w_{b} \dots$ corresponds to external lines. The numerical indices
$1,2,\dots$ will always be understood as $i_{1}, i_{2}, \dots$, their number
in Eq. (\ref{V1}) equals to the number of the letter indices $a,b,\dots$ and
is determined by the number of ``rays,''  that is, the number of lines
that attach to the vertex of the operator. These lines are given by products
of the propagators $\langle \theta \theta' \rangle _0$ from Eq. (\ref{Lines})
with $\epsilon_{\k} =\nu k^{2}$ (see above) and are connected by the lines
$vv$ from Eq. (\ref{3}). As an example, we present the general form of
a four-ray diagram $\Gamma$:
\begin{equation}
\raisebox{-4.8ex}{\psfig{file=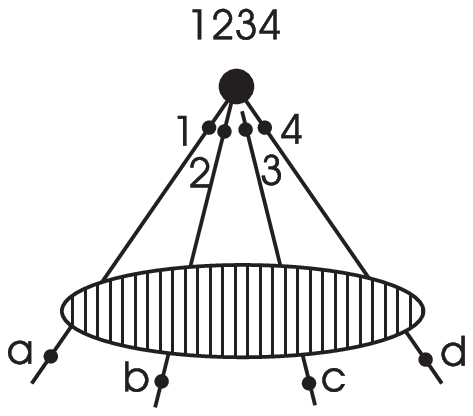,height=3cm,width=3cm}} =
V_{1234} \, I^{abcd}_{1234} \, w_{a} w_{b} w_{c} w_{d} .
\label{V3}
\end{equation}
The thick dot on the top represents the vertex of the composite operator,
it corresponds to the vertex factor $V_{1234}$ in Eq.~(\ref{V2}). The dots
on the lines denote the differential operation $\partial_{i}\equiv \partial
/ \partial x_{i}$; their indices are shown explicitly on the diagram.
The lower external lines of the diagram correspond to factors $\theta$;
the derivatives at the external vertices (\ref{vertex}), denoted by dots with
the indices $a,b,c,d$, act on these fields and turn them into the product
$w_{a}w_{b}w_{c}w_{d}$ with $w_{i}\equiv\partial\theta/\partial x_{i}$.
After all these differentiations have been performed, all the external
momenta in the diagrams are set to zero; the IR regularization is provided by
the parameter $m$ in the function $\langle vv \rangle $ from Eq.~(\ref{3}).

The diagrams are calculated in the time-momentum representation. The
integration momenta are assigned to all internal lines;
the number of independent momenta is equal to the number of $vv$ lines.
The dots with the numerical indices $1,2,\dots$ on the
upper lines in Eq.~(\ref{V3}) correspond to the vector factors $\pm {\rm i}
k_{s}$, coming from the vertex of the composite operator. Here
$\k$ is the integration momentum flowing via the line, with the index shown
near the dot ($s=1,2,3,4$); the coefficient equals $-{\rm i}$ if the momentum
``flows into the dot'' and $+{\rm i}$ if the momentum ``flows out.''
In general, similar factors are also present inside the diagram, that is,
inside the shaded block in Eq.~(\ref{V3}). Collecting all such factors
$\pm {\rm i}$ from the whole diagram gives certain ``sign factor'' $\pm 1$,
which, of course, should be taken into account in the calculations.

Since the vertex factor (\ref{V2}) and the product $w_{a} w_{b} \dots$
are symmetrical with respect to any permutations of their indices, the
quantity $I^{ab\dots}_{12\dots}$ in Eq.~(\ref{V1}) is automatically
symmetrized with respect to any permutations of the letter indices
$a,b,\dots$ and the numerical indices $1,2,\dots$. In what follows,
such symmetrization will always be denoted by the symbol $Sym$.

For any fixed number of rays $k$, the quantity $Sym\, I$ is
represented as a linear combination
\begin{equation}
Sym\, I = \sum_{i} B_{i}\, S_{i}
\label{V4}
\end{equation}
of certain basis tensor structures $S_{i} = (S_{i})^{ab\dots}_{12\dots}$
with certain numerical coefficients $B_{i}$. The structures for the
$k$-ray diagrams with $k=2,3,4$ have the forms (there are two structures
for $k=2,3$ and three structures for $k=4$)
\begin{eqnarray}
k&=&2: \qquad S_{1} = Sym\, [\delta_{1a}\delta_{2b}], \qquad
            S_{2} = Sym\, [\delta_{12}\delta_{ab}];
\nonumber \\
k&=&3: \qquad S_{1} = Sym\, [\delta_{1a}\delta_{2b}\delta_{3c}], \qquad
S_{2} = Sym\, [\delta_{12}\delta_{ab}\delta_{3c}]; \qquad
\nonumber \\
k&=&4: \qquad S_{1} = Sym\, [\delta_{1a}\delta_{2b}\delta_{3c}
\delta_{4d}], \qquad
S_{2} =  Sym\, [\delta_{12}\delta_{ab}\delta_{3c}\delta_{4d}], \qquad
S_{3} =  Sym\, [\delta_{12}\delta_{34}\delta_{ab}\delta_{cd}].
\label{V5}
\end{eqnarray}
This is sufficient for the three-loop calculation, because the diagrams
with $k=5,6$ in Fig.~1 factorize into products of the blocks with $k=2,3$.

The quantities which will be directly calculated from the diagrams are not
the coefficients $B_{i}$ themselves, but the following scalar quantities
related to them:
\begin{equation}
A_{i} = {\rm tr}\, \bigl[(S_{i})^{ab\dots}_{12\dots} \,
Sym\, I^{ab\dots}_{12\dots}\bigr] = {\rm tr}\, [S_{i} \cdot Sym\, I].
\label{V6}
\end{equation}
Here and below, the symbol tr denotes the contraction with respect to all
repeated indices, which will not be shown explicitly.

It is therefore necessary to express the coefficients $B_{i}$ in
Eq.~(\ref{V4}) in terms of the quantities (\ref{V6}). This is easily done:
substituting Eq.~(\ref{V4}) into (\ref{V6}) gives
\begin{equation}
A_{i} = \sum _{k} M_{ik} \, B_{k}, \qquad {\rm where} \qquad
M_{ik} \equiv {\rm tr}\, [S_{i} \cdot S_{k}].
\label{V7}
\end{equation}
In a compact notation, $A=MB$ (matrix $M$ acts onto vector $B$ and gives
vector $A$). The symmetrical matrix $M$ defined in Eq.~(\ref{V7}) is easily
calculated for any given set of structures of the form (\ref{V5}); then the
corresponding inverse matrix $M^{-1}$ is found, and the desired expressions
for $B$ in terms of $A$ follow from the relation $B=M^{-1}A$.

Below we give the explicit expressions for matrix elements $M_{ik}=M_{ki}$
for the structures (\ref{V5}):
\begin{eqnarray}
k=2: \qquad M_{11} &=&d(d+1)/2, \quad M_{12} =d, \quad  M_{22} =d^{2};
\nonumber \\
k=3:\qquad  M_{11} &=&d(d+1)(d+2)/6, \quad M_{12} =d(d+2)/3,
\quad M_{22} =d(d+2)^{2}/9;
\nonumber \\
k=4:\qquad  M_{11} &=&d(d+1)(d+2)(d+3)/24, \quad M_{12} =d(d+2)(d+3)/12,
\quad M_{13} =d(d+2)/3,
\nonumber \\
M_{22} &=& d(d+2)(d^{2}+7d+16)/72, \quad M_{23} =d(d+2)^{2}/9,
\quad M_{33} = d^{2}(d+2)^{2} /9.
\label{V9}
\end{eqnarray}
It is worth noting that in the calculation of quantities like $M_{ik}$ in
Eq.~(\ref{V7}) or $A_{i}$ in (\ref{V6}), it is sufficient to retain the
symbol $Sym$ only in one of the cofactors, because the second is symmetrized
automatically and can be replaced with one of its terms, like the expressions
in square brackets in Eq.~(\ref{V5}).

Inverting the matrices $M$ in Eq.~(\ref{V9}) gives the following explicit
expressions of the coefficients $B$ in terms of $A$:
\begin{mathletters}
\label{V10}
\begin{eqnarray}
k=2: \quad B_{1} &=& 2\alpha [dA_{1}-A_{2}],
\nonumber \\
B_{2} &=& \alpha [-2A_{1}+(d+1)A_{2}] \ {\rm with} \
\alpha \equiv [(d-1)d(d+2)]^{-1},
\label{V10a}\\
k=3: \quad B_{1} &=& 6\alpha [(d+2)A_{1}-3A_{2}],
\nonumber \\
B_{2} &=& 9\alpha [-2A_{1}+(d+1)A_{2}]
\ {\rm with} \ \alpha \equiv [(d-1)d(d+2)(d+4)]^{-1},
\label{V10b} \\
k=4: \quad B_{1} &=& 24\alpha [(d+2)(d+4)A_{1}-6(d+2)A_{2}+3A_{3}],
\nonumber \\
B_{2} &=& 72\alpha [-2(d+2)A_{1} + (d^{2}+3d+6)A_{2}-(d+3)A_{3}],
\nonumber \\
B_{3} &=& 9\alpha [8A_{1}-8(d+3)A_{2}+(d+3)(d+5)A_{3}]
\ {\rm with} \ \alpha \equiv [(d-1)d(d+1)(d+2)(d+4)(d+6)]^{-1}.
\label{V10c}
\end{eqnarray}
\end{mathletters}

It should be emphasized that the above relations (\ref{V9}) and (\ref{V10})
are independent of the specific choice of a composite operator built of the
gradients $w_{i}=\partial_{i}\theta$, which only determines the explicit
form of the vertex factor (\ref{V2}).

\subsection{Contractions of basic tensor structures} \label{sec:Contr}

Consider the procedure of the contraction of the quantities
$I^{ab\dots}_{12\dots}$ in Eq. (\ref{V1}) with external factors:
the vertex factor $V_{12\dots}$ of the composite operator and the product
$ w_{a} w_{b} \dots$. General rules given below are valid for any local
monomial $F$ built of the gradients $w_{i}=\partial_{i}\theta$. The operator
$F$ may have vector indices; as a rule, they will not be shown explicitly.

Substituting the decomposition (\ref{V4}) into Eq. (\ref{V1}) gives:
\begin{equation}
\Gamma = \sum_{i} B_{i} \, V_{12\dots} \, (S_{i})^{ab\dots}_{12\dots}
\, w_{a} w_{b} \dots = \sum_{i} B_{i} \Gamma _{i},
\label{V11}
\end{equation}
where $\Gamma _{i}$ are the contractions of the quantities (\ref{V5}) with the
external factors:
\begin{equation}
\Gamma _{i} = V_{12\dots} \, (S_{i})^{ab\dots}_{12\dots}
\, w_{a} w_{b} \dots.
\label{V12}
\end{equation}
Consider first the contractions
\begin{equation}
(T_{i})_{12\dots} =  (S_{i})^{ab\dots}_{12\dots} \, w_{a} w_{b} \dots
\label{V13}
\end{equation}
of the structures $S_{i}$ with the external factors $w$. For quantities
(\ref{V5}), they are easily calculated:
\begin{eqnarray}
k&=&2: \qquad (T_{1})_{12} = w_{1} w_{2}, \quad
(T_{2})_{12} = w^2 \delta_{12};
\nonumber \\
k&=&3: \qquad (T_{1})_{123} = w_{1} w_{2} w_{3}, \quad
(T_{2})_{123} = w^2\, Sym\, [\delta_{12} w_{3}];
\nonumber \\
k&=&4: \qquad (T_{1})_{1234} = w_{1} w_{2} w_{3} w_{4}, \quad
(T_{2})_{1234} = w^2\, Sym\, [\delta_{12} w_{3} w_{4}], \quad
(T_{3})_{1234} = w^4\, Sym\, [\delta_{12} \delta_{34}],
\label{V14}
\end{eqnarray}
where the symbol $Sym$ denotes the symmetrization
with respect to the {\it numerical} indices.

From Eqs. (\ref{V12}) and (\ref{V13}) one obtains:
\begin{equation}
\Gamma _{i} = V_{12\dots} \,  (T_{i})_{12\dots},
\label{V15}
\end{equation}
where the symbol $Sym$ in the quantities (\ref{V14}) can be omitted owing
to the fact that the vertices $V_{12\dots}$ are symmetrical.

For a general $k$-ray diagram we need to calculate contractions of the
vertex factors $V_{1\dots k}$ with the structures $T_{1\dots k}$ of the
form (\ref{V14}). Their basis is provided by the set
\begin{equation}
w_{1} \dots w_{k}, \quad w^2 \delta_{12} w_{3} \dots w_{k}, \quad
w^4 \delta_{12}\delta_{34} w_{5} \dots w_{k}, \quad {\rm and\ so\ on.}
\label{V16}
\end{equation}

Consider first the contraction of the vertex factor (\ref{V2}) with the
first monomial in Eq.~(\ref{V16}), that is, the expression $L_{k}\,F$ with
the operator
\begin{equation}
L_{k}\equiv w_{1} \dots w_{k}\, \partial_{1} \dots \partial_{k}\, ,
\qquad  \partial_{i} \equiv \partial / \partial w_{i}.
\label{Lk}
\end{equation}
The key role in the calculation of the quantity $L_{k}\,F$ is played by the
following consideration: by permutations of the factors $w$ and
$\partial/\partial w$, the operator $L_{k}$ in Eq.~(\ref{Lk})
can be represented in the form of a polynomial in the operation
${\cal D}\equiv w_{i}\partial / \partial w_{i}$. The action of the latter
onto any monomial $F$ is easily found, namely, ${\cal D} F=nF$, where
$n$ is the total number of the fields $w$ in the monomial $F$. By
permutations of the factors $w$ and $\partial / \partial w$, one can easily
obtain the recurrent relation $L_{k+1} = L_{k} ({\cal D}-k)$ for the operator
$L_{k}$, which along with the relation $L_{1} = {\cal D}$ gives
\begin{equation}
L_{k} = {\cal D}({\cal D}-1)({\cal D}-2)\dots ({\cal D}-k+1), \qquad
{\cal D}\equiv w_{i}\partial / \partial w_{i}.
\label{V17}
\end{equation}
When this operation acts onto $F$, each symbol ${\cal D}$ is replaced with
the number $n$, which gives the desired coefficient:
\begin{equation}
L_{k}F = n(n-1)(n-2) \dots (n-k+1) F,
\label{V18}
\end{equation}
where $n$ is the total number of the factors $w$ in the operator $F$.

Consider now the contraction of the  vertex factor (\ref{V2}) with the
second structure in Eq.~(\ref{V16}). The latter includes the factor
$\delta_{12}$, which after the contraction with $\partial_{1}\partial_{2}$
gives the Laplace operator $\partial^{2}\equiv\partial_{i}\partial_{i}$
with $\partial_{i} \equiv \partial / \partial w_{i}$. This operator
commutes with the other derivatives $\partial_{i}$ and acts directly onto
$F$. Therefore the first task is the calculation of the quantity
$\partial^{2}F$, which is easily done for any specific monomial $F$
built of the gradients $w_{i}=\partial_{i}\theta$.

From now on, we shall confine ourselves with the family of operators
$F_{nl}$ from Eq.~(\ref{Fnp}), which in the new notation take on the form
\begin{equation}
F_{nl} =  {\cal IRP}\, \bigr[ w_{1} \cdots w_{l} \,
w^{2p} \bigl], \qquad   n=l+2p.
\label{Wnp}
\end{equation}
The quantities $\partial^{2}F$ for such operators are easily calculated:
\begin{equation}
\partial^{2}F_{nl} = \lambda _{nl} F_{n-2,l}, \qquad
\lambda _{nl} = (n-l)(d+n+l-2).
\label{V20}
\end{equation}
It is worth noting that in the calculation of the quantity $\partial^{2}F$,
there is no need to take into account contributions in which the both
derivatives act onto the monomial $w_{1}\dots w_{l}$ in Eq.~(\ref{Wnp}):
in this case, two factors $w$ are necessarily replaced with the symbol
$\delta_{is}$ (the number of free indices is preserved, and the number of
factors $w$ is reduced by 2 under the action of $\partial^{2}$), and any
contribution with $\delta_{is}$ disappears under the action of the operation
${\cal IRP}$ in Eq.~(\ref{Wnp}).

It is clear that the combination of the relations (\ref{V18}) and (\ref{V20})
gives a complete solution to the problem of the calculation of the
contractions of vertex factor (\ref{V2}) with basis structures of the form
(\ref{V16}) for our specific operators (\ref{Wnp}), and that the above
considerations can be directly generalized to the case of any polynomial
in $w$ operator $F$. If the basis structure includes several delta
symbols (e.g., two delta symbols in the third monomial in Eq. (\ref{V16})),
the operation $\partial^{2}$ should be applied repeatedly; one should also
remember that each next Laplacian acts onto the operator which has already
been modified (replacement $n\to n-2$ in Eq. (\ref{V20})). After all the
operations $\partial^{2}$ have been applied, only the operation $L_{k}$
(\ref{Lk}) with the reduced number of factors remains; it acts onto the
properly modified operator $F$ according to the rule (\ref{V18}). For our
operators $F_{nl}$, additional scalar factors $w^2$, $w^4,\dots$ in
monomials (\ref{V16}) restore the original scalar factor $w^{2p}$ in Eq.
(\ref{Wnp}). Therefore, all the quantities $\Gamma_{i}$ in Eq. (\ref{V15})
are proportional to the original operator (\ref{Wnp}): $\Gamma_{i} =
k_{i} F$, with some numerical coefficients $k_{i}$. They are easily found
using the rules discussed above. In particular, for the 2-ray diagrams
$\Gamma_{1} = V_{12}\, w_{1}w_{2} = L_{2}F=n(n-1)F$,
$\Gamma_{2} = V_{12}\, w^2 \delta_{12} = w^2 \partial^{2}F =
\lambda_{nl}F$, and similarly for the 3-ray and 4-ray diagrams.

Substituting the relation $\Gamma_{i} = k_{i} F$ into Eq. (\ref{V11}) gives:
\begin{equation}
\Gamma = F \overline\Gamma, \qquad  \overline\Gamma =\sum_{i} k_{i} B_{i} ,
\label{V21}
\end{equation}
where $k_{i}$ are the coefficients in front of $F$ in the contractions of the
vertex factor $V_{1\dots k}$ with structures (\ref{V16}), numbered in the
same order ($k_{1}$ corresponds to the structure without delta symbols,
$k_{2}$ corresponds to the structure with one delta symbol, and so on):
\begin{mathletters}
\label{V22}
\begin{eqnarray}
k=2: \qquad k_{1} &=&n(n-1), \quad k_{2}= \lambda _{nl}, \quad
{\rm where} \quad \lambda _{nl}= (n-l)(d+n+l-2);
\label{V22a} \\
k=3: \qquad k_{1}&=&n(n-1)(n-2), \quad k_{2}= (n-2)\lambda _{nl};
\label{V22b}   \\
k=4: \qquad k_{1}&=&n(n-1)(n-2)(n-3), \quad k_{2}= (n-2)(n-3)\lambda _{nl},
\quad k_{3} = \lambda _{n-2,l}\lambda _{nl};
\label{V22c} \\
k=5: \qquad k_{1}&=&n(n-1)(n-2)(n-3)(n-4),
\quad k_{2}= (n-2)(n-3)(n-4)\lambda _{nl},
\quad k_{3} =(n-4) \lambda _{n-2,l}\lambda _{nl};
\label{V22d} \\
k=6: \qquad k_{1}&=&n(n-1)(n-2)(n-3)(n-4)(n-5),
\quad k_{2}= (n-2)(n-3)(n-4)(n-5)\lambda _{nl},
\nonumber \\
k_{3} &=& (n-4)(n-5) \lambda _{n-2,l}\lambda _{nl},
\quad k_{4} = \lambda _{n-4,l}\lambda _{n-2,l}\lambda _{nl}.
\label{V22e}
\end{eqnarray}
\end{mathletters}
The coefficients $k_{i}$ with $k=5$ and 6 will be needed only for the
calculation of the factorizable diagrams Nos~51,61 with five and
six rays, repectively.

The quantities $A_{i}$ are calculated directly from the diagrams, then
one calculates the quantities $B_{i}$ using the relations (\ref{V10}),
and then, finally, one calculates the desired contribution of the
diagram $\Gamma$ using the relations (\ref{V21}) and (\ref{V22}).

\subsection{Calculation of the quantities $A_{i}$}
\label{sec:A}

The quantities $I^{ab\dots}_{12\dots}$ for the diagrams $\Gamma$ are
calculated at zero external momenta; the independent
integration momenta will always be denoted by $\k$ in the one-loop diagrams,
$\k$, $\q$ in all two-loop diagrams, and $\k$, $\q$, $\l$ in all three-loop
diagrams. They are always assigned to the horizontal lines $vv$; in the normal
(not factorizable) diagrams the order is the following: $\k$ flows via the
uppermost line, $\q$ flows via the next line, and $\l$ flows via the lowest
line.

These rules are illustrated by Fig.~2, where the one-loop and two-loop
diagrams are presented with all their momenta; the directions of the
latter are shown by arrows. We denote all the propagators by solid lines,
because the horizontal $vv$ lines cannot be confused with the ``rays''
built of the $\theta\theta'$ lines.
The letter indices $a,b,\dots$ of the $vv$ lines are always free
(they will be contracted with the indices of the external factors
$w_{a}w_{b}\dots$), while the numerical indices of the $vv$ lines are always
contracted with the indices of derivatives (that is, the momentum factors
$\pm {\rm i} p_{s}$), shown by dots on the $\theta\theta'$ lines.

\begin{figure}
\centerline{ \psfig{file=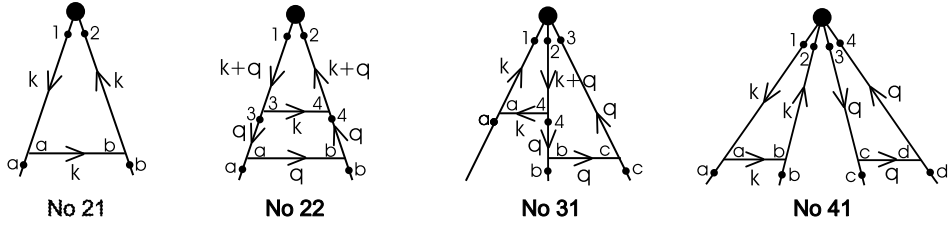,height=20ex}}
\bigskip
\caption{Diagrams of the function $\Gamma$ in the one-loop
and two-loop approximations in the detailed notation.}
\end{figure}

We shall calculate the diagrams in the time--momentum ($t$--$\k$)
representation.
First, the elementary integrations over all times are performed. This gives
certain ``energy denominator'' $\phi_{E}$, that is, the factor
$\phi_{E}^{-1}$ in the integrand. The remaining integrations are those over
$d$-dimensional momenta ($\k$ in one-loop diagrams, $\k$, $\q$ in two-loop
diagrams, and $\k$, $\q$, $\l$ in three-loop diagrams). Each integration is
accompanied by the factor $D_{0}(2\pi)^{-d}$ with $D_{0} =
g\nu\mu^{\eps}$, and all $\nu$'s will later be cancelled out with analogous
factors in the energy denominators. As a result, the quantity
$I^{ab\dots}_{12\dots}$ from Eq.~(\ref{V1}) is represented as follows:
\begin{mathletters}
\label{V24}
\begin{eqnarray}
&1&\ {\rm loop}:\  \qquad I^{ab\dots}_{12\dots} = \left[D_{0} (2\pi)^{-d}
\right] \, \int d{\bf k}\, N_{k}\, \phi ^{ab\dots}_{12\dots} \, ,
\label{V24a} \\
&2&\ {\rm loops}: \qquad I^{ab\dots}_{12\dots} = \left[D_{0} (2\pi)^{-d}
\right]^{2} \,
\int\int d{\bf k}d{\bf q}\, N_{k}N_{q}\, \phi ^{ab\dots}_{12\dots}\, ,
\label{V24b} \\
&3&\ {\rm loops}: \qquad I^{ab\dots}_{12\dots} = \left[D_{0} (2\pi)^{-d}
\right] ^{3} \, \int\int\int d{\bf k}d{\bf q}d{\bf l}\, N_{k}N_{q}N_{l} \,
\phi ^{ab\dots}_{12\dots}\, ,
\label{V24c}
\end{eqnarray}
\end{mathletters}
with $N_{k}$ from Eq.~(\ref{3}). After these scalar factors have been
isolated, only the transverse projectors $P$ remain in the $vv$ lines.
The nontrivial parts $\phi ^{ab\dots}_{12\dots}$ of the integrands in Eq.
(\ref{V24}) are represented as follows:
\begin{equation}
\phi ^{ab\dots}_{12\dots} =C \cdot \phi_{E}^{-1} \cdot \phi_{s} \cdot
\bar I ^{ab\dots}_{12\dots}\,  ,
\label{V25}
\end{equation}
where $C$ is a numerical factor, $\phi_{E}$ is the energy denominator,
$\phi_{s}$ is an additional ``scalar factor,''  which can arise due to
contractions over the internal indices, and $\bar I ^{ab\dots}_{12\dots}$
is the nontrivial ``index structure'' of the integrand, which remains after
$\phi_{s}$ has been isolated.

Let us explain in more detail the calculation of the factors entering into
Eq. (\ref{V25}), starting with the energy factors $\phi_{E}^{-1}$. Since the
velocity correlation function (\ref{3}) contains the delta function in time,
the dots on the rays connected by such correlators have coincident times $t$.
If we assign the time $t=0$ to the uppermost vertex of a diagram (that is,
the vertex of the composite operator), the independent variables will be the
times $t_{i}<0$ of the horizontal $vv$ lines. On any of the
rays, the times $t_{i}$ decrease from above to below owing to the retardation
property of the functions $\langle \theta\theta'\rangle_{0}$ in
Eq.~(\ref{Lines}). For the normal (not factorizable)
one and two-loop diagrams in Fig.~1, the order of the times $t_{i}<0$
is determined by the form of the diagram in a unique way, that is, there is
only one ``temporal version.'' This is also true for all the three-loop
diagrams with two and three rays. However, for two of the three-loop diagrams
(Nos 43,44 in Fig.~1) there are two equivalent temporal versions (the
exchange $t_{2}\leftrightarrow t_{3}$ in diagram No 43 and
$t_{1}\leftrightarrow t_{2}$  in diagram No 44). This gives the additional
factor 2 in the coefficient $C$ in Eq. (\ref{V25}) for these two diagrams.

The energy denominator $\phi_{E}$, obtained as the result of the integrations
over all time variables, is given by the product of the factors
corresponding to the ``temporal cross-sections'' of the diagrams between the
horizontal $vv$ lines; to each cross-section corresponds
the sum of ``energies'' $\epsilon_{\p} = \nu p^{2}$ for all intersected
$\theta\theta'$ lines. For example, one has
$\phi_{E}=2\epsilon_{\k}$ for diagram No~21 in Fig.~2,
$\phi_{E}=2\epsilon_{\q}\cdot2\epsilon_{\k+\q}$ for diagram No~22, and
$\phi_{E}=2\epsilon_{\q}\bigl[\epsilon_{\k}+\epsilon_{\q}+\epsilon_{\k+\q}
\bigr]$ for diagram No~31.

The scalar factors $\phi_{s}$ in Eq. (\ref{V25}) are generated by the
$ vv$ lines whose vector indices
are contracted with the indices of {\it two} integration momenta
in the diagram. This gives rise to expressions of the form
$(\p P\p') \equiv p_{i} P_{ij} p_{j}'$, where $P$ is the projector that
corresponds to the $vv$ line and $\p$, $\p'$ are the momenta
contracted to the line (the factors $\pm {\rm i}$ are not included).
In what follows, we denote as $P_{\k}$, $P_{\q}$ and $P_{\l}$ the transverse
projectors for the momenta $\k$, $\q$ and $\l$, respectively. Then the
scalar factors from the diagrams in Fig.~2 take on the form
$\phi_{s}=(\q P_{\k}\q)$ for No~22 and $\phi_{s}=1$ for Nos 21,31,
because the latter two diagrams do not contain lines $ vv$
contracted to two momenta.

The numerical coefficient $C$ in Eq. (\ref{V25}) is represented as the
product $C=C_{1}C_{2}C_{3}$, where $C_{1}$ is the symmetry coefficient
of the diagram, shown in Fig.~1, $C_{2}=\pm 1$ is the sign factor
(see Sec.~\ref{sec:Scal}) and $C_{3}=1$ or 2 is the number of the temporal
versions. For the most of the diagrams one has $C_{2}=+ 1$, but there are
two diagrams with $C_{2}=- 1$ (Nos~44,45). The factor $C_{3}$ is equal
to 1 except for the diagrams Nos~43,44 with $C_{3}=2$. In what follows,
we shall
always present the coefficient $C$ as the product $C=C_{1}C_{2}C_{3}$,
for example, $C= 1/2 \cdot 1 \cdot 1$ for diagrams Nos~21,22 in Fig.~2
and $C= 1 \cdot 1 \cdot 1$ for No~31.

Now let us turn to the quantities  $\bar I ^{ab\dots}_{12\dots}$ in
Eq.~(\ref{V25}), which remain in the integrands after the scalar factors
$\phi_{s}$ have been isolated. They have the same stucture for all the
diagrams with two and three rays and any number of loops, and two possible
structures for the four-ray diagrams, namely:
\begin{eqnarray}
k=2: \qquad {\rm n}1 &=& \bar I ^{ab}_{12} = x_{1} x_{2} P_{ab} ;
\nonumber \\
k=3: \qquad {\rm n}2 &=& \bar I ^{abc}_{123} = x_{1} (x+y)_{2}
y_{3} \alpha_{a} P_{bc} ;
\nonumber \\
k=4: \qquad {\rm n}3 &=& \bar I ^{abcd}_{1234} = x_{1} y_{2} z_{3} (x+y+z)_{4}
\alpha_{a}\beta_{b} P_{cd} ;
\nonumber \\
k=4: \qquad {\rm n}4 &=& \bar I ^{abcd}_{1234} = x_{1} y_{2} z_{3} (x+y+z)_{4}
P_{ab} P_{cd} '.
\label{V26}
\end{eqnarray}
We have introduced the numbering n1--n4 and in what follows, when describing
the diagrams we shall only give the number of the corresponding structure,
e.g., $\bar I = n1$ with the specification of the vectors
$\x,\y,\z,\balpha,\bbeta$ and projectors $P,P'$. In particular, we have
$\bar I = n1$ with $\x=\k$, $P=P_{\k}$ for diagram No~21 in Fig.~2,
$\bar I = n1$ with $\x=\k+\q$, $P=P_{\q}$ for diagram No~22, and
$\bar I = n2$ with $\x=\k$, $\y=\q$, $\balpha =P_{\k}\q$, $P=P_{\q}$
for diagram No~31 (quantities like $\balpha =P_{\k}\q$ are undestood
as vectors obtained by the action of the matrix $P_{\k}$ onto the
vector $\q$).

Our aim is the calculation of the quantities $A_{i}$ in Eq. (\ref{V6})
for each diagram in Fig.~1. They are given by the contractions of the
integrals (\ref{V24}) with the structures $S_{i}$ from Eq. (\ref{V5}),
which leads to the replacement of the factors $\phi ^{ab\dots}_{12\dots}$
in Eq. (\ref{V24}) with the quantities
$\phi_{i} = {\rm tr}\, \bigl[ (S_{i})^{ab\dots}_{12\dots} \cdot
Sym\,\phi^{ab\dots}_{12\dots} \bigr]$. Substituting expression (\ref{V25})
into this construction gives $\phi_{i}=C\phi_{E}^{-1}\phi_{s}\bar A_{i}$,
where
\begin{equation}
\bar A_{i} = {\rm tr}\, \bigl[ S_{i} \cdot Sym\, \bar I \bigr]
\label{V27}
\end{equation}
are the analogs of quantities (\ref{V6}) with the replacement $I\to\bar I$.
As a result, from the relations (\ref{V6}), (\ref{V24}), (\ref{V25}) and
(\ref{V27}) one obtains the following explicit formulas for the calculation
of the coefficients $A_{i}$:
\begin{mathletters}
\label{V28}
\begin{eqnarray}
&1&\ {\rm loop}:\  \qquad A_{i} = \left[D_{0} (2\pi)^{-d}
\right] \, \int d{\bf k}\, N_{k}\, C\, \phi_{E}^{-1} \,
\phi_{s}\, \bar A_{i} ,
\label{V28a} \\
&2&\ {\rm loops}: \qquad A_{i}  = \left[D_{0} (2\pi)^{-d} \right]^{2} \,
\int\int d{\bf k}d{\bf q}\, N_{k}N_{q}\, C\, \phi_{E}^{-1}\,\phi_{s}
\,\bar A_{i} ,
\label{V28b}   \\
&3&\ {\rm loops}: \qquad A_{i}   = \left[D_{0} (2\pi)^{-d}\right]
^{3} \, \int\int\int d{\bf k}d{\bf q}d{\bf l}\, N_{k}N_{q}N_{l}\,
C\, \phi_{E}^{-1}\,\phi_{s}\,\bar A_{i}
\label{V28c}
\end{eqnarray}
\end{mathletters}
with $\bar A_{i}$ from Eq.~(\ref{V27}).

Calculation of these quantities for structures n1 and n2 from Eq. (\ref{V26})
and $S_{i}$ from (\ref{V5}) gives:
\begin{eqnarray}
k = 2\ (\bar I= {\rm n1}): \qquad
&{\bar A}&_{1} = (\x P\x), \qquad \bar A_{2} = (d-1) x^{2} ;
\label{V29} \\
k = 3\ (\bar I={\rm n2}): \qquad
&3&\bar A_{1} = (\y\balpha)(\x P\x) + (\x\balpha)(\y P\y) +
((\x+\y)\balpha)(\x P\y),
\nonumber \\
&9&\bar A_{2} = \bigl[x^{2}+2(\x\y)\bigr] \bigl[(d-1)(\y\balpha) +2
(\y P\balpha)\bigr] + \bigl[y^{2}+2(\x\y)\bigr]
\bigl[(d-1)(\x\balpha) +2 (\x P\balpha)\bigr].
\label{V30}
\end{eqnarray}

Analogous expressions can also be obtained for the structures n3 and n4
in Eq. (\ref{V26}) for the four-ray diagrams; they are rather cumbersome
are will be given later in Sec.~\ref{sec:Blin}.

\section{Calculation in the one-loop and two-loop approximations}
\label{sec:Oneloop}

In the one-loop approximation, we need the only diagram No~21 , and in
the two-loop approximation, all the four diagrams from Fig.~2 are needed.
Using the general rules discussed in the previous Section, for the
normal diagrams in the notation (\ref{V25}), (\ref{V26}) we obtain:
\begin{eqnarray}
{\rm No}\ 21: \qquad C&=&1/2 \cdot 1 \cdot 1, \quad \phi_{s} =1,
\quad \phi_{E} =2\epsilon_{\k},
\nonumber \\
\bar I &=& {\rm n}1 \quad {\rm with} \quad \x=\k, \quad P=P_{\k}
\quad [P\x=0];
\label{V31} \\
{\rm No}\ 22: \qquad C&=&1/2 \cdot 1 \cdot 1, \quad \phi_{s} = (\q P_{\k}\q),
\quad \phi_{E} =4\epsilon_{\q}\epsilon_{\k+\q},
\nonumber \\
\bar I &=& {\rm n}1 \quad {\rm with} \quad \x=\k+\q, \quad P=P_{\q} ;
\label{V32}  \\
{\rm No}\ 31: \qquad C&=&1 \cdot 1 \cdot 1, \quad \phi_{s} = 1,
\quad \phi_{E} = 2\epsilon_{\q} (\epsilon_{\k}+\epsilon_{\q}+
\epsilon_{\k+\q}), \nonumber \\
\bar I &=& {\rm n}2 \quad {\rm with} \quad \x=\k, \quad \y=\q, \quad
\balpha =P_{\k}\q, \quad P=P_{\q} \quad
\bigl[(\x\balpha)=0, \quad P\y=0\bigr];
\label{V33}
\end{eqnarray}
in the square brackets we give simple relations specific of a given set
of momenta and projectors, which are useful in the calculations.
We also recall that $\epsilon_{\p} =\nu p^{2}$.

Substituting the explicit expressions (\ref{V31})--(\ref{V33}) into the
general relations (\ref{V29}), (\ref{V30}) gives:
\begin{eqnarray}
{\rm No}\ 21&:& \qquad \bar A_{1} = 0, \quad \bar A_{2} = (d-1) k^{2} ;
\label{V34} \\
{\rm No}\ 22&:& \qquad \bar A_{1} = (\k P_{\q}\k), \quad
                  \bar A_{2} =  (d-1) (\k+\q)^{2} ;
\label{V35} \\
{\rm No}\ 31&:& \qquad 3\bar A_{1} = (\q P_{\k}\q)\, (\k P_{\q}\k), \quad
9\bar A_{2} = (d-1)\,(\q P_{\k}\q) \bigl[ k^{2} +2 (\k\q)\bigr] +
2 (\k P_{\q}P_{\k}\q) \bigl[ q^{2} +2 (\k\q)\bigr].
\label{V36}
\end{eqnarray}

\subsection{One-loop approximation} \label {sec:subsub1}

The one-loop approximation $\overline\Gamma^{(1)}$ in Eq. (\ref{useries})
is determined by the only diagram No~21 in Fig.~2. Substituting the
quantities known from Eqs. (\ref{V31}), (\ref{V34}) and the relation
$\epsilon_{\k} =\nu k^{2}$ into Eq. (\ref{V28a}) gives (we recall that
the symmetry coefficient 1/2 is included into $C$):
\begin{equation}
{\rm No}\ 21: \qquad  A_{1} = 0, \quad A_{2} = [(d-1)/4] g\mu^{\eps} \, U
\label{V37}
\end{equation}
with the integral $U \equiv (2\pi)^{-d} \int d\k N_{k}$ from Eq. (\ref{V38}).
Substituting that expression into (\ref{V37}) gives:
\begin{equation}
{\rm No}\ 21: \qquad A_{1} = 0, \quad A_{2} = u\, [(d-1)/4\eps]
(\mu/m)^{\eps}, \qquad u =gC_{d}
\label{V39}
\end{equation}
with coefficient $C_{d}$ from Eq.~(\ref{V38})

From the relations (\ref{V10a}) and (\ref{V39}) we find
$B_{1} = -2\alpha A_{2}$, $ B_{2} = \alpha (d+1)A_{2}$ with $\alpha$ from
(\ref{V10a}) and $A_{2}$ from (\ref{V39}); then from Eqs. (\ref{V21}) and
(\ref{V22a}) we obtain the quantity $\overline\Gamma$ for our diagram:
$\overline\Gamma({\rm No}\ 21)=\alpha A_{2} [-2n(n-1)+(d+1)\lambda_{nl}]$.
Substituting the known expressions for $\alpha$ and $A_{2}$ gives:
\begin{equation}
\overline\Gamma^{(1)} = \overline\Gamma({\rm No}\ 21)= u \, (\mu/m)^{\eps}\,
\frac {(d+1)\lambda_{nl}-2n(n-1)} {4\eps d(d+2)}.
\label{V40}
\end{equation}
Then the relation (\ref{deltaZ1}) gives the final result for the $O(u)$
contribution to the renormalization constant:
\begin{equation}
Z_{F}^{-1} = 1 + \left[Z_{F}^{-1}\right]_{1}, \qquad
\left[ Z_{F}^{-1}\right]_{1} =
u\, \frac {2n(n-1)-(d+1)\lambda_{nl}} {4\eps d(d+2)} \, .
\label{V41}
\end{equation}

\subsection{Two-loop approximation}

Consider now the two-loop diagrams Nos~22,31 in Fig.~2; the
corresponding coefficients $\bar A_{i}$ are known from Eqs. (\ref{V35})
and (\ref{V36}). They include the quantities
$(\k P_{\q}\k) =k^{2} \sin^{2} \vartheta $,
$(\q P_{\k}\q) =q^{2} \sin^{2} \vartheta $,
$(\k P_{\q} P_{\k}\q) = - (\k\q) \sin^{2} \vartheta $,
$(\k\q)^{2} = k^{2} q^{2} \cos ^{2} \vartheta $, where $\vartheta $ is
the angle between the vectors $\k$ and $\q$. Substitutuing these expressions
into Eqs. (\ref{V35}) and (\ref{V36}) gives:
\begin{eqnarray}
{\rm No}\ 22: \qquad\ \ \bar A_{1} &=& k^{2} \sin^{2} \vartheta,
\quad \bar A_{2} = (d-1) (\k+\q)^{2};
\label{V42}  \\
{\rm No}\ 31: \qquad 3\bar A_{1} &=& k^{2}q^{2} \sin^{4} \vartheta,
\nonumber \\
\quad 9\bar A_{2} &=&
\sin^{2} \vartheta \left\{ (d-1) q^{2} [k^{2}+ 2(\k\q)] - 2 (\k\q)
[q^{2}+ 2(\k\q)] \right\}
\nonumber \\
&=&  q^{2}\sin^{2} \vartheta \left\{ (d-1) [k^{2}+ 2(\k\q)]
- 2(\k\q) -4k^{2} \cos ^{2} \vartheta \right\}.
\label{V43}
\end{eqnarray}

Substituting these expressions along with the other needed information from
Eqs. (\ref{V32}), (\ref{V33}) and the relation $\epsilon_{\p} =\nu p^{2}$
into Eq. (\ref{V28b}) gives
\begin{eqnarray}
{\rm No}\ 22: \qquad A_{1} &=& [(g\mu^{\eps})^{2}/8] (2\pi)^{-2d}
\int\int d{\bf k}d{\bf q}\, N_{k}N_{q} \,
\frac {k^{2}\sin^{4} \vartheta} {(\k+\q)^{2}} \, , \nonumber \\
A_{2} &=& [(d-1)(g\mu^{\eps})^{2}/8] (2\pi)^{-2d}
\int\int d{\bf k}d{\bf q}\, N_{k}N_{q} \, \sin^{2} \vartheta;
\label{V44}
\end{eqnarray}
\begin{eqnarray}
{\rm No}\ 31: \qquad 3A_{1} &=& [(g\mu^{\eps})^{2}/4] (2\pi)^{-2d}
\int\int d{\bf k}d{\bf q}\, N_{k}N_{q} \,
\frac {k^{2}\sin^{4} \vartheta} {k^{2}+q^{2}+(\k\q)} \, , \nonumber \\
9A_{2} &=& [(g\mu^{\eps})^{2}/4] (2\pi)^{-2d}
\int\int d{\bf k}d{\bf q}\, N_{k}N_{q} \, \frac
{\sin^{2} \vartheta \left\{(d-1) [k^{2}+2(\k\q)]
- 2(\k\q) -4k^{2} \cos ^{2} \vartheta \right\} }
{k^{2}+q^{2}+(\k\q)} \,.
\label{V45}
\end{eqnarray}
We note that in the coefficients $A_{1,2}$ for diagram No~31 the factor
$q^{2}$ from $\epsilon_{\q}$ in the denominators cancels out with the
analogous factor in the numerators, while in the coefficient $A_{2}$
for diagram No~22 a similar cancellation of the factor $(\k+\q)^{2}$
takes place.

The denominators in the integrals (\ref{V44}), (\ref{V45}) are symmetrical
with respect to the permutation $\k \leftrightarrow\q$. This allows one to
perform the analogous symmetrization in the numerators, which is equivalent
to the replacement $k^{2} \to (k^{2}+q^{2})/2$ in the latter. Then the
expressions simplify and reduce to linear combinations of the following
standard integrals:
\begin{equation}
H_{2p} = (2\pi)^{-2d} \int\int d{\bf k}d{\bf q}\, N_{k}N_{q} \,
\sin^{2p} \vartheta, \quad p=1,2,
\label{V46}
\end{equation}
\begin{equation}
h=(2\pi)^{-2d} \int\int d{\bf k}d{\bf q}\, N_{k}N_{q} \, (\k\q)\,
\sin^{4} \vartheta \, / (\k+\q)^{2},
\label{V47}
\end{equation}
\begin{equation}
h_{2p} = (2\pi)^{-2d} \int\int d{\bf k}d{\bf q}\, N_{k}N_{q} \, (\k\q)\,
\sin^{2p} \vartheta \, / [k^{2}+q^{2}+(\k\q)], \quad p=1,2.
\label{V48}
\end{equation}

The coefficients $A_{i}$ in Eqs. (\ref{V44}), (\ref{V45}) are expressed
in terms of these integrals as follows:
\begin{eqnarray}
{\rm No}\ 22&:& \qquad A_{1} = [(g\mu^{\eps})^{2}/8] \, [H_{4}/2-h],
\qquad A_{2} = [(g\mu^{\eps})^{2}/8] \, [ (d-1) H_{2}];
\label{V49}  \\
{\rm No}\ 31&:& \quad 3  A_{1} = [(g\mu^{\eps})^{2}/4] \,
[H_{4}/2-h_{4}/2],
\qquad 9 A_{2} = [(g\mu^{\eps})^{2}/4] \, [ (d-5) H_{2}/2
+2H_{4} +3(d-1)h_{2}/2  -2h_{4}].
\label{V50}
\end{eqnarray}

These expressions along with Eq. (\ref{V10a}) for diagram No~22 and
Eq. (\ref{V10b}) for diagram No~31 give the coefficients $B_{i}$,
then using Eqs. (\ref{V21}), (\ref{V22a}) and (\ref{V22b}) one obtains
the corresponding quantities $\overline\Gamma({\rm No}\ 22)$ and
$\overline\Gamma({\rm No}\ 31)$; the answers are represented as linear
combinations of the standard integrals (\ref{V46})--(\ref{V48}).

The two-loop quantity $\overline\Gamma^{(2)}$ also includes the contribution
from the factorizable diagram No~41 with the symmetry coefficient 1/8;
see Fig.~1. It reduces to the contraction of the vertex factor $V_{1234}$
with the product $T_{12}T_{34}$ of two identical one-loop blocks
similar to diagram No~21. In the notation of Eqs. (\ref{V11})--(\ref{V15})
this contribution has the form
\begin{equation}
\Gamma = C \, V_{1234} \,
\bigl[B_{1} w_{1}w_{2} + B_{2} w^{2}\delta_{12} \bigr]
\bigl[B_{1} w_{3}w_{4} + B_{2} w^{2}\delta_{34} \bigr] \equiv
F \overline\Gamma,
\label{V51}
\end{equation}
where $B_{1,2}$ are the known coefficients for the one-loop diagram No~21
(see Sec.~\ref{sec:subsub1}) and $C$ is an additional symmetry
coefficient; in the case at hand, $C=1/2$. Let us explain its origin.
By definition, the quantities $B_{1,2}$ for diagram No~21 already include
its symmetry coefficient $1/2$ (in the factor $C$ in Eq. (\ref{V25})),
which gives $1/4$ for the product of two such blocks. Therefore, in order
to obtain the correct symmetry coefficient $1/8$ for diagram No~41,
the additional coefficient $C=1/2$ should be included.

Multiplying the expressions in the square brackets in Eq. (\ref{V51})
and taking into account the definition of the coefficients $k_{i}$ in Eq.
(\ref{V22}) gives
\begin{equation}
\overline\Gamma({\rm No}~41) = [1/2] \, [k_{1}B_{1}^{2} + 2k_{2} B_{1}B_{2}+
k_{3}B_{2}^{2}]
\label{V52}
\end{equation}
with coefficients $k_{1,2,3}$ from Eq. (\ref{V22c}) and known quantities
$B_{1,2}$ for the one-loop diagram No~21. In what follows, the same
scheme will be used for the calculation of the contributions of the
three-loop factorizable diagrams Nos~46,51,61; the coefficients
$k_{i}$ from Eqs. (\ref{V22d}) and (\ref{V22e}) will be involved.

We have found all terms in the two-loop expression
\begin{equation}
\overline\Gamma^{(2)} = \overline\Gamma({\rm No}~22)+
\overline\Gamma({\rm No}~31) +\overline\Gamma({\rm No}~41).
\label{V53}
\end{equation}
Now Eq. (\ref{deltaZ2}) is used to determine the two-loop contribution
$\bigl[Z_{F}^{-1}\bigr]_{2}$ in the renormalization constant $Z_{F}^{-1}$,
with $\overline\Gamma^{(1)}$ from Eq. (\ref{V40}), $\overline\Gamma^{(2)}$ from
Eq. (\ref{V53}) and $\bigl[Z_{F}^{-1}\bigr]_{1}$ from Eq.~(\ref{V41}).
It is also necessary to set $\mu=m$ in all the diagrams of
$\overline\Gamma^{(1,2)}$ in order to eliminate contributions with the
logarithms $\ln (\mu/m)$ (they would be cancelled with the contributions
of the diagrams with self-energy insertions, which we already omitted).

All the quantities in Eq. (\ref{deltaZ2}) are known explicitly, except for
the contributions of diagrams Nos~22,31, for which we only know the
expressions of the coefficients $A_{i}$ in terms of the standard integrals
(\ref{V46})--(\ref{V48}). In order to calculate the quantity (\ref{deltaZ2}),
we need only their divergent parts $\sim 1/\eps^{2}$ and $1/\eps$, but for
the three-loop calculation we shall also need the zero-order in $\eps$
(constant) terms.

Calculation of these standard integrals is a separate task and will be
discussed in the next Subsection.

\subsection{Calculation of the two-loop integrals
(\protect\ref{V46})--(\protect\ref{V48})} \label{sec:Integrals}

In this Subsection we denote $\n_{k} \equiv \k/k$ for any vector $\k$,
$d\n_{k}$ is the area element of the unit $d$-dimensional sphere and
$\langle \dots \rangle$ is the averaging over the sphere. In particular,
\begin{equation}
\int d\k N_{k} \dots =  \int _{m}^{\infty} \frac
{dk}{k^{1+\eps}} \, \int d \n_{k} \dots = S_{d}
\int _{m}^{\infty} \frac {dk}{k^{1+\eps}} \,
\bigl\langle \dots \bigr\rangle
\label{L1}
\end{equation}
with $N_{k}$ from Eq.~(\ref{3}); $S_{d}=2\pi ^{d/2}/\Gamma (d/2)$
is the surface area of the unit sphere.

For any two vectors with the angle $\vartheta$ between them,
the following formulas will be useful:
\begin{equation}
\alpha_{2n} \equiv \bigl\langle \cos^{2n} \vartheta
\bigr\rangle = \frac{(2n-1)!!} {d(d+2)\dots(d+2n-2)}
\label{L2}
\end{equation}
with $n=1,2,\dots$ (obviously, $\alpha_{0} \equiv \langle 1 \rangle  =1$).
From Eq. (\ref{L2}) one easily obtains:
\begin{equation}
\bigl\langle \sin^{2}\vartheta \cos^{2n} \vartheta \bigr\rangle =
\frac{(d-1)} {(d+2n)}\, \alpha_{2n}, \qquad
\bigl\langle \sin^{4}\vartheta \cos^{2n} \vartheta \bigr\rangle =
\frac{(d^{2}-1)} {(d+2n)(d+2n+2)}\, \alpha_{2n}
\label{L3}
\end{equation}
with $n=0,1,2,\dots$ and $\alpha_{2n}$ from (\ref{L2}).

In the notation introduced above, the integrals $H_{2p}$ in Eq. (\ref{V46})
are written in the form
\begin{equation}
H_{2p}= C_{d}^{2}\, \int _{m}^{\infty} \frac {dk}{k^{1+\eps}} \,
\int _{m}^{\infty} \frac {dq}{q^{1+\eps}} \,
\langle \sin^{2p}\vartheta \rangle, \qquad p=1,2.
\label{L4}
\end{equation}
Then Eq. (\ref{L3}) gives:
\begin{equation}
H_{2}= C_{d}^{2}\, \frac{m^{-2\eps}(d-1)}{d\eps^{2}}\, ,  \qquad
H_{4}= C_{d}^{2}\, \frac{m^{-2\eps}(d^{2}-1)}{d(d+2)\eps^{2}}.
\label{L5}
\end{equation}

Now let us turn to the integrals $h_{2p}$ in Eq. (\ref{V48}).
Expanding their integrands in $(\k\q)$ gives:
\begin{equation}
h_{2p} = C_{d}^{2}\, \sum_{l=0}^{\infty}\, (-1)^{l} \,
\int _{m}^{\infty}\frac {dk}{k^{1+\eps}} \,\int _{m}^{\infty}
\frac {dq}{q^{1+\eps}}\, \left( \frac{kq} {k^{2}+q^{2}}\right)^{l+1} \,
\bigl\langle \sin^{2p}\vartheta \cos^{l+1} \vartheta \bigr\rangle.
\label{L6}
\end{equation}
The average $\langle \sin^{2p}\vartheta \cos^{l+1} \vartheta \rangle$
differs from zero only for odd $l$, and the series
in Eq. (\ref{L6}) can be written as
\begin{equation}
h_{2p} = - C_{d}^{2}\, \sum_{n=0}^{\infty}\, I_{n} (m) \,
\bigl\langle \sin^{2p}\vartheta \cos^{2n+2} \vartheta \bigr\rangle
\label{L7}
\end{equation}
with the integral
\begin{equation}
I_{n} (m) \equiv \int _{m}^{\infty}\frac {dk}{k^{1+\eps}} \,
\int _{m}^{\infty}\frac {dq}{q^{1+\eps}}\,
\left( \frac{kq} {k^{2}+q^{2}}\right)^{2n+2} = m^{-2\eps}\,I_{n} (1).
\label{L8}
\end{equation}
Using the identity
\begin{equation}
I_{n} (m) = - \frac{1}{2\eps} {\cal D}_{m} I_{n} (m), \qquad
{\cal D}_{m} \equiv m \partial / \partial m,
\label{L9}
\end{equation}
which follows from the last equality in Eq.~(\ref{L8}), the integral
$I_{n} (m)$ can be represented in the form
\begin{equation}
I_{n} (m) = \frac{m^{-2\eps}}{\eps} \, \int^{\infty}_{1}
\frac {dk}{k^{1+\eps}} \, \left( \frac{k} {k^{2}+1}\right)^{2n+2},
\label{L10}
\end{equation}
where the number of integrations is reduced and the pole in $\eps$ is
isolated explicitly. Expanding the integrand in Eq. (\ref{L10}) in $\eps$
and neglecting the terms of order $O(\eps)$ and higher, with the desired
accuracy we obtain
\begin{equation}
I_{n} (m) \simeq m^{-2\eps} \, \int^{\infty}_{1} \, \frac{dk
\,k^{2n+1}} {(k^{2}+1)^{2n+2}} \left( \eps^{-1} - \ln k \right) =
m^{-2\eps} \, \frac{(n!)^2}{4(2n+1)!} \, \left( \eps^{-1} - \B_n
\right) , \label{L11}
\end{equation}
where the quantities $\B_n$, needed only for the three-loop calculation,
can be represented as finite sums:
\begin{equation}
\B_n = \ln 2 + \frac{(2n+1)!}{n!} \, \sum^{n}_{k=0} \, \frac
{(-1)^{k}\, \left[(1/2)^{n+k}-1\right] } {k!(n-k)!(n+k+1)^{2}}\, .
\label{L12}
\end{equation}

For the integrals $h_{2p}$ in Eq. (\ref{V48}) this gives:
\begin{mathletters}
\label{L13}
\begin{eqnarray}
h_{2} &=& m^{-2\eps}\, C_{d}^{2} \, (d-1) \, \sum_{n=0}^{\infty}\, \frac
{n!\,2^{-n-2}} {d(d+2)\dots(d+2n+2)} \, \left(- \eps^{-1} + \B_n \right);
\label{L13a} \\
h_{4} &=& m^{-2\eps}\, C_{d}^{2} \, (d^{2}-1) \, \sum_{n=0}^{\infty}\,
\frac {n!\,2^{-n-2}}{d(d+2)\dots(d+2n+4)} \, \left(-\eps^{-1} + \B_n \right).
\label{L13b}
\end{eqnarray}

Expanding the integrand in Eq. (\ref{V47}) in $2(\k\q)$ and proceeding as
above for $h_{2p}$, we obtain analogous expression for the integral $h$:
\begin{equation}
h =  m^{-2\eps}\, C_{d}^{2} \, (d^{2}-1) \, \sum_{n=0}^{\infty}\, \frac
{n!\,2^{n-1}} {d(d+2)\dots(d+2n+4)} \, \left(- \eps^{-1} + \B_n \right).
\label{L13c}
\end{equation}
\end{mathletters}

The $O(\eps^{-1})$ terms in Eqs. (\ref{L13}) can be expressed in terms
of the hypergeometric series (see, e.g., \cite{Gamma})
\begin{equation}
F(a,b;c;z)\equiv 1+\frac{ab}{c}\, z+ \frac{a(a+1)b(b+1)}{c(c+1)}
\cdot \frac{z^{2}}{2!}+\dots
\label{hyper}
\end{equation}
as follows:
\begin{mathletters}
\label{L14}
\begin{eqnarray}
h_{2} &=& \frac{m^{-2\eps}\, C_{d}^{2} \, (d-1)} {4d(d+2)} \, \left[
- \eps^{-1} F(1,1;d/2+2;1/4) + (d+2) \C_2(d) \right],
\label{L14a} \\
h_{4} &=& \frac{m^{-2\eps}\, C_{d}^{2} \, (d^{2}-1)} {4d(d+2)(d+4)} \, \left[
- \eps^{-1} F(1,1;d/2+3;1/4) + (d+4) \C_4(d) \right],
\label{L14b} \\
h &=& \frac{m^{-2\eps}\, C_{d}^{2} \, (d^{2}-1)} {2d(d+2)(d+4)} \, \left[
- \eps^{-1} F(1,1;d/2+3;1) + (d+4) \C(d) \right],
\label{L14c}
\end{eqnarray}
\end{mathletters}
and for the $O(1)$ terms one has:
\begin{mathletters}
\label{L15}
\begin{eqnarray}
\C_2(d) &=& \sum_{n=0}^{\infty}\, \frac {n!\,\B_{n}} {4^{n}(d/2+1)\dots
(d/2+1+n)}\, ,
\label{L15a} \\
\C_4(d) &=& \sum_{n=0}^{\infty}\, \frac {n!\,\B_{n}} {4^{n}(d/2+2)\dots
(d/2+2+n)} = \C_2(d+2),
\label{L15b} \\
\C(d) &=& \sum_{n=0}^{\infty}\, \frac {n!\,\B_{n}} {(d/2+2)\dots
(d/2+2+n)} .
\label{L15c}
\end{eqnarray}
\end{mathletters}
It is worth noting that the series entering into Eqs. (\ref{L14}) and
(\ref{L15}) are convergent; this fact is nontrivial for $h$ since $|z|=1$
is the convergence radius for the corresponding series.

 For $F(\dots)$ in Eqs.
(\ref{L14c}) one has $F(1,1;d/2+3;1)= (d+4)/(d+2)$  for any $d$,
while the expressions for $F(\dots)$ in Eqs. (\ref{L14a}),
(\ref{L14b}) simplify only for integer $d$; see \cite{Gamma}. In
particular, for $d=2$ and $d=3$ one has

\begin{mathletters}
\label{Numbers}
\begin{eqnarray}
d=2: \qquad &F&(1,1;d/2+2;1/4) = 8 [-3 \ln (4/3)+1] \simeq 1.0956,
\nonumber \\
&F&(1,1;d/2+3;1/4) = 6[18\ln (4/3)-5] \simeq 1.0696,
\nonumber \\
&{\C_2}&\, \simeq 0.3671, \quad \C_{4}\simeq 0.2411 , \quad \C \simeq 0.2917;
\label{Numbers2} \\
d=3: \qquad &F&(1,1;d/2+2;1/4) =10 (\pi\sqrt3-16/3) \simeq 1.0806,
\nonumber \\
&F&(1,1;d/2+3;1/4) =14 (-15\pi\sqrt3 +82)/5 \simeq 1.0613,
\nonumber \\
&{\C_2}&\, \simeq 0.2912, \quad \C_{4}\simeq 0.2056 , \quad \C \simeq 0.2412;
\label{Numbers3}
\end{eqnarray}
\end{mathletters}
and for the other integer $d$ analogous expressions
can be obtained from the recurrent relation
\begin{equation}
3 F(1,1;d/2+2;1/4) + (d+2) F(1,1;d/2+3;1/4)/ (d+4)=4,
\label{Juha}
\end{equation}
valid for all $d$. In Eq. (\ref{Numbers})
we also included the numerical values for the coefficients
$\C_{2,4}$ and $\C$ obtained from the series (\ref{L15}).

Substituting all these expressions for the integrals
into Eqs. (\ref{V49}) and (\ref{V50}) gives the final answers
for the coefficients $A_{i}$ of the diagrams Nos~22,31:
\begin{eqnarray}
{\rm No}\ 22: \qquad A_{1} &=& \frac {u^{2}\, (\mu/m)^{2\eps} \, (d^{2}-1)}
{16d(d+2)}\, \left[ \frac{1}{\eps^{2}}+ \frac{1}{(d+2)\eps} -
\frac{1}{2}\C(d)\right],
\nonumber  \\
A_{2} &=& \frac {u^{2}\, (\mu/m)^{2\eps} \, (d-1)^{2}}
{16d\eps^{2}};
\label{L16}  \\
{\rm No}\ 31:
\qquad A_{1} &=& \frac {u^{2}\, (\mu/m)^{2\eps}} {24d(d+2)}\,
\left[ \frac{1}{\eps^{2}}+ \frac{F(1,1;d/2+3;1/4)}{4(d+4)\eps}
- \frac{1}{8}\C_{4}(d) \right],
\nonumber  \\
\qquad A_{2} &=& \frac {u^{2}\, (\mu/m)^{2\eps} \, (d-1)} {24d(d+2)}\,
\left[ \frac{(d-2)(d+3)}{3\eps^{2}}-\frac{(d-1)F(1,1;d/2+2;1/4)}{4\eps}+
\right.
\nonumber  \\
&+& \left. \frac{(d+1)F(1,1;d/2+3;1/4)}{3(d+4)\eps}+
\frac{1}{8} (d-1)(d+2)\C_{2}(d) - \frac{1}{6} (d+1)\C_{4}(d)\right]
\label{L17}
\end{eqnarray}
with $F(\dots)$ from Eq.~(\ref{hyper}) and $\C,\C_{2,4}$ from~(\ref{L15}).
From these expressions, using the standard general scheme, one obtains
the contributions of the diagrams Nos 22,31 into Eq. (\ref{V53})
and then the two-loop contribution $\left[Z_{F}^{-1}\right]_{2}$
into the renormalization constant $Z_{F}^{-1}$, presented earlier
in Ref.~\cite{RG}.

\section{Three-loop approximation} \label{sec:Three}
\subsection{Scalarization of the three-loop diagrams} \label{sec:Blin}

All the needed three-loop diagrams with the symmetry coefficients are given
in Fig.~1. Below we shall describe them in more detail and give in a compact
form the complete set of relations which allow one to express the
coefficients $A_{i}$ in Eq. (\ref{V6}) for any given diagram in terms of the
scalar integrals (\ref{V28c}). We recall that all the external momenta in
the diagrams are set equal to zero.

We begin with the normal (not factorizable) diagrams. For these, the
integration momenta $\k$, $\q$, $\l$ are always assigned to the horizontal
lines $\langle vv\rangle$ in the following order: $\k$ flows via the
uppermost line, $\q$ flows via the middle line, and $\l$ flows via the
lowest line. Then in order to determine the momenta for all lines in the
diagram, it is sufficient to show (by an arrow) the chosen direction of
the momentum in each line. The needed derivatives $\partial$ on the lines
are always implied (for the one-loop and two-loop diagrams, they were shown
explicitly by dots in Fig.~2). The numerical indices $1,2,\dots$ of the upper
dots are always chosen to increase from the left to the right (like in Eq.
(\ref{V3}) and in the diagrams in Fig.~2); the positions of the letter
indices $a,b,\dots$ in the diagram are always shown explicitly. This
information allows one to completely restore the configuration of the
momenta and the form of the index structures (\ref{V26}) for any diagram.

Now we turn to the description of specific diagrams in this notation.
For any diagram, we give the directions of the momenta,
positions of the indices $a,b,\dots$ and forms of all cofactors in
Eq.~(\ref{V25}) in the same form as in Eqs. (\ref{V31})--(\ref{V33}).

The two-ray diagram No~23:
\begin{equation}
\raisebox{-7ex}{\psfig{file=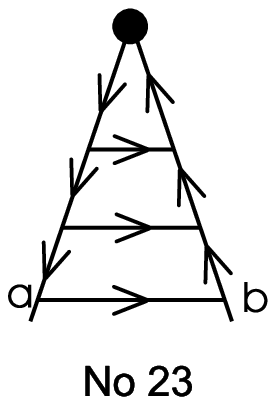,height=18ex}}
\qquad\qquad\quad
\begin{array}{l}
C= 1/2 \cdot 1 \cdot 1, \quad \phi_{s} = (\l
P_{\q}\l)((\q+\l)P_{\k} (\q+\l)),\\
{\phi_{E}}= 8
\epsilon_{\l}\epsilon_{\q+\l}\epsilon_{\k+\q+\l},\\
{\bar I}=
{\rm n}1 \quad {\rm with} \quad \x=\k+\q+\l, \ P=P_{\l}.
\end{array}
\label{V55}
\end{equation}

The three-ray normal diagrams:
\begin{equation}
\raisebox{-8ex}{\psfig{file=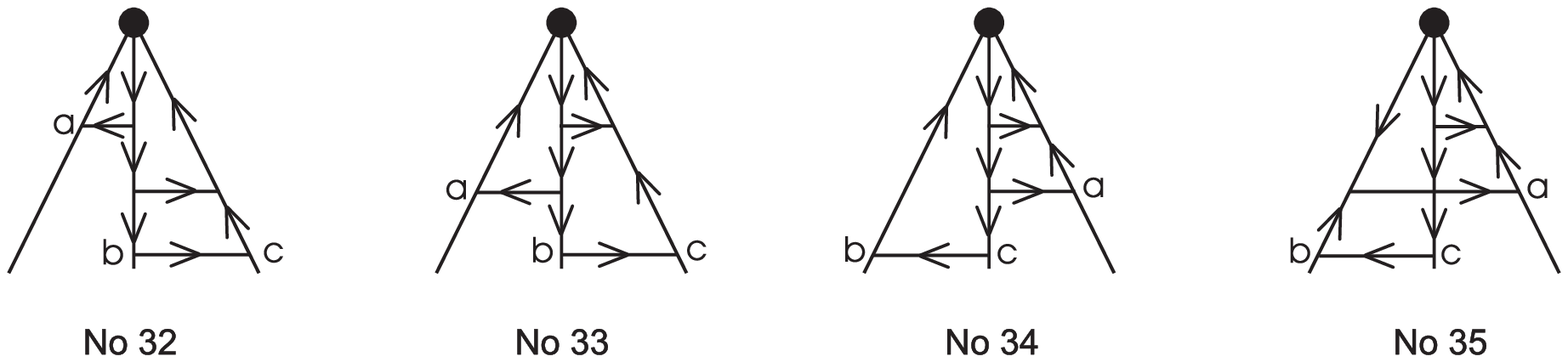,height=18ex}}\,\,. \label{V56}
\end{equation}
For these, one has:
\begin{eqnarray}
{\rm No}~32: \qquad
&C& = 1 \cdot 1 \cdot 1, \quad \phi_{s} = (\l P_{\q}\l),
\nonumber \\
&{\phi_{E}}&= 4 \epsilon_{\l}\epsilon_{\q+\l} (\epsilon_{\k}+
\epsilon_{\q+\l} + \epsilon_{\k+\q+\l}),
\nonumber \\
&{\bar I} &= {\rm n}2 \quad {\rm with} \quad \x=\k, \ \y=\q+\l, \
\balpha = P_{\k} (\q+\l), \ P=P_{\l}  \quad [(\x\balpha)=0];
\label{V57} \\
{\rm No}~33: \qquad
&C& = 1 \cdot 1 \cdot 1, \quad \phi_{s} = (\l P_{\k}(\q+\l)),
\nonumber \\
&{\phi_{E}}&= 2 \epsilon_{\l}
(\epsilon_{\q}+ \epsilon_{\l} + \epsilon_{\q+\l})
(\epsilon_{\q}+ \epsilon_{\k+\l} + \epsilon_{\k+\q+\l}),
\nonumber \\
&{\bar I} &= {\rm n}2 \quad {\rm with} \quad \x=\q, \ \y=\k+\l, \
\balpha = P_{\q}\l, \ P=P_{\l}  \quad [(\x\balpha)=0];
\label{V58} \\
{\rm No}~34: \qquad
&C& = 1 \cdot 1 \cdot 1, \quad \phi_{s} = (\q P_{\k}(\q+\l)),
\nonumber \\
&{\phi_{E}}&= 2 \epsilon_{\l}
(\epsilon_{\q}+ \epsilon_{\l} + \epsilon_{\q+\l})
(\epsilon_{\l}+ \epsilon_{\k+\q} + \epsilon_{\k+\q+\l}),
\nonumber \\
&{\bar I} &= {\rm n}2 \quad {\rm with} \quad \x=\l, \ \y=\k+\q, \
\balpha = P_{\q}\l, \ P=P_{\l}  \quad [P\x=0];
\label{V59} \\
{\rm No}~35: \qquad
&C& = 1 \cdot 1 \cdot 1, \quad \phi_{s} = (\q P_{\k} \l),
\nonumber \\
&{\phi_{E}}&= 2 \epsilon_{\l}
(\epsilon_{\q}+ \epsilon_{\l} + \epsilon_{\q-\l})
(\epsilon_{\k+\q}+ \epsilon_{\k+\l} + \epsilon_{\q-\l}),
\nonumber \\
&{\bar I} &= {\rm n}2 \quad {\rm with} \quad \x=\k+\l, \ \y=\q-\l, \
\balpha = P_{\q}\l, \ P=P_{\l}.
\label{V60}
\end{eqnarray}

The structure ${\bar I}= {\rm n}2$ from Eq. (\ref{V26}) is the same for all
these diagrams; the corresponding quantities $\bar A_{i}$ are determined by
the general formula (\ref{V30}) in which the specific values of the vectors
$\x,\y,\z,\balpha$ and projectors $P$ for any given diagram should be
substituted from Eqs.~(\ref{V57})--(\ref{V60}).

The four-ray normal diagrams:
\begin{equation}
\raisebox{-8ex} {\psfig{file=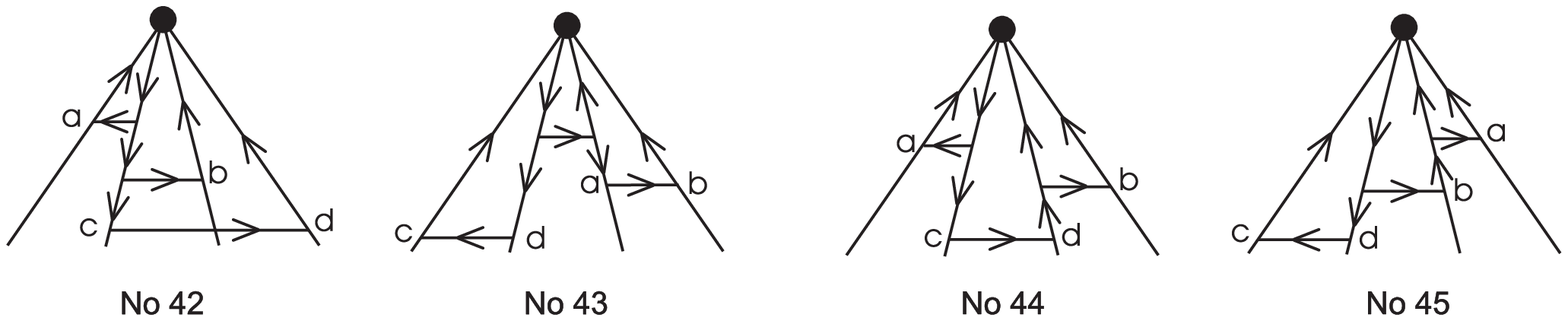,height=18ex}}\,\, .
\label{V61}
\end{equation}
For these, one has:
\begin{eqnarray}
{\rm No}~42: \qquad
&C& = 1 \cdot 1 \cdot 1, \quad \phi_{s} = 1,
\nonumber \\
&{\phi_{E}}&= 2 \epsilon_{\l}
(\epsilon_{\q}+\epsilon_{\l}+ \epsilon_{\q+\l})
(\epsilon_{\k}+\epsilon_{\q}+\epsilon_{\l}+ \epsilon_{\k+\q+\l}),
\nonumber \\
&{\bar I} &= {\rm n}3 \quad {\rm with} \quad \x=\k, \ \y=\q, \
\z = \l, \ \balpha = P_{\k} (\q+\l), \  \bbeta = P_{\q} \l, \
P=P_{\l}
\nonumber \\
&{}&[(\x\balpha)=0, (\y\bbeta)=0, P\z=0];
\label{V62}\\
{\rm No}~43: \qquad
&C& = 1/2 \cdot 1 \cdot 2, \quad \phi_{s} = (\q P_{\k}\l),
\nonumber \\
&{\phi_{E}}&= 4 \epsilon_{\l}(\epsilon_{\q}+\epsilon_{\l})
(\epsilon_{\q}+\epsilon_{\l}+\epsilon_{\k-\q}+ \epsilon_{\k+\l}),
\nonumber \\
&{\bar I} &= {\rm n}4 \quad {\rm with} \quad \x=\q, \ \y=\l, \
\z = \k-\q,\  P=P_{\l}, \ P' = P_{\q}
\nonumber \\
&{}&[P\y=0, P'\x=0];
\label{V63}\\
{\rm No}~44: \qquad
&C& = 1/2 \cdot (-1) \cdot 2, \quad \phi_{s} = 1,
\nonumber \\
&{\phi_{E}}&= 2 \epsilon_{\l}
(\epsilon_{\q}+\epsilon_{\l}+ \epsilon_{\l-\q})
(\epsilon_{\k}+\epsilon_{\q}+\epsilon_{\k+\l}+ \epsilon_{\l-\q}),
\nonumber \\
&{\bar I} &= {\rm n}3 \quad {\rm with} \quad \x=\k, \ \y=\q, \
\z = \l-\q, \ \balpha = P_{\k} \l, \  \bbeta = P_{\q} \l, \
P=P_{\l}
\nonumber \\
&{}&[(\x\balpha)=0, (\y\bbeta)=0, P(\y+\z)=0];
\label{V64}\\
{\rm No}~45: \qquad
&C& = 1 \cdot (-1) \cdot 1, \quad \phi_{s} = 1,
\nonumber \\
&{\phi_{E}}&= 2 \epsilon_{\l}
(\epsilon_{\q}+\epsilon_{\l}+ \epsilon_{\q+\l})
(\epsilon_{\k}+\epsilon_{\l}+\epsilon_{\q-\k}+ \epsilon_{\q+\l}),
\nonumber \\
&{\bar I} &= {\rm n}3 \quad {\rm with} \quad \x=\k, \ \y=\l, \
\z = \q-\k, \ \balpha = P_{\k} \q, \  \bbeta = P_{\q} \l, \
P=P_{\l}
\nonumber \\
&{}&[(\x\balpha)=0, ((\x+\z)\bbeta)=0, P\y=0].
\label{V65}
\end{eqnarray}
Three diagrams have the index structure $\bar I={\rm n}3$ from Eq.
(\ref{V26}) and one diagram has the structure $\bar I={\rm n}4$;
$(\x\balpha)=0$ for all the diagrams with $\bar I={\rm n}3$.

Below we give the expressions for the coefficients $\bar A_{i}$ for the
structures $\bar I={\rm n}3,{\rm n}4$, analogous to
Eqs. (\ref{V29}), (\ref{V30}).

For $\bar I={\rm n}3$ from Eqs. (\ref{V26}) and (\ref{V27}) using
the relation $(\x\balpha)=0$ one obtains:
\begin{eqnarray}
12 \bar A_{1} &=& (\x P\x) [(\y\balpha)(\z\bbeta)+(\y\bbeta)(\z\balpha)]+
(\y P\y)(\z\balpha)(\x\bbeta)+(\z P\z)(\x\bbeta)(\y\balpha)+
\nonumber \\
&+& 2(\x P\y) \bigl[(\y\balpha)(\z\bbeta)+(\z\balpha)(\x+\y+\z,\bbeta)\bigr]
+2(\x P\z) \bigl[(\y\bbeta)(\z\balpha)+(\y\balpha)(\x+\y+\z,\bbeta)\bigr]+
\nonumber \\
&+& 2(\y P\z)(\x\bbeta) (\z+\y,\balpha);
\nonumber \\
72\bar A_{2} &=& U_{1} \Bigl\{ (d-1) \bigl[(\y\balpha)(\z\bbeta)
+(\y\bbeta)(\z\balpha)\bigr] + 2(\y P\balpha)(\z\bbeta)+
2(\z P\balpha)(\y\bbeta) +
\nonumber \\
&+& 2(\y P\bbeta)(\z\balpha) + 2 (\z P\bbeta)(\y\balpha) +
2(\y P\z)(\balpha\bbeta) \Bigr\} +
\nonumber \\
&+& U_{2} \Bigl\{(d-1) (\z\balpha)(\x\bbeta)+2(\x P\balpha)(\z\bbeta)+
2(\z P\balpha)(\x\bbeta) +
2 (\x P\bbeta)(\z\balpha) + 2(\x P\z)(\balpha\bbeta) \Bigr\} +
\nonumber \\
&+& U_{3} \Bigl\{(d-1)(\y\balpha)(\x\bbeta) +
2(\x P\balpha)(\y\bbeta)+2(\y P\balpha)(\x\bbeta)+
2(\x P\bbeta)(\y\balpha)+2(\x P\y)(\balpha\bbeta) \Bigr\} +
\nonumber \\
&+& 2(d-1) \bigl[(\z\balpha)(\z\bbeta)(\x\y)+(\y\balpha)(\y\bbeta)(\x\z)
\bigr] +
\nonumber \\
&+& 2 \bigl[ (\z P\balpha)(\z\bbeta)(\x\y)+(\y P\balpha)(\y\bbeta)(\x\z)+
(\x P\balpha)(\x\bbeta)(\y\z)+ (\z P\bbeta)(\z\balpha)(\x\y)+
(\y P\bbeta)(\y\balpha)(\x\z) \bigr]+
\nonumber \\
&+&2(\balpha\bbeta)\bigl[ (\x P\x)(\y\z)+(\y P\y)(\x\z)+(\z P\z)(\x\y)\bigr];
\nonumber \\
9 \bar A_{3} &=& U_{4} \Bigl\{(d-1)(\balpha\bbeta)+2(\balpha P\bbeta)\Bigr\}.
\label{V66}
\end{eqnarray}
Here and below in Eq.~(\ref{V68})
\begin{eqnarray}
U_{1} &\equiv& x^{2}+2(\x\y)+2(\x\z), \quad U_{2} \equiv y^{2}+2(\x\y)+2(\y\z),
\quad U_{3} \equiv z^{2}+2(\x\z)+2(\y\z),
\nonumber \\
U_{4} &\equiv& x^{2}(\y\z)+y^{2}(\x\z)+z^{2}(\x\y)+ 2 (\x\y)(\x\z)+
2(\x\y)(\y\z)+2(\x\z)(\y\z).
\label{V67}
\end{eqnarray}

For $\bar I={\rm n}4$ from Eq. (\ref{V26})
using the relations $P\y=0$, $P'\x=0$ in
Eq.~(\ref{V63}) one obtains:
\begin{eqnarray}
6 \bar A_{1} &=& (\x P\x)(\y P'\z) +2 (\x P\z)(\y P'\z) +
(\x P\z)(\y P'\y),
\nonumber \\
36 \bar A_{2} &=& (d-1)\Bigl\{U_{1}(\y P'\z)+U_{2}(\x P\z)+
(\x P\x)(\y\z)+(\y P'\y)(\x\z)+ (\z(P+P')\z)(\x\y)\Bigr\}+
\nonumber \\
&+& 2\Bigl\{U_{1} (\y P'P\z)+U_{2}(\x PP'\z)+U_{3}(\x PP'\y)+
(\z(PP'+P'P)\z)(\x\y)\Bigr\},
\nonumber \\
9 \bar A_{3} &=& U_{4} \Bigl\{ (d-1)^{2} +2 {\rm tr}\, (PP')\Bigr\}
\label{V68}
\end{eqnarray}
with $U_{1}$--$U_{4}$ from Eq.~(\ref{V67}); ${\rm tr}\, (PP')$ is the trace
of a matrix product.

Substituting the specific values for the vectors and projectors for any given
diagram from Eqs. (\ref{V62})--(\ref{V65}) into expressions
(\ref{V66})--(\ref{V68}) gives explicit expressions for the corresponding
coefficients $\bar A_{i}$. Then, using the additional information from
(\ref{V62})--(\ref{V65}) one obtains the desired expressions for
the coefficients $A_{i}$ for any diagram in the form of scalar integrals
(\ref{V28c}). All these simple but cumbersome technical operations are easily
performed by means of a computer.

Now let us turn to the factorizable diagrams Nos~46,51,61. Diagram No~46
factorizes to a product of two blocks Nos~21,22; diagram No~51
factorizes to a product of the blocks Nos~21,31; diagram No~61
factorizes to a product of three blocks No~21. Since the quantities
$B_{i}$ for diagrams Nos~21,22 already include their own symmetry
coefficient $1/2$, we conclude that in order to obtain the needed symmetry
coefficient in Fig.~1 for diagrams Nos~46,51 an additional symmetry
coefficient (like $C$ in Eq. (\ref{V51})) is not needed, while for
diagram No~61 it should be taken to be $1/6$.

Therefore, taking into account Eqs. (\ref{V4}) and (\ref{V14}), we obtain
for diagrams Nos 46,51,61:
\begin{mathletters}
\label{V69}
\begin{eqnarray}
{\rm No}\ 46&:& \qquad \Gamma = V_{1234}
\bigl[B_{1}w_{1}w_{2}+B_{2} w^{2}\delta_{12}\bigr]
\bigl[B_{1}'w_{3}w_{4}+B_{2}' w^{2}\delta_{34}\bigr],
\label{V69a} \\
{\rm No}\ 51&:& \qquad \Gamma = V_{12345}
\bigl[B_{1}w_{1}w_{2}+B_{2} w^{2}\delta_{12}\bigr]
\bigl[B_{1}'w_{3}w_{4}w_{5}+B_{2}' w^{2}\delta_{34}w_{5}\bigr],
\label{V69b} \\
{\rm No}\ 61&:& \qquad \Gamma = [1/6]\, V_{123456}
\bigl[B_{1}w_{1}w_{2}+B_{2} w^{2}\delta_{12}\bigr]
\bigl[B_{1}w_{3}w_{4}+B_{2} w^{2}\delta_{34}\bigr]
\bigl[B_{1}w_{5}w_{6}+B_{2} w^{2}\delta_{56}\bigr],
\label{V69c}
\end{eqnarray}
\end{mathletters}
where $B_{1,2}$ are the known coefficients for diagram No~21, $B'_{1,2}$ in
Eq. (\ref{V69a}) are the analogous coefficients for diagram No~22, and
$B'_{1,2}$ in Eq. (\ref{V69b}) are the coefficients for diagram No~31.

Using the relations (\ref{V22}), from Eqs. (\ref{V69}) one obtains:
\begin{eqnarray}
\overline\Gamma({\rm No}\ 46) &=& k_{1}B_{1}B_{1}'+ k_{2}
(B_{1}B_{2}'+B_{2}B_{1}') + k_{3}B_{2}B_{2}' ,
\nonumber \\
\overline\Gamma({\rm No}\ 51) &=& k_{1}B_{1}B_{1}'+ k_{2}
(B_{1}B_{2}'+B_{2}B_{1}') + k_{3}B_{2}B_{2}' ,
\nonumber \\
\overline\Gamma({\rm No}\ 61) &=& [1/6]\, [k_{1}B_{1}^{3} +
3k_{2}B_{1}^{2}B_{2}+3k_{3}B_{1}B_{2}^{2}+k_{4}B_{2}^{3}]
\label{V70}
\end{eqnarray}
with the coefficients $k_{i}$ known from Eq. (\ref{V22c}) for diagram No~46,
from Eq. (\ref{V22d}) for diagram No~51, and from Eq. (\ref{V22e}) for
diagram No~61. The relations (\ref{V70}) give the desired answers for the
three-loop factorizable diagrams.

\subsection{Calculation of the three-loop integrals and anomalous
dimensions: General scheme} \label{sec:Nerav}

Substituting the specific values for the vectors and projectors, given
in (\ref{V55}), (\ref{V57})--(\ref{V60}), (\ref{V62})--(\ref{V65})
for all normal three-loop diagrams, into the general formulas (\ref{V29}),
(\ref{V30}), (\ref{V66})--(\ref{V68}) gives explicit expressions for the
quantities $\bar A_{i}$ in Eq.~(\ref{V28c}) for any diagram in the form of
polynomials in the scalar products $(\k\q)$, $(\k\l)$, $(\q\l)$ and moduli
$k,q,l$ of the vectors $\k,\q,\l$.

It is worth noting that for all the normal three-loop diagrams, the products
$\phi_{s}\bar A_{i}$ have the form of sums of monomials in $\k,\q,\l$ of
order six. In the variables ``moduli--angles,'' each of these monomials
contains the factor $l^{2}$, which is canceled out by the analogous factor
in the energy denominator $\epsilon_{\l}=\nu l^{2}$, present in each of these
diagrams. It is also worth noting that these products involve the modulus $k$
in the power 2 or less (otherwise the integrals over $k$ would be divergent).

In the three-loop diagrams, we only need the poles in $\eps$, that is, the
contributions of order $\eps^{-3}$, $\eps^{-2}$ and $\eps^{-1}$. The
following general scheme is used in their calculation:

1. The integrands are expanded in the set of scalar products $(\k\q)$,
$(\k\l)$, $(\q\l)$; the results are represented as multiple series in these
quantities.

2. Then, the angular averaging $\langle\dots\rangle$ is performed with
respect to the directions of the vectors $\k,\q,\l$, that is, the following
quantities are calculated:
\begin{equation}
T_{n_{1}n_{2}n_{3}} = \bigl\langle (\k\q)^{n_{1}}\,
(\k\l)^{n_{2}} \, (\q\l)^{n_{3}}
\bigr\rangle
\label{V71}
\end{equation}
with arbitrary integer exponents $n_{i} \ge 0$.

3. The next step is the integration over the moduli $k,q,l$. All the needed
integrals reduce to the form:
\begin{equation}
\int^{\infty}_{1} \frac{dk}{k^{1+\eps}} \int^{\infty}_{1}
\frac{dq}{q^{1+\eps}} \int^{\infty}_{1} \frac{dl}{l^{1+\eps}}\,
\frac{k^{2n_1}q^{2n_2}l^{2n_3}}{(k^2+q^2+l^2)^{n_4}(q^2+l^2)^{n_5}},
\label{Integral}
\end{equation}
where the integer exponents $n_{i} \ge0$ satisfy the relation
$n_1+n_2+n_3=n_4+n_5$, so that the integrals contain only logarithmic UV
divergencies for $\eps\to0$.
Only the pole parts are extracted from the integrals (\ref{Integral}).

4. The last step is the summation of the resulting series. They have the
forms of double infinite series with the coefficients given by $n$-fold
finite sums ($n\le5$); the number of terms in the latter increases rapidly
with the order of the coefficient. This summation is the only operation
which cannot be performed exactly (analytically) and is therefore the only
source of errors in numerical coefficients in expressions like
Eqs.~(\ref{Qnp3}).

Of course, the straightforward but cumbersome operations listed above
have been performed with the aid of a computer.

The first step, the expansion in the scalar products $(\k\q)$, $(\k\l)$,
$(\q\l)$, contains some conceptual subtleties, and we shall discuss it in
more detail. The problem is that the plain expansion of the integrands in
the powers of the scalar products in some cases leads to divergent series.

The non-polynomial dependence of the integrands on the cosines of the angles
appears only from the energy factors $\phi_{E}^{-1}$, whose explicit
forms are given in Eqs. (\ref{V55}), (\ref{V57})--(\ref{V60}),
(\ref{V62})--(\ref{V65}).
For all the diagrams, the nontrivial factors in $\phi_{E}$ have the forms
$q^{2}+l^{2}+ {\rm const}\, (\q\l)$ and $k^{2}+q^{2}+l^{2}+$ some linear
combination of all scalar products, and in the expansions in powers of the
scalar products, the denominators will contain powers of the quantities
$(q^{2}+l^{2})$ and $(k^{2}+q^{2}+l^{2})$. Such expansions converge for all
cofactors in $\phi_{E}^{-1}$, which do not contain the ``energy''
$\epsilon_{\k+\q+\l}$ with the sum of all three momenta. The cofactors with
$\epsilon_{\k+\q+\l}$ require special consideration. They are present in
diagrams
Nos 23,32,33,34,42 and are proportional to the following factors:
\begin{eqnarray}
{\rm No}\ 23: \qquad \phi^{-1}_{E}& \propto &
\epsilon_{\k+\q+\l}^{-1} \propto [Q+2S]^{-1}
\nonumber \\
{\rm No}\ 32: \qquad \phi^{-1}_{E}& \propto &
[\epsilon_{\k}+\epsilon_{\q+\l}+\epsilon_{\k+\q+\l}]^{-1}
\propto [Q+S+(\q\l)]^{-1},
\nonumber \\
{\rm No}\ 33: \qquad \phi^{-1}_{E}& \propto &
[\epsilon_{\q}+\epsilon_{\k+\l}+\epsilon_{\k+\q+\l}]^{-1}
\propto [Q+S+(\k\l)]^{-1},
\nonumber \\
{\rm No}\ 34: \qquad \phi^{-1}_{E}& \propto &
[\epsilon_{\l}+\epsilon_{\k+\q}+\epsilon_{\k+\q+\l}]^{-1}
\propto [Q+S+(\k\q)]^{-1},
\nonumber \\
{\rm No}\ 42: \qquad \phi^{-1}_{E}& \propto &
[\epsilon_{\k}+\epsilon_{\q}+\epsilon_{\l}+\epsilon_{\k+\q+\l}]^{-1}
\propto [Q+S]^{-1},
\label{Energy}
\end{eqnarray}
where $Q \equiv k^{2}+q^{2}+l^{2}$ and $S \equiv (\k\q)+(\k\l)+(\q\l)$.

From the obvious identities $(\k+\q+\l)^{2} \ge 0$,
$(\k-\q)^{2}+(\k-\l)^{2}+(\q-\l)^{2}\ge 0$, $|(\k\q)| \le kq$,
$|(\k\l)| \le kl$ and $|(\q\l)| \le ql$ it follows that
\begin{eqnarray}
-Q/2 \le S \le Q, \quad |(\k\q)| \le Q/2,
\quad |(\k\l)| \le Q/2, \quad  |(\q\l)| \le Q/2.
\label{INEQ}
\end{eqnarray}

It then follows that the factor (\ref{Energy}) in diagram No 42 can be
expanded in the powers of the ratio $S/Q$ with $|S/Q|\le1$, while for No 23
this is impossible: in the integration region, there is a subregion of the
same dimension (namely, $3\times d$) in which $2|S|> Q$. This difficulty
can be circumvented by means of the shift of the point around which the
expansion is made in diagram No~23:
\begin{eqnarray}
[Q+2S]^{-1} = [ 3Q/2 +(4S-Q)/2] ^{-1} =
[ 2/3Q] \sum^{\infty}_{n=0} [(1-4S/Q)/3]^{n}.
\label{23}
\end{eqnarray}
The convergence of the series in (\ref{23}) is ensured by the
inequality $|(1-4S/Q)/3|\le1$, which follows from Eq. (\ref{INEQ}).
The equality takes place only in the $2\times d$-dimensional subregion
$\k+\q+\l=0$, which has zero measure in the $3\times d$-dimensional
integration region and does not spoil the convergence of the corresponding
series for the integral (similar considerations ensure the convergence
of the series (\ref{L13c}) for the two-loop integral $h$).

For the factor (\ref{Energy}) in diagram No 32 the following shift
can be used:
\[ [Q+S+(\q\l)]^{-1} = \{ 5Q/4 + [S+(\q\l)-Q/4] \}^{-1} \]
with the subsequent expansion in the powers of the ratio
$[S+(\q\l)-Q/4] / (5Q/4)$, whose modulus is $\le1$ owing to inequalities
(\ref{INEQ}). Similar expansions (with obvious modifications) can be
written for diagrams Nos~33,34.

Such infinite series contain additional finite sums originated from the
powers of the expansion parameters, which have the forms ``constant + linear
combination of the scalar products.''  It is clear that this summation
should be performed in the first place, before all the other summations:
only this order of summations ensures the convergence of the series.

Calculation of the angular (\ref{V71}) and momentum (\ref{Integral})
integrals is a separate task; it is discussed in the next two Subsections.
The results for the coefficients $A_{i}$ for all the normal three-loop
diagrams are presented in the Appendix. The $\eps^{-3}$ contributions in
all diagrams have been found analytically for general space dimensionality
$d$ (in order to calculate them, it is sufficient to neglect all the scalar
products in the energy denominators $\phi_{E}$), while the $\eps^{-2}$ and
$\eps^{-1}$ parts have been found for the physical cases $d=2,3$ and in the
large $d$ limit; the results have been presented in Ref. \cite{cube} (of
course, this calculation can be made for any fixed given value of $d$).

Now, using the standard scheme (see Sec.~\ref{sec:Operators}), one finds the
three-loop contribution $\left[ Z^{-1}_{F} \right]_{3}$ in the expansion
(\ref{3*}) for the renormalization constant $Z^{-1}_{F}$. The anomalous
dimension $\gamma_{F} \equiv \gamma_{nl}$ is given by the relation
$\gamma_{F} = -\beta \partial _{u}\ln Z_{F}^{-1}$; see Eq. (\ref{Matrix2}).
Substituting Eqs. (\ref{RGF2}) and (\ref{3*}) into the
last relation expresses the anomalous dimension in the coefficients
$\left[ Z^{-1}_{F} \right]_{k}$. Within our accuracy one obtains
\begin{equation}
\gamma_{F}(u)= \bigl[\eps-u\,(d-1)/2d\bigr] \left\{
\left[Z^{-1}_{F}\right]_{1} +
2 \left[Z^{-1}_{F}\right]_{2} -
\left[Z^{-1}_{F}\right]_{1} ^{2} +
3\left[Z^{-1}_{F}\right]_{3} -
3\left[Z^{-1}_{F}\right]_{1}\left[Z^{-1}_{F}\right]_{2} +
 \left[Z^{-1}_{F}\right]_{1} ^{3} \right\}.
\label{logarifm}
\end{equation}
Substituting the known expressions for $\left[ Z^{-1}_{F} \right]_{k}$
into Eq. (\ref{logarifm})
gives the anomalous dimension $\gamma_{F}(u)$ to order $u^{3}$.

Since the quantity $\gamma_{F}$ is UV finite, that is, finite at $\eps=0$,
the pole parts must cancel each other in Eq. (\ref{logarifm}). This implies
some exact relations between the senior poles in the quantities
$\left[ Z^{-1}_{F} \right]_{k}$ with $k\ge2$
($\eps^{-2}$ in $\left[ Z^{-1}_{F} \right]_{2}$ and $\eps^{-2}$,
$\eps^{-3}$ in $\left[ Z^{-1}_{F} \right]_{3}$) and the $\eps^{-1}$
parts of the previous orders in $u$.
Such relations provide an additional possibility to control
the absence of calculational errors. In fact, the knowledge of senior poles
($\eps^{-k}$ with $k\ge2$) is needed only to check this cancellation;
the nonvanishing contributions in Eq. (\ref{logarifm}) are completely
determined by the $\eps^{-1}$ terms in the quantities
$\left[ Z^{-1}_{F} \right]_{k}$. In the MS scheme, the anomalous dimension
$\gamma_{F}(u)$ appears independent of $\eps$; this is a consequence of the
relation $Z_{F}=1+$ only poles in $\eps$.

Finally, the relation (\ref{Matrix2}) with $u_{*}$ from Eq. (\ref{FP}) gives
the $O(\eps^{3})$ results for the critical dimensions presented in
Ref. \cite{cube} and Eqs.~(\ref{Qnp3}).

\subsection{Angular integrations in the three-loop diagrams}
\label{sec:Angular}

The two-loop integrals (\ref{V28b}) involve only two vectors $\k$ and $\q$
and one angle $\vartheta$ between them, so that the procedure of the
angular averaging reduces there to the only standard formula (\ref{L2}).
The three-loop integrals (\ref{V28c}) involve three vectors $\k$, $\q$
and $\l$ and three angles between them, so that the calculation of the
quantities (\ref{V71}) is not all that simple.
Below we present the results of this calculation.

Obviously, the quantity (\ref{V71}) differs from zero only if all the three
numbers $n_{1,2,3}$ are simultaneously even or odd. It is also clear that
the quantity (\ref{V71}) is symmetrical with respect to any permutation of
the exponents $n_{1,2,3}$, and with no loss of generality it can be assumed
that $n_{2}$ is the minimal exponent:
\begin{equation}
n_{2} = {\rm min}\, \{ n_{1},n_{2},n_{3} \}
\label{V72}
\end{equation}
(the notation in the formulas below is consistent with the
assumption (\ref{V72})).

The straightforward calculation (first, the averaging over the direction of
one momentum, say $\k$, and then the averaging over the angle between the
two remaining vectors $\q$ and $\l$) leads to the following result for
the quantity (\ref{V71}):
\begin{equation}
T_{n_{1}n_{2}n_{3}} = k^{n_{1}+n_{2}}\,q^{n_{1}+n_{3}}\,l^{\,n_{2}+n_{3}}\,
\overline T_{n_{1}n_{2}n_{3}}
\label{V73}
\end{equation}
where the moduli of the vectors are isolated explicitly, and
\begin{equation}
\overline T_{n_{1}n_{2}n_{3}} = \alpha_{n_{1}+n_{2}} \alpha_{n_{2}+n_{3}}\,
K(n_{1}, n_{2})\, {\widetilde T} _{n_{1}n_{2}n_{3}}
\label{V74}
\end{equation}
with coefficients $\alpha_{2n}$ from Eq.~(\ref{L2}) and
\begin{equation}
K(n_{1}, n_{2})= \frac {2^{n_{2}}\,n_{1}!\,[(n_{1}+n_{2})/2]!}
{(n_{1}+n_{2})!\, [(n_{1}-n_{2})/2]!} =
\frac {(n_{1}-n_{2}+2)(n_{1}-n_{2}+4)\dots(n_{1}+n_{2})}
{(n_{1}+1)(n_{1}+2)\dots(n_{1}+n_{2})} \, ,
\label{V75}
\end{equation}
\begin{equation}
{\widetilde T} _{n_{1}n_{2}n_{3}} =\sum_{k=0}^{n_{2}/2} \, Y_{k} , \qquad
Y_{k}= \frac {2^{-2k} \,n_{2}!\, [(n_{1}-n_{2})/2]! \,
\alpha_{n_{2}+n_{3}-2k}} {k!\,(n_{2}-2k)!\, [(n_{1}-n_{2})/2+k]! \,
\alpha_{n_{2}+n_{3}}} \, .
\label{V76}
\end{equation}
We recall that the numbers $n_{1,2,3}$ in these expressions have the same
parity, and that $n_{2}$ is the minimal one according to Eq.~(\ref{V72}).
The upper limit in the sum (\ref{V76}) for odd $n_{2}$ is understood as
the integer part of $n_{2}/2$.

The following special values and recurrent relations for the quantity
$K$ in Eq. (\ref{V75}) are useful in the computer calculations:
\[ K(m,0)=K(m,1)=1, \quad K(m,m)=m!/(2m-1)!!, \quad
K(m+1,m+1)/K(m,m) = (m+1)/(2m+1), \]
\begin{eqnarray}
K(m+2,n)/K(m,n) = (m+1)(m+2)\, / \, (m-n+2)(m+n+1).
\label{V77}
\end{eqnarray}
For the terms $Y_{k}$ in the sum (\ref{V76}) one has:
\begin{equation}
Y_{0}=1, \quad  Y_{k+1}/Y_{k}= \frac{(n_{2}-2k)(n_{2}-2k-1)
(d+n_{2}+n_{3}-2k-2)} {2(k+1)(n_{1}-n_{2}+2k+2)(n_{2}+n_{3}-2k-1)} \, .
\label{V78}
\end{equation}
Using these relations, the quantities (\ref{V75}), (\ref{V76}) are easily
calculated by a computer.

\subsection{Modular integrations in the three-loop diagrams} \label{sec:Mom}

Let us turn to the calculation of the three-loop integrals (\ref{Integral})
over the moduli $k,q,l$, which arise as coefficients in the expansion of the
quantities $A_{i}$ in Eq.~(\ref{V28c}) in the scalar products $(\k\q),
(\k\l), (\q\l)$; see Sec.~\ref{sec:Nerav}. For the complete
three-loop calculation, it is sufficient to know some special cases of the
general integral (\ref{Integral}), which we denote $I_{1}$--$I_{9}$ in what
follows. Below we give the explicit answers for these intergals, with the
precise specification of the indices $n_{1}$--$n_{5}$, and then turn to the
derivation. Only the pole parts $\eps^{-3}$, $\eps^{-2}$ and $\eps^{-1}$ of
these integrals are given, which is sufficient for the calculation of the
quantities $A_{i}$ within our accuracy. We shall use the notation
\begin{equation}
I \bigl\{ F \bigr\} \equiv
\int^{\infty}_{1} \frac{dk}{k^{1+\eps}}
\int^{\infty}_{1} \frac{dq}{q^{1+\eps}}
\int^{\infty}_{1} \frac{dl}{l^{1+\eps}} \, F
\label{LL}
\end{equation}
for any function $F=F(k,q,l)$.
The integrals $I_{1}$--$I_{9}$ are as follows:
\begin{equation}
I_{1} \equiv I  \left\{ \frac{k^{2n_{1}}q^{2n_{2}}l^{2n_{3}}}
{(k^{2}+q^{2}+l^{2})^{n_{4}}(q^{2}+l^{2})^{n_{5}}} \right\} =
\frac{(n_{1}-1)!(n_{2}-1)!(n_{3}-1)!(n_{4}-n_{1}-1)!}
{12\eps\,(n_{4}-1)!(n_{2}+n_{3}-1)!} \label{I1}
\end{equation}
with $n_{1}+n_{2}+n_{3}=n_{4}+n_{5}$, $n_{4}>n_{1}$, $n_{1,2,3,4}>0$
and $n_{5}\ge 0$;
\begin{equation}
I_{2} \equiv I \left\{ \frac{q^{2n_{2}}l^{2n_{3}}}
{(k^{2}+q^{2}+l^{2})^{n_{4}}(q^{2}+l^{2})^{n_{5}}} \right\} =
\frac{(n_{2}-1)!(n_{3}-1)!}{12(n_{2}+n_{3}-1)!}\, \left[
\frac{1}{\eps^{2}} + \frac{1}{\eps} \left(
\sum^{n_{2}+n_{3}-1}_{k=1} \frac{1}{k} -
\sum^{n_{4}-1}_{k=1} \frac{1}{k} -  \frac{1}{2}
\sum^{n_{2}-1}_{k=1} \frac{1}{k} -  \frac{1}{2}
\sum^{n_{3}-1}_{k=1} \frac{1}{k} \right)\right]
\label{I2}
\end{equation}
with $n_{2}+n_{3}=n_{4}+n_{5}$, $n_{2,3,4}>0$ and $n_{5}\ge 0$ (in Eq.
(\ref{I2}) and all the formulas below, any sum with the upper limit lesser
than the lower one is understood as equal to zero);
\begin{equation}
I_{3} \equiv  I\left\{ \frac{k^{2n_{1}}q^{2n_{2}}}
{(k^{2}+q^{2}+l^{2})^{n_{4}}} \right\} =
\frac{(n_{1}-1)!(n_{2}-1)!}{12(n_{4}-1)!}\, \left[
\frac{1}{\eps^{2}} - \frac{1}{2\eps} \left(
\sum^{n_{1}-1}_{k=1} \frac{1}{k} + \sum^{n_{2}-1}_{k=1} \frac{1}{k}
\right)\right]
\label{I3}
\end{equation}
with $n_{1}+n_{2}=n_{4}$ and  $n_{1,2}>0$;
\begin{equation}
I_{4} \equiv I \left\{ \frac{k^{2n_{1}}q^{2n_{2}}}
{(k^{2}+q^{2}+l^{2})^{n_{4}}(q^{2}+l^{2})} \right\} =
\frac{(n_{1}-1)!(n_{2}-2)!}{12\eps\,(n_{4}-1)!}\, \left[
\frac{1}{\eps^{2}} - \frac{1}{2\eps} \left(
\frac{2}{(n_{2}-1)} +
\sum^{n_{1}-1}_{k=1} \frac{1}{k} + \sum^{n_{2}-2}_{k=1} \frac{1}{k}
\right)\right]
\label{I4}
\end{equation}
with $n_{1}+n_{2}=n_{4}+1$, $n_{1}>0$ and  $n_{2}>1$;
\begin{equation}
I_{5} \equiv  I\left\{ \frac{k^{2}}{(k^{2}+q^{2}+l^{2})}
\left(\frac{ql}{q^{2}+l^{2}}\right)^{2n} \right\} =
\frac{[(n-1)!]^{2}}{12(2n-1)!} \, \left[ \frac{2}{\eps^{2}} -
\frac{1}{\eps} \left( 3{\cal B}_{n-1} +
\sum_{k=n}^{2n-1}\frac{1}{k} \right)\right] \label{I5}
\end{equation}
with $n>0$ and ${\cal B}_{n}$ from Eq.~(\ref{L12});
\begin{equation}
I_{6} \equiv  I\left\{ \frac{k^{2}q^{2(n+1)}l^{2n}}
{(k^{2}+q^{2}+l^{2})(q^{2}+l^{2})^{2n+1}} \right\} =
\frac{1}{2} \, I_{5};
\label{I6}
\end{equation}
\begin{equation}
I_{7} \equiv  I\left\{ \frac{k^{2}} {k^{2}+q^{2}+l^{2}} \right\} =
\frac{1}{3\eps^{3}} ;
\label{I7}
\end{equation}
\begin{equation}
I_{8} \equiv  I\left\{ \frac{k^{2}q^{2}}
{(k^{2}+q^{2}+l^{2})(q^{2}+l^{2})} \right\} =
\frac{1}{6\eps^{3}};
\label{I8}
\end{equation}
\begin{equation}
I_{9} \equiv  I\left\{ \frac{q^{4}}
{(k^{2}+q^{2}+l^{2})(q^{2}+l^{2})} \right\} =
\frac{1}{3\eps^{3}} -  \frac{1}{12\eps^{2}} - \frac{1}{12\eps}.
\label{I9}
\end{equation}

Now let us turn to the derivation of expressions (\ref{I1})--(\ref{I9}).
The integrals $I_{1,2}$ can be obtained from the generation function
\begin{equation}
R(a,b;m) \equiv \int^{\infty}_{m} \frac{dk}{k^{1+\eps}}
\int^{\infty}_{m} \frac{d\rho}{\rho^{1+2\eps}} \frac {\rho^{2}}
{(ak^{2}+b\rho^{2})} = \frac{m^{-3\eps}}{6b} \left[
\frac{1}{\eps^{2}} +  \frac{1}{\eps} \, \ln (b/a) + O(1) \right] .
\label{GenF}
\end{equation}
We shall also need the first terms of the $\eps$ expansion of the integral
\begin{eqnarray}
J(p,q) &\equiv& \int^{\pi/2}_{0} d\varphi \, (\cos \varphi)^{2p-1-\eps}
(\sin \varphi)^{2q-1-\eps}  = \frac{\Gamma(p-\eps/2)
\Gamma(q-\eps/2)} {2\Gamma(p+q-\eps)} =
\nonumber\\ &=& \frac{\Gamma(p) \Gamma(q)} {2\Gamma(p+q)}
\left\{ 1+ \eps \left[ \psi (p+q) -  \frac{1}{2} \psi (p)
-  \frac{1}{2} \psi (q) \right]\right\} +O(\eps^{2}) =
\nonumber\\ &=&
\frac{(p-1)!(q-1)!}{2(p+q-1)!} \left[ 1+\eps \left(
\sum^{p+q-1}_{k=1} \frac{1}{k} - \frac{1}{2}
\sum^{p-1}_{k=1} \frac{1}{k} -  \frac{1}{2}
\sum^{q-1}_{k=1} \frac{1}{k} \right)\right]+O(\eps^{2}),
\label{GenF2}
\end{eqnarray}
where $\Gamma(z)$ is the Euler Gamma function and
$\psi(z) = d \ln \Gamma(z) /dz$.

The calculation of the quantity $R\equiv R(a,b;m)= m^{-3\eps}R(a,b;1)$
is similar to the derivation of Eq. (\ref{L11}). The operation ${\cal D}_{m}
\equiv m\partial/\partial m$ is applied to the double integral in Eq.
(\ref{GenF}), which reduces it to the sum of two single integrals and
explicitly isolates the pole factor $\eps^{-1}$:
\[ R = -\frac{1}{3\eps} {\cal D}_{m} R = \frac{m^{-3\eps}}{3\eps} \left[\,\,
\int_{1}^{\infty} dk \, \frac {1}{k^{1+\eps}(ak^{2}+b)} +
\int_{1}^{\infty} \, \frac {d\rho}{\rho^{1+2\eps}}
\frac {\rho^{2}} {(a+b\rho^{2})} \right]. \]
Now one can set $\eps=0$ in the integral over $k$ (it remains finite at
$\eps=0$). The pole in $\eps$ in the integral over $\rho$ comes
from large $\rho$ and it can be isolated explicitly:
\[ \int_{1}^{\infty}  \, \frac {d\rho}{\rho^{1+2\eps}}
\frac {\rho^{2}} {(a+b\rho^{2})}  = \frac{1}{b}
\int_{1}^{\infty}  \, \frac {d\rho}{\rho^{1+2\eps}} - \frac{a}{b}
\int_{1}^{\infty}  \, \frac {d\rho}{\rho^{1+2\eps}}
\frac {1} {(a+b\rho^{2})} = \frac{1}{2b\eps}- \frac{a}{b}
\int_{1}^{\infty} \frac{d\rho}{\rho(a+b\rho^{2})} +O(\eps) , \]
which immediately leads to the answer (\ref{GenF}).

In the integral $I_{1}$ the pole contribution comes from the part of the
integration region where all three momenta $k,q,l$ simultaneously tend to
infinity, and in $I_{2}$ there is also a pole contribution coming from the
subregion where $q,l$ simultaneously tend to infinity at fixed finite $k$.
This means that the pole parts of the integrals $I_{1,2}$ are not affected
if one changes to the polar coordinates $\rho$, $\varphi$
($q=\rho\cos\varphi$, $l=\rho\sin\varphi$) with the integration region
$0\le\varphi\le\pi/2$, $1\le\rho<\infty$, that is, the following
replacement is performed:
\[ \int_{1}^{\infty} dq \int_{1}^{\infty} dl \dots \to
 \int_{0}^{\pi/2} d\varphi \int_{1}^{\infty}\rho\, d\rho\, \dots\, . \]
The integration regions in these two expressions are not identical, but the
values of integrals differ only by an unessential contribution,
finite at $\eps\to0$. For $I_{1}$ this gives:
\[ I_{1} = \int_{0}^{\pi/2} d\varphi\, (\cos\varphi)^{2n_{2}-1-\eps}
(\sin\varphi)^{2n_{3}-1-\eps} \int_{1}^{\infty} \frac{dk}{k^{1+\eps}}
\int_{1}^{\infty}
\frac{d\rho}{\rho^{1+2\eps}} \, \frac{k^{2n_{1}}\rho^{2n_{4}-2n_{1}}}
{(k^{2}+\rho^{2})^{n_{4}}}. \]
This integral can be expressed using $J(p,q)$ from Eq. (\ref{GenF2})
and the derivative of the generating function $R(a,b;m)$ from Eq.
(\ref{GenF}) as follows:
\begin{equation}
I_{1} = [(n_{4}-1)!]^{-1}\,  J(n_{2},n_{3}) (- \partial _{a})^{n_{1}}
(- \partial _{b})^{n_{4}-n_{1}-1} R(a,b;1) |_{a=b=1} \, ,
\label{lohi}
\end{equation}
with $\partial _{a} = \partial /\partial a$,
$\partial _{b} = \partial /\partial b$, which immediately leads to the
answer (\ref{I1}). Note that the $\eps^{-2}$ contribution in $R$ depends
only on $b$ (see Eq. (\ref{GenF})) and therefore vanishes in the above
expression owing to the differentiation with respect to $a$ (we recall
that $n_{1}>0$ for $I_{1}$). Although the integral $I_{2}$ is formally
given by expression (\ref{lohi}) with $n_{1}=0$,
\begin{equation}
I_{2} = [(n_{4}-1)!]^{-1}\,   J(n_{2},n_{3})
(- \partial _{b})^{n_{4}-1} R(1,b;1) |_{b=1} \, ,
\label{lohi2}
\end{equation}
the $\eps^{-2}$ part of the function (\ref{GenF}) survives in Eq.
(\ref{lohi2}) owing to the absense of the derivative $\partial_{a}$,
so that the $O(\eps)$ contributions should be taken into account
in the integral (\ref{GenF2}).
This explains the sharp difference between the final expressions
(\ref{I1}), (\ref{I2}) for the integrals $I_{1}$ and $I_{2}$.

The relations (\ref{I3})--(\ref{I6}) follow easily from
(\ref{I1}), (\ref{I2}). The result (\ref{I3}) for $I_{3}$ is
obtained from Eq. (\ref{I2}) at $n_{5}=0$ using the symmetries of
the integrand. In its turn, the result (\ref{I4}) for $I_{4}$
follows from Eqs. (\ref{I1}), (\ref{I3}) if the obvious identity $
q^{2n_{2}} /(q^{2}+l^{2}) = q^{2(n_{2}-1)} - q^{2(n_{2}-1)}l^{2}
/(q^{2}+l^{2})$ is substituted on its left-hand side. The result
(\ref{I5}) for $I_{5}$ is obtained using the substitution $k^{2} /
(k^{2}+q^{2}+l^{2}) =1- (q^{2}+l^{2})/(k^{2}+q^{2}+l^{2})$: in the
first term, the integral over $k$ is trivial and the remaining
integrations over $q,l$ are performed using Eqs. (\ref{L8}),
(\ref{L11}), (\ref{L12}), while the second term reduces to
$I_{2}$. The integral $I_{6}$ in Eq. (\ref{I6}) reduces to $I_{5}$
after the symmetrization of the integrand with respect to $q,l$.

The results (\ref{I7}), (\ref{I8}) follow from the obvious relation $I\{1\}=
\eps^{-3}$ in Eq. (\ref{LL}) after the symmetrization of the integrands
with respect to $k,q,l$. The last relation (\ref{I9}) is obtained using
the substitution $q^{2}/(q^{2}+l^{2}) =1 -l^{2}/(q^{2}+l^{2})$: the first
term coincides with $I_{7}$ and the second one is given by $I_{2}$ with
$n_{2}=n_{3}=n_{4}=n_{5}=1$.

\subsection{Anomalous exponents at large ${\lowercase{d}}$}
\label{sec:larged}

In this Section, we shall briefly discuss the behavior of the coefficients
$\Delta^{(k)}_{nl}$ in the expansion (\ref{epsilon}) for $d\to\infty$.
The model (\ref{1})--(\ref{3}) has no finite upper critical dimension,
above which the anomalous scaling vanishes. It disappears for
infinite $d$ \cite{Infty}, but reveals itself already in the $O(1/d)$
approximation \cite{Falk1}. This fact confirms the relevance of the large
$d$ expansions for the issue of anomalous scaling; see also Refs.
\cite{FFR,AR} for the discussion of the Navier--Stokes problem.

It follows from Eqs. (\ref{L2}) that for $d\to\infty$, the angular averages
$\alpha_{2n} \equiv \bigl\langle \cos^{2n} \vartheta \bigr\rangle$
behave as $\alpha_{2n}\propto d^{-n}$ (each additional factor
$\cos^{2}\vartheta$ introduces additional smallness in $1/d$).
Then from Eqs. (\ref{V73})--(\ref{V76}) for the averages (\ref{V71})
one obtains $T_{n_{1}n_{2}n_{3}} \propto d^{-(n_{1}+n_{2}+n_{3})/2}$
for $n_{1,2,3}$ even and
$T_{n_{1}n_{2}n_{3}} \propto d^{-(n_{1}+n_{2}+n_{3}+1)/2}$
for $n_{1,2,3}$ odd. This means
that, in order to find the behavior of the coefficients $A_{i}$ for
large $d$ to any given finite order in $1/d$, one needs to take into
account only {\it finite} number of terms in the expansion of the
integrands in Eq.~(\ref{V24b}) in the scalar product $(\k\q)$ and
integrands in Eq.~(\ref{V24c}) in the set of scalar products
$(\k\q)$, $(\k\l)$, $(\q\l)$.

The $d$ dependence of the coefficient $\Delta^{(1)}_{nl}$ in Eq.
(\ref{epsilon}) is known from Eq. (\ref{Qnp}), while the quantities
$\Delta^{(2)}_{nl}$, $\Delta^{(3)}_{nl}$ can be found as series in $1/d$
to any given order as explained above. For general $n$ and $l$ to order
$1/d^{2}$ we have obtained
\begin{eqnarray}
\Delta_{nl} &=& \eps \big[ -n(n-2) (1-2/d) /2d + (l/2) (1-2/d+l/d
+2/d^{2}) \big] + 3\eps^{2} (n-2)(n-l) /4d^{2} +
\nonumber \\
&+& \eps^{3} (n-l) \big[ 1.74988(n-2)-0.624916\,l \big] /d^{2}
+  O(\eps^{4}).
\label{Qd}
\end{eqnarray}

Note that the $\eps^{2}$ and $\eps^{3}$ contributions decay for
$d\to\infty$ faster than $1/d$ in agreement with the $O(1/d)$ result
obtained in Ref. \cite{Falk1} for $\Delta_{n0}$. Moreover, from Eq.
(\ref{Qd}) it follows that the leading $O(1/d^{2})$ terms in these
contributions vanish for $n=l$, so that the decay at $d\to\infty$
becomes even faster. For $n=l$ we have obtained
\begin{eqnarray}
\Delta_{nn} &=& \eps n/2 + n(n-1)\, \bigl\{ \eps\, /(d-1)(d+2) -
  \eps^{2} \big[ 1+ (2n-7)/d \big] / d^{3} -
\eps^{3} (3n-8) /2d^{4} \bigr\} +  O(\eps^{4})
\label{Qdd}
\end{eqnarray}
with the accuracy of $O(1/d^{4})$.

\section{Convergence of the $\eps$ expansion, inverse $\eps$ expansion,
and comparison with nonperturbative results} \label {sec:Inverse}

An important issue which can be discussed on the example of the rapid-change
model is that of the nature and convergence of $\eps$ expansions in models
of turbulence and the possibility of their extrapolation to finite values of
$\eps\sim1$. The knowledge of the three terms of the $\eps$ expansion in
model (\ref{1})--(\ref{3}) allows one to discuss its convergence properties
and to obtain improved predictions for finite $\eps$ in reasonable agreement
with the existing nonperturbative results: analytical solution of the
zero-mode equations for $n=2$ \cite{Falk1}, numerical solutions for $n=3$
\cite{Pumir} and numerical experiments for $n=4$ \cite{VMF1,VMF2} and $n=6$
\cite{MM}.

In Figs.~3{\it a} and 3{\it b}, we show the anomalous dimension
$\gamma\equiv\gamma_{22}^{*}\equiv\gamma_{22}(u_{*})$ for $d=3$ ({\it a})
and 2 ({\it b}) in the $O(\eps)$, $O(\eps^{3})$ and $O(\eps^{2})$
approximations (from above to below); the latter two obtained as simple
sums of two and three terms of the $\eps$ expansion, respectively.
The dashed line corresponds to the exact solution by Ref. \cite{Falk1};
see also Ref. \cite{Gat} for the special cases $d=2$ and~3. Analogous
diagrams for the cases $n=3$, $l=1$ and $n=4$, $l=0$ can be found in
Ref.~\cite{cube}.

All these figures show that the agreement between the $\eps$ expansion and
nonperturbative results for small $\eps$ improves when the higher-order
terms are taken into account, but the deviation becomes remarkable for
$\eps\sim1$ and decreasing $d$. Furthermore, the convergence of the $\eps$
series appears more irregular for $d=2$, while the forms of the
nonperturbative results are not much affected by the choice of $d$.

Such behavior can be naturally understood on the example of the exact
analytical result for $\gamma\equiv\gamma_{22}^{*}$ \cite{Falk1}, which
can be written in the form
\begin{equation}
2 \gamma = -(d+2+\eps) + \sqrt{(\eps+\eps_{+})(\eps+\eps_{-})},
\qquad
\eps_{\pm} = \left(d^{2}+d+2 \pm \sqrt {8d(d+1)}\,\right)/(d-1)
\label{koren}
\end{equation}
with $\eps_{+}\eps_{-}=(d+2)^{2}$. It is useful
to rewrite the smaller quantity $\eps_{-}$ in the form
\begin{equation}
\eps_{-}=(d-1)\left\{1+\left(1+3d+\sqrt{8d(d+1)}\right)^{-1}\right\}.
\label{koren2}
\end{equation}

From these expressions it follows that the corresponding $\eps$ expansion
has the finite radius of convergence $\eps_{-}$, ranging from 0 to $\infty$
when $d$ varies from 1 to $\infty$ (in particular, $\eps_{-}\simeq1.1$,
$\eps_{+}\simeq14.5$ for $d=2$ and $\eps_{-}\simeq2.1$, $\eps_{+}\simeq11.9$
for $d=3$). Hence, the naive summation of the $\eps$ expansion for
$\gamma$ works only in the interval $\eps<\eps_{-}$, which decreases
almost linearly with $(d-1)$.
Since the singularity in Eq. (\ref{koren}) occurs for negative $\eps$, it
affects strongly the convergence of the $\eps$ expansion but is not
``visible'' in the form of the exact curve for positive $\eps$, in contrast
with the singularities occurring at $\eps=2$ in higher-order critical
dimensions \cite{Pumir,Siggia,VMF1,VMF2}.

Therefore, in order to recover the behavior of $\gamma$ from its $\eps$
series for larger $\eps$, it is necessary to isolate explicitly the
singularity at $\eps=-\eps_{-}$ in Eq. (\ref{koren}), thus changing to
a kind of improved $\eps$ expansion, whose radius of convergence is
determined by a more distant singularity. This can be done, for example,
by introducing the new expansion parameter
$x = \sqrt{\eps+\eps_{-}} - \sqrt{\eps_{-}}$, that is,
$\eps= x^{2} +2x \sqrt{\eps_{-}}$.
Then Eq. (\ref{koren}) can be written in the form
\begin{equation}
2 \gamma = - \left(d+2+2x \sqrt{\eps_{-}}+x^2\right) +
\left(x+\sqrt{\eps_{-}}\right)\, \sqrt{ x^{2} + 2x \sqrt{\eps_{-}} +
{\eps_{+}}}   \,.
\label{koren3}
\end{equation}
The convergence radius of the expansion in $x$ is found from the equation
$x^{2} + 2x \sqrt{\eps_{-}} + {\eps_{+}}=0$ with the solutions
$x_{\pm} = - \sqrt{\eps_{-}} \pm {\rm i} \sqrt{\eps_{+}-\eps_{-}}$
and is therefore equal to $|x_{\pm}| =\sqrt{\eps_{+}}$.
For positive $x$ and $\eps$, this corresponds to the convergence for
$0<x<\sqrt{\eps_{+}}$ or, equivalently,
$0<\eps < \eps_{+} + 2\sqrt{\eps_{-}\eps_{+}} = \eps_{+}+ 2(d+2)$.

The improvement of the convergence is illustrated by Fig.~3{\it c}, where
the exact exponent $\gamma$ for $d=2$ is shown as a function of $\eps$
along with its first ($x$), second ($x$ and $x^{2}$) and third ($x$,
$x^{2}$ and $x^{3}$) orders of the improved $x$ expansion, in which the
variable $x$ is also expressed as a function of $\eps$.
One can see that the convergence of the $x$ expansion appears
more regular and that its third-order approximation is hardly distinguishable
from the exact result for all $0<\eps<2$ (for $d=3$, the agreement is even
better and for this reason is not shown). It should be stressed that crucial
for the improvement was not the existence of the exact solution or explicit
form of the substitution $x(\eps)$ but the knowledge of the character and
location of the singularity which determines the convergence properties of
the plain $\eps$ expansion.

The difference with the models of critical phenomena, where $\eps$
series are always asymptotical, can be traced back to the fact that in
the rapid-change models, there is no factorial growth of the number of
diagrams in higher orders of the perturbation theory. The divergence
for $d\to1$ is naturally explained by the fact that the transverse vector
field does not exist in one dimension. We also recall that the RG
fixed point diverges at $d=1$; see Eq. (\ref{FP}), so that the coefficients
of the $\eps$ series diverge for all dimensions $\Delta_{nl}$.

It is then natural to assume that the $\eps$ series for all $\Delta_{nl}$
also have finite radii of convergence with the behavior similar to that of
$\eps_{-}$ in Eq. (\ref{koren2}). Therefore, in order to improve their
convergence and to obtain reasonable predictions for finite values of $\eps$,
one should augment plain $\eps$ expansions by the information about the
character of the singularities and their location in the complex $\eps$ plane.
Such information can be extracted from the asymptotical behavior of the
coefficients $\Delta^{(k)}_{nl}$ in Eq. (\ref{epsilon}) at large $k$.
To our knowledge, the large $k$ behavior of the $\eps$ series remains an
open problem for any dynamical model; the instanton analysis developed in
Refs. \cite{instanton} has mostly been concentrated on the behavior of the
exponents in the limit $n\to\infty$. One can hope that the implementation
of the instanton calculus within the RG framework will give the solution of
this important problem.

It turns out, however, that certain elementary considerations allow one to
improve the convergence of the $\eps$ series and, at the same time, to
achieve a better agreement with the nonperturbative results. Let us explain
the idea on the example of the exact solution (\ref{koren}). We express
$\eps$ as a function of the exponent $\gamma$ using Eq. (\ref{koren})
and expand the right-hand side of the resulting exact relation
\begin{equation}
\eps = \gamma\,(d-1)(d+2+\gamma)/ [2-\gamma(d-1)]
\label{koren4}
\end{equation}
in $\gamma$; this gives the ``inverse $\gamma$ expansion.'' It is easy to see
that, for $\eps\sim1$ and physical dimensions $d=2$ and 3, the respective
values of $\gamma$ lie within the region of convergence of the inverted
series: $\gamma < 2/(d-1)$. The improvement of the convergence is also seen
from Fig.~3{\it d}, where the exponent $\gamma=\gamma^{*}_{22}$ for $d=2$
is shown as a function of $\eps$ along with the first ($\gamma$), second
($\gamma$ and $\gamma^{2}$) and third ($\gamma$, $\gamma^{2}$ and
$\gamma^{3}$) orders of the $\gamma$ expansion, expressed in the original
variable $\eps$: the approximate curves approach the
exact curve from the same side and represent the exact result (\ref{koren})
much better than the corresponding approximations of the plain $\eps$
expansion. The improvement is even better for $d=3$.

A simple explanation of the improvement follows. The convergence radius for
the direct $\eps$ series is determined by the singularity in the right-hand
side of Eq. (\ref{koren}), closest to the origin, that is,
$\eps_{c}=-\eps_{-}$. This {\it square-root} singularity disappears in the
inverse relation (\ref{koren4}), that is, the dependence of $\eps$ on
$\gamma$ in the vicinity of the corresponding point
$2\gamma_{c}=-d-2-\eps_{c}=-d-2+\eps_{-}$ becomes analytic. This would also
happen for any singularity of the form $\gamma - \gamma_{c} \propto
(\eps-\eps_{c})^{1/k}$ with any integer $k\ge0$.

We assume that these features are also typical to the
higher-order dimensions $\Delta_{nl}$ and construct the corresponding
$\gamma$ expansions. It is important here that $\gamma\equiv\gamma_{nl}=
O(\eps)$, so that there is a one-to-one correspondence between these two
expansions and three terms of the $\gamma$ expansion can be immediately
obtained from Eqs. (\ref{HZ3}), (\ref{Qnp2}) and (\ref{Qnp3}). This simple
procedure leads to remarkable improvement of the convergence and, at the same
time, the agreement with the numerical results, as is easily seen from
Figs.~4 and~5.

In Fig.~4, we show the anomalous dimension $\gamma\equiv\gamma_{31}^{*}$
for $d=3$ (left) and $d=2$ (right): the $O(\eps)$ approximation, the
third-order approximation of the inverse $\gamma$ expansion, and the
third-order approximation of the plain $\eps$ expansion (from above to
below); the dashed lines represent the exact numerical solution by Refs.
\cite{Pumir} ($\gamma=\lambda-3$ in the notation of \cite{Pumir}).

In Figs.~5{\it a} and 5{\it b}, we show the quantities
$\Delta_{n}\equiv \Delta_{n0}$ in three dimensions for $n=6$ ({\it a}) and
$n=4$ ({\it b}): the $O(\eps)$ slope, the third-order approximation of the
$\gamma$ expansion, and the third-order approximation of the plain $\eps$
expansion (from above to below); the dots connected by dashed lines
represent the results of the numerical simulations by
Refs. \cite{MM} ($n=6$) and \cite{VMF1,VMF2} ($n=4$).

In all these cases, the improvement in the agreement with nonperturbative
results is obvious. It should be emphasized, however, that even the plain
$\eps$ expansion captures some qualitative features of the dimensions
$\Delta_{n}$ established in the numerical simulations \cite{VMF1,VMF2,MM}:
the quantity $|\Delta_{n}|$ increases with $\eps$, achieves a maximum at some
point inside the interval $0<\eps<2$ and then decreases to zero; the height
of the maximum increases and its position moves to the left as $n$ grows from
4 to 6 or $d$ decreases from 3 to 2 ($\zeta_{2n}-n\zeta_{2}=\Delta_{2n}$
in the notation of Refs. \cite{VMF1,VMF2,MM}).

It is no surprise, of course, that the disagreement between the perturbative
and nonperturbative results becomes rather strong for $\eps>1$; this can be
explained, for example, by the effect of the singularity at $\eps=2$ in the
exact solutions \cite{Pumir,Siggia,VMF1,VMF2} and by insufficient number
of the known terms in the $\eps$ and $\gamma$ series. For the case $n=4$,
$d=3$, the situation can be improved by an interpolation formula which takes
into account the first terms of the $\eps$ expansion along with the
asymptotical behavior of the dimension $\Delta_{4}$ in the vicinity of the
opposite edge $\eps=2$, known from the numerical simulation \cite{VMF1,VMF2}:
$\Delta_{4} = - \left[ 0.06\,(2-\eps)+1.13\,(2-\eps)^{3/2}\right]$.
In particular, one can choose:
\begin{equation}
\Delta_{4} = - \, \Bigl[c_{1} \eps + c_{2}\eps ^2+ c_{3}\eps^3 + \dots
+ c_{k}\eps ^k+ \dots \Bigr]
\Bigl[ 0.06\,(2-\eps)+1.13\,(2-\eps)^{3/2}\Bigr].
\label{interpol}
\end{equation}
The first coefficients $c_{1}$--$c_{p}$ are determined by the requirement
that the expansion in $\eps$ of the right-hand side of Eq. (\ref{interpol})
reproduce the first $p$ terms of the $\eps$ expansion for $\Delta_{4}$,
known from the RG (therefore, in practice one can only take $p\le3$). The
values of these coefficients, once determined, will not change if one takes
a larger value of $p$. The remaining coefficients $c_{k}$ with $k>p$ should
be chosen to reproduce the correct behavior at $\eps\to2$. The simplest
possibility is to set $c_{k}=0$ for all $k>p+1$; then the last remaining
coefficient $c_{p+1}$ is unambiguously determined by the relation
$\left(2c_{1}+4c_{2}+ 8c_{3}+ \dots + 2^{p+1}c_{p+1} \right)=1\,$
[the $O(2-\eps)$ terms in the expression in the first square brackets
produce only corrections of order $(2-\eps)^2$ to the behavior of the
right-hand of Eq. (\ref{interpol}) side at $\eps\to2$ and thus should
be neglected]. This procedure gives $c_{1}=0.241$, $c_{2}=0.129$ for $p=1$
(upper curve in Fig.~5{\it c}) and $c_{1}=0.241$, $c_{2}=0.168$,
$c_{3}=-0.225$, $c_{4}=0.103$ for $p=3$ (lower curve). One can see that the
inclusion of the higher orders of the $\eps$ expansion improves remarkably
the agreement with the ``experimental'' results of Refs. \cite{VMF1,VMF2},
shown in Fig.~5{\it c} by the dots connected by dashed lines. It is also
worth noting that the value of the coefficient $c_{2}$ for $p=1$ (determined
by the behavior at $\eps\to2$) appears rather close to its value for $p=3$
(determined by the $\eps$ expansion), which demonstrates the robustness of
the results obtained by the above procedure.
\bigskip

\begin{figure}
\centerline{
\psfig{file=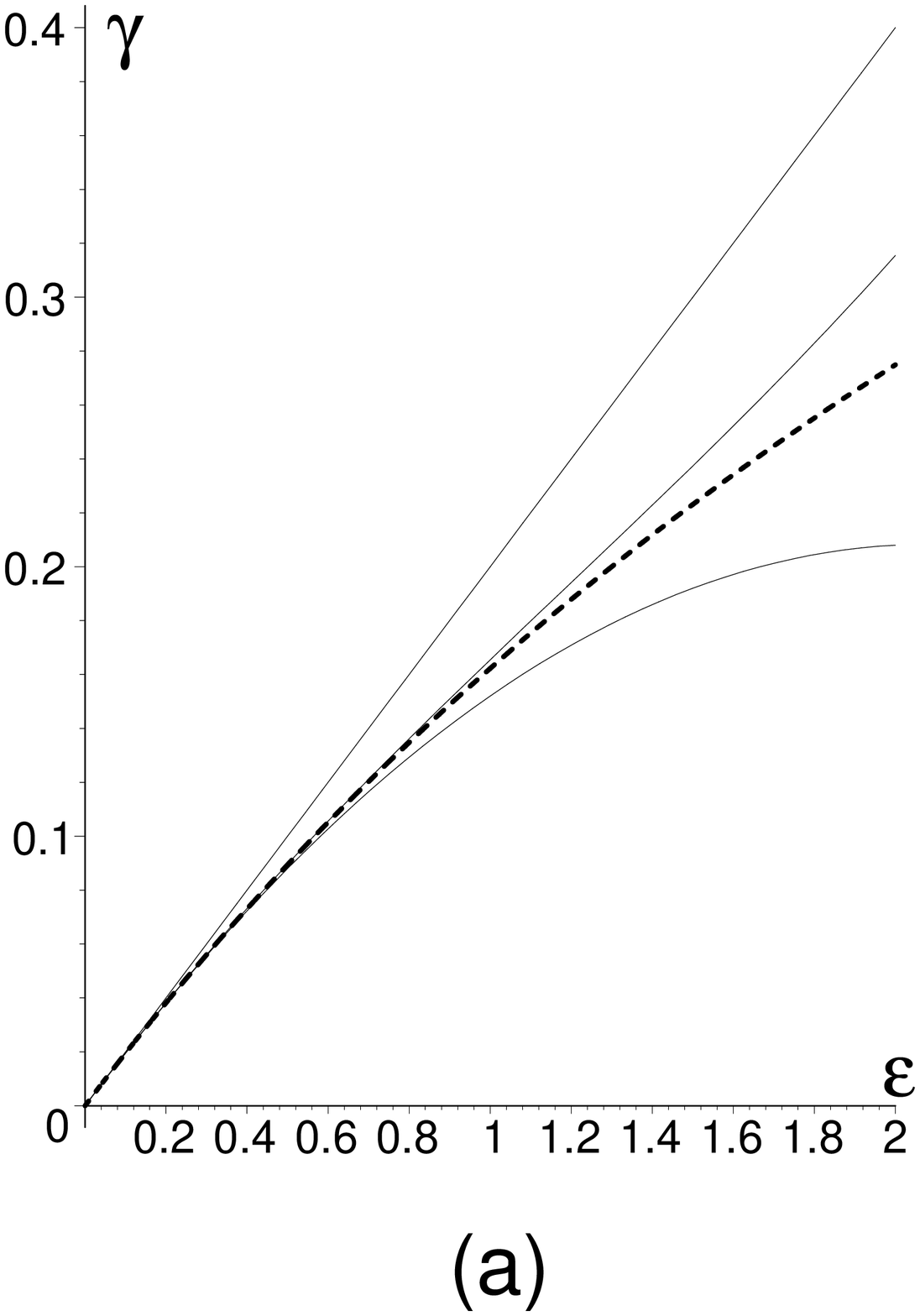,height=4cm,width=4cm}
\psfig{file=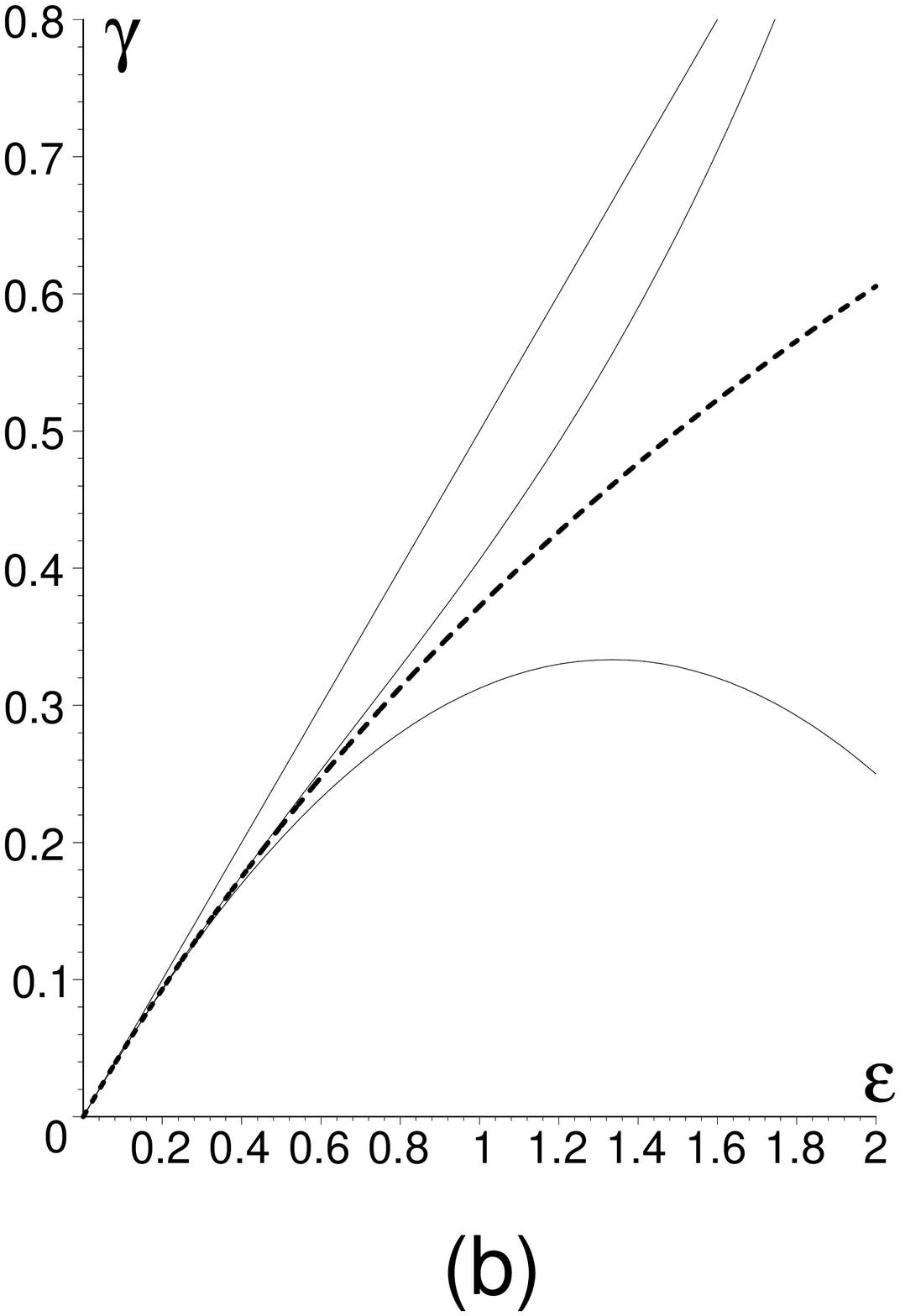,height=4cm,width=4cm}
\psfig{file=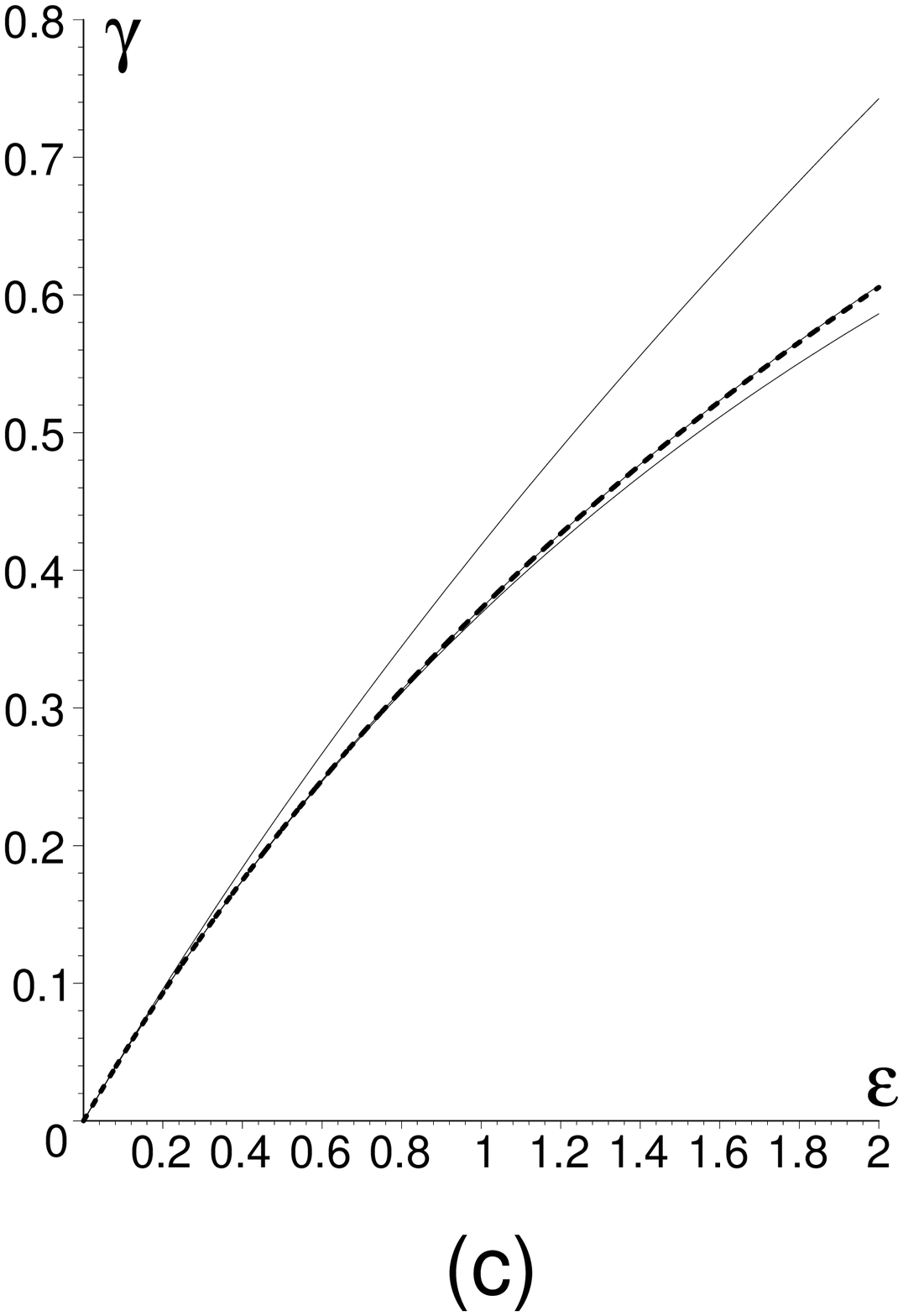,height=4cm,width=4cm}
\psfig{file=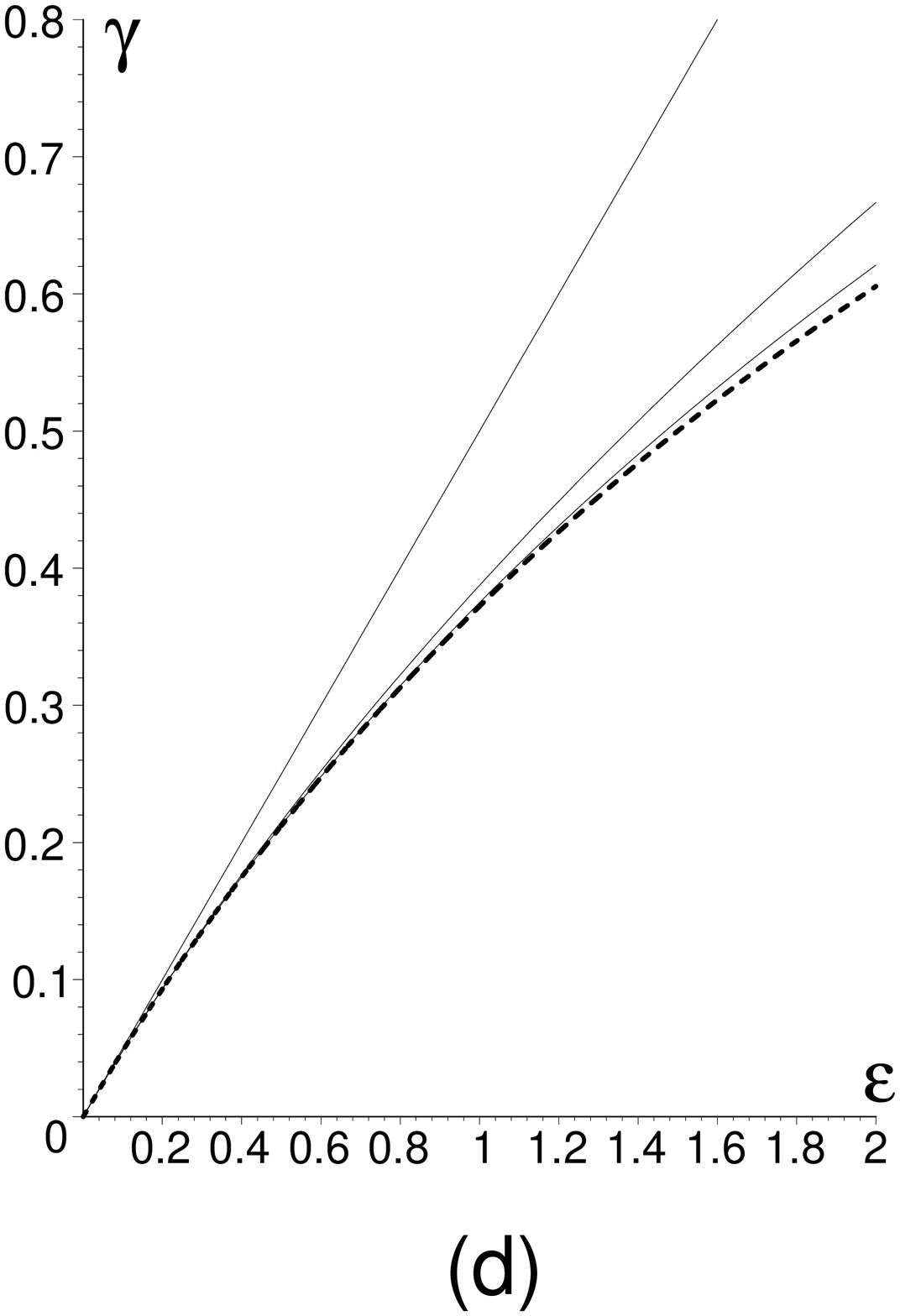,height=4cm,width=4cm} }
\bigskip
\caption{
Anomalous dimension $\gamma\equiv\gamma_{22}^{*}$ for $d=3$ ({\it a}) and
$d=2$ ({\it b,c,d}). Dashed line: exact solution by Refs.
\protect\cite{Falk1,Gat}. Solid lines in Figs.~3{\it a} and 3{\it b}:
first, third and second approximations of the plain $\varepsilon$ expansion
(from above to below).
Solid lines in Fig.~3{\it c}: first, third and second approximations of
the improved $x$ expansion (from above to below; the third approximation
is practically indistinguishable from the exact solution for all
$0<\varepsilon<2$). Solid lines in Fig.~3{\it d}: first, second and third
approximations of the inverse $\gamma$ expansion (from above to below).}
\end{figure}

\begin{figure}
\centerline{ \psfig{file=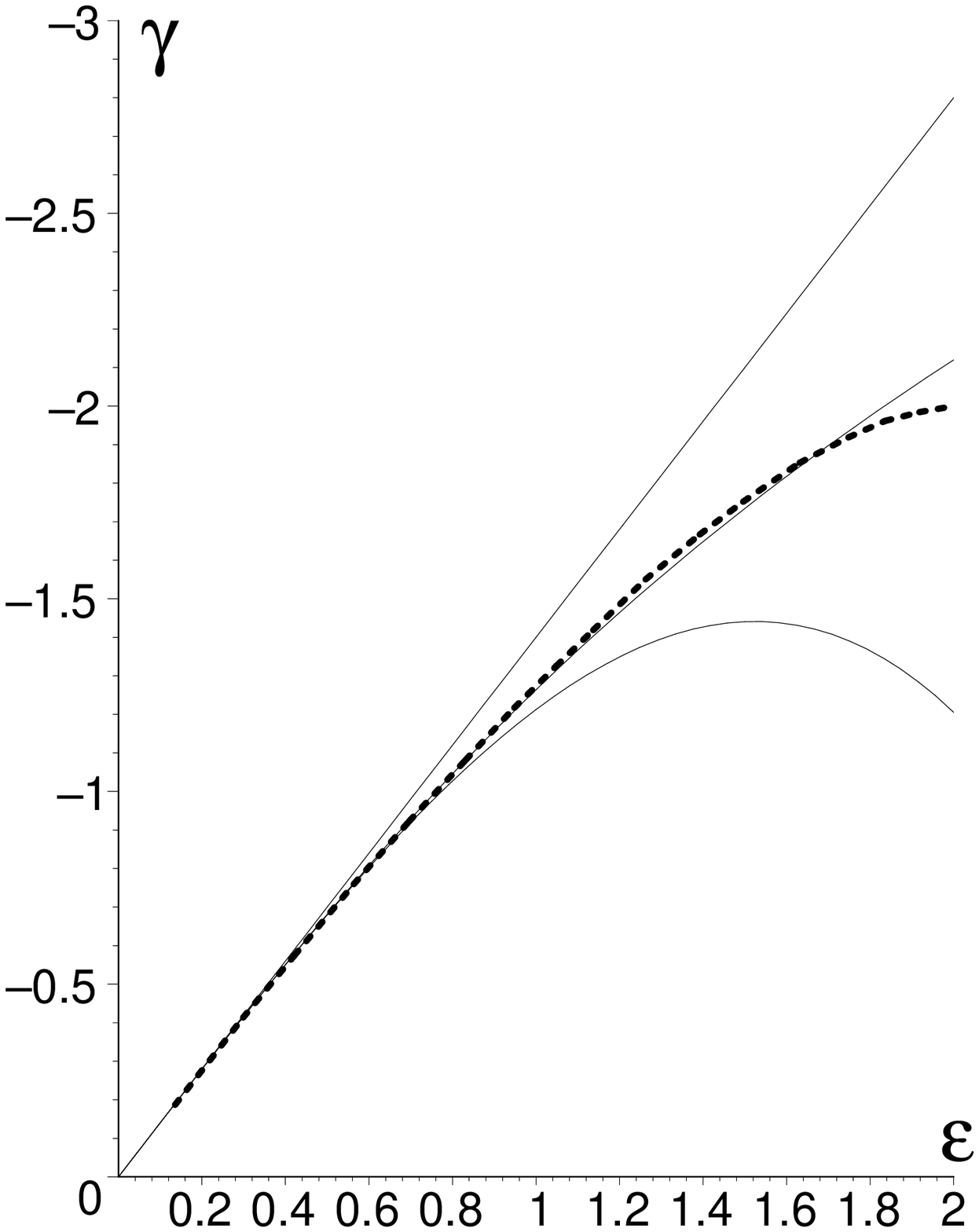,height=4cm,width=4cm}
\psfig{file=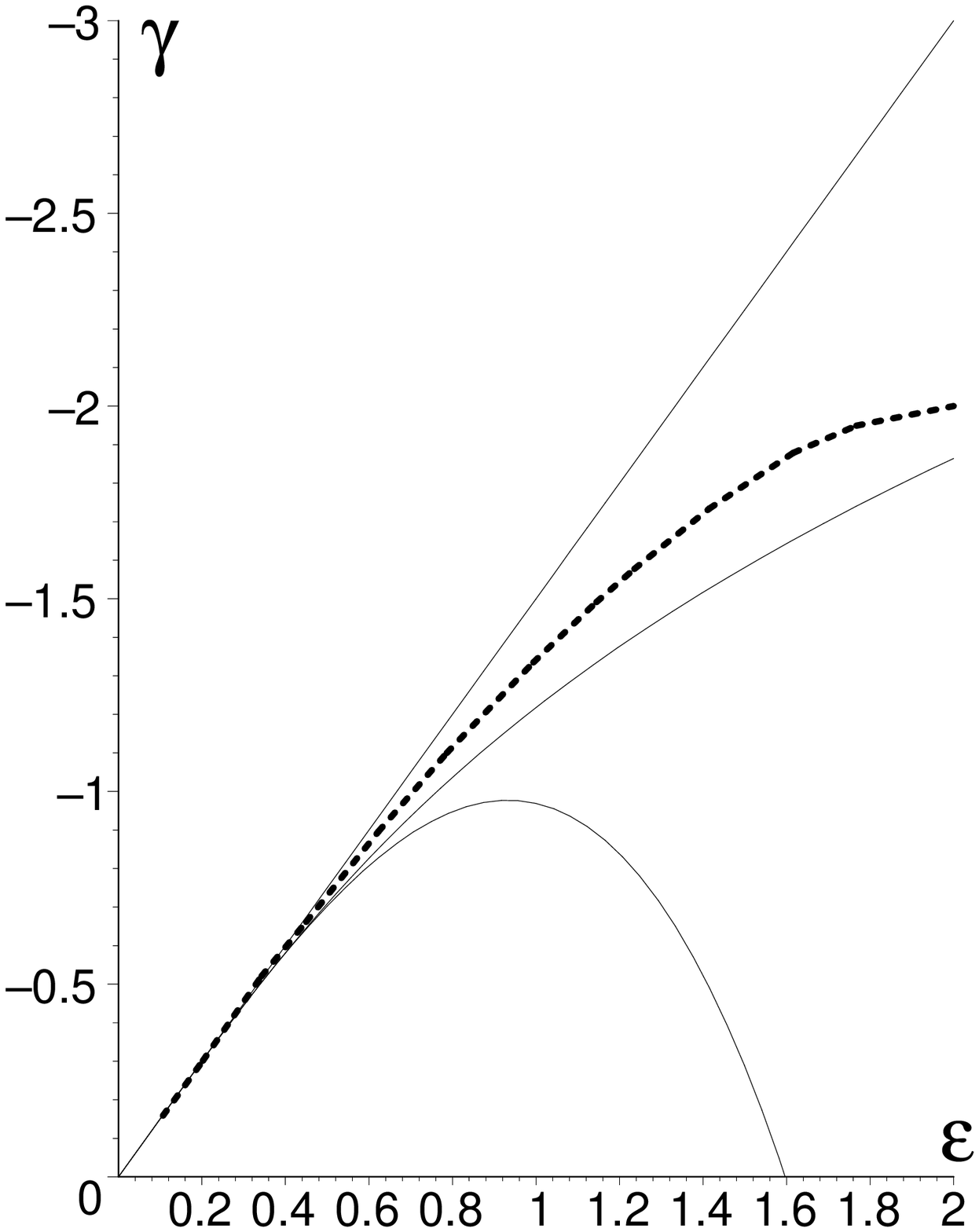,height=4cm,width=4cm} }
\bigskip
\caption{
Anomalous dimension $\gamma\equiv\gamma_{31}^{*}$ for $d=3$ (left) and $d=2$
(right): the $O(\eps)$ approximation, the third-order approximation of the
inverse $\gamma$ expansion, and the third-order approximation of the plain
$\eps$ expansion (from above to below). Dashed line: numerical solution by
Refs.~\protect\cite{Pumir}.}
\end{figure}

\begin{figure}
\centerline{
\psfig{file=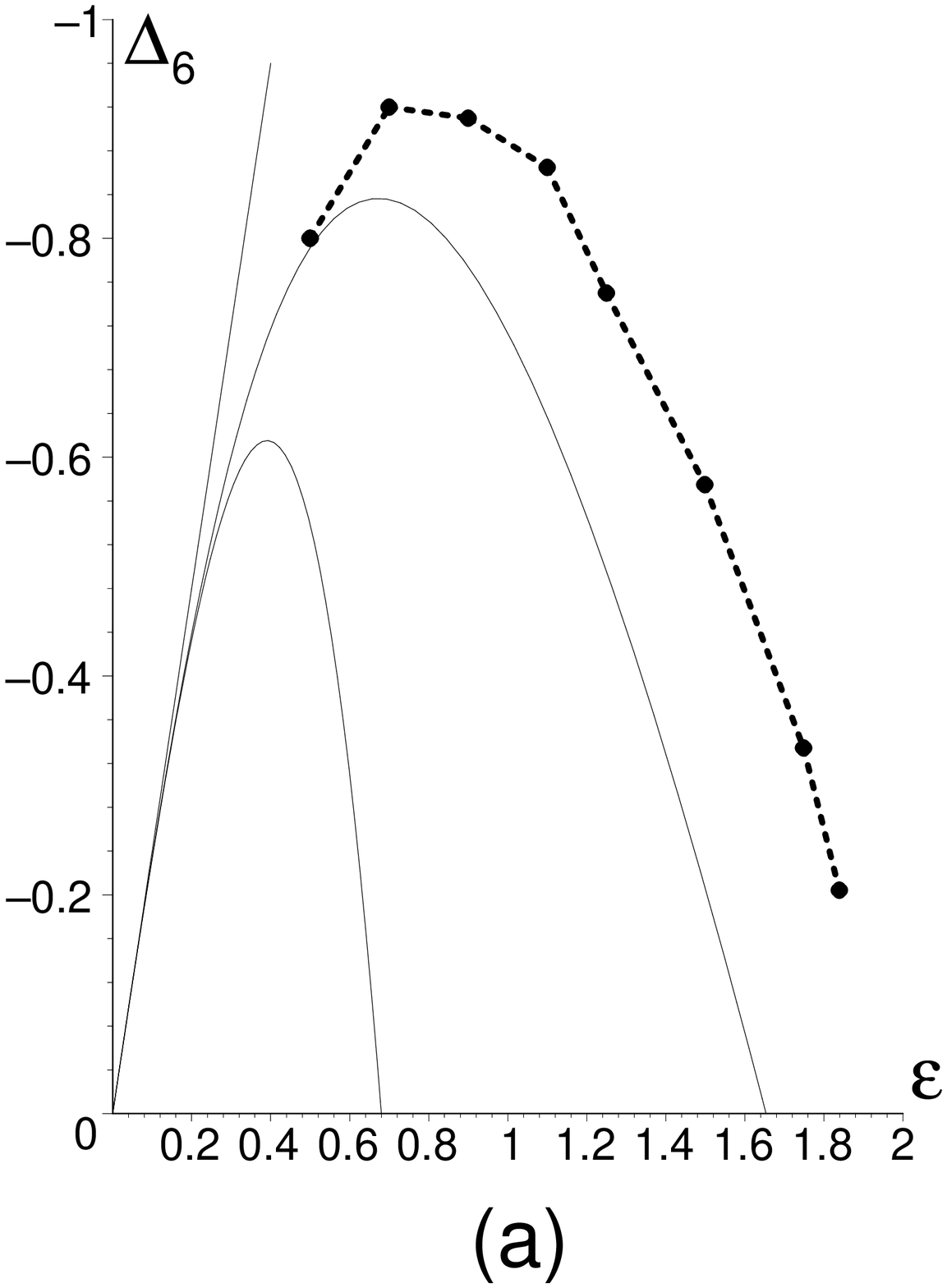,height=4cm,width=4cm}
\psfig{file=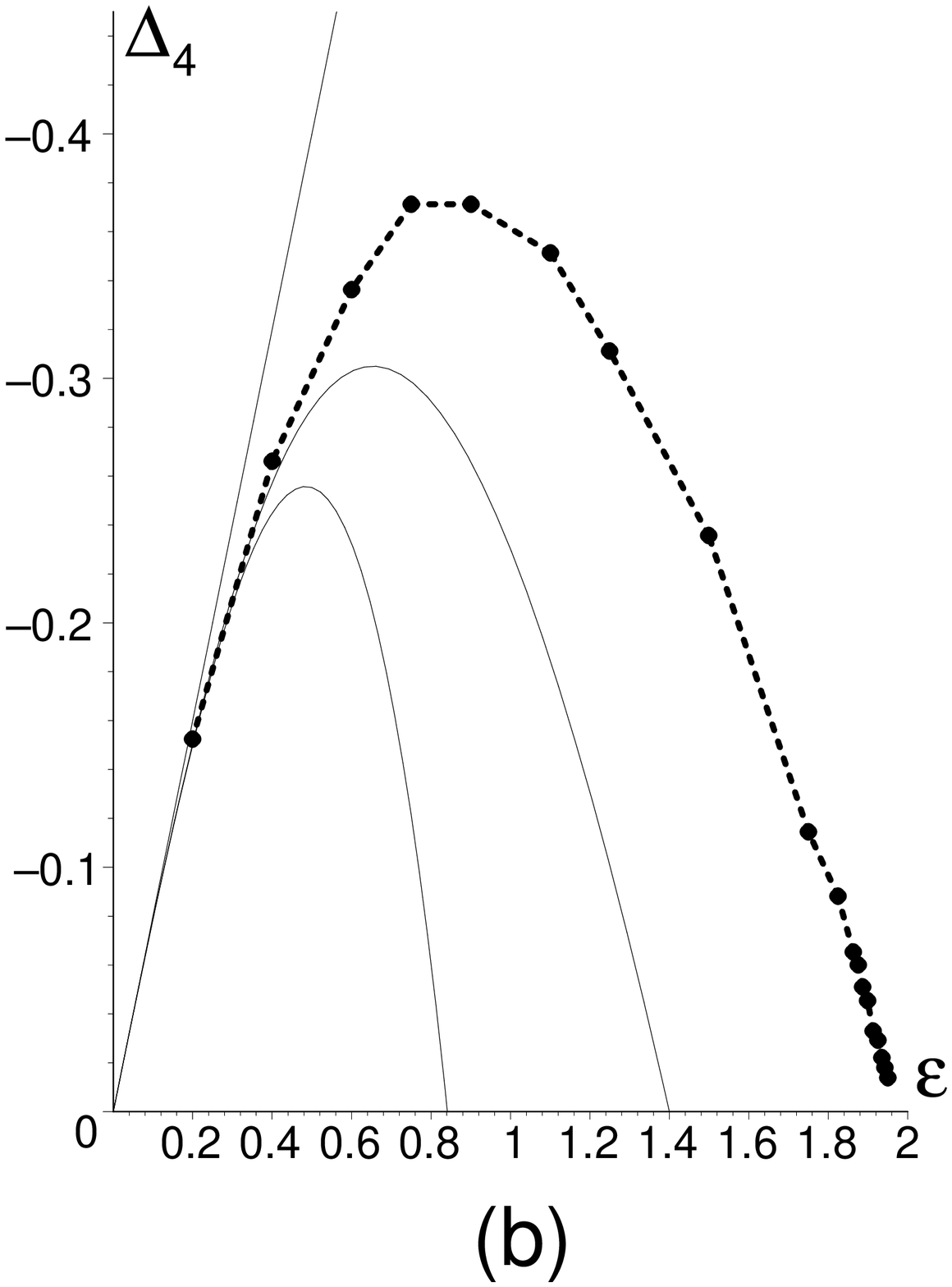,height=4cm,width=4cm}
\psfig{file=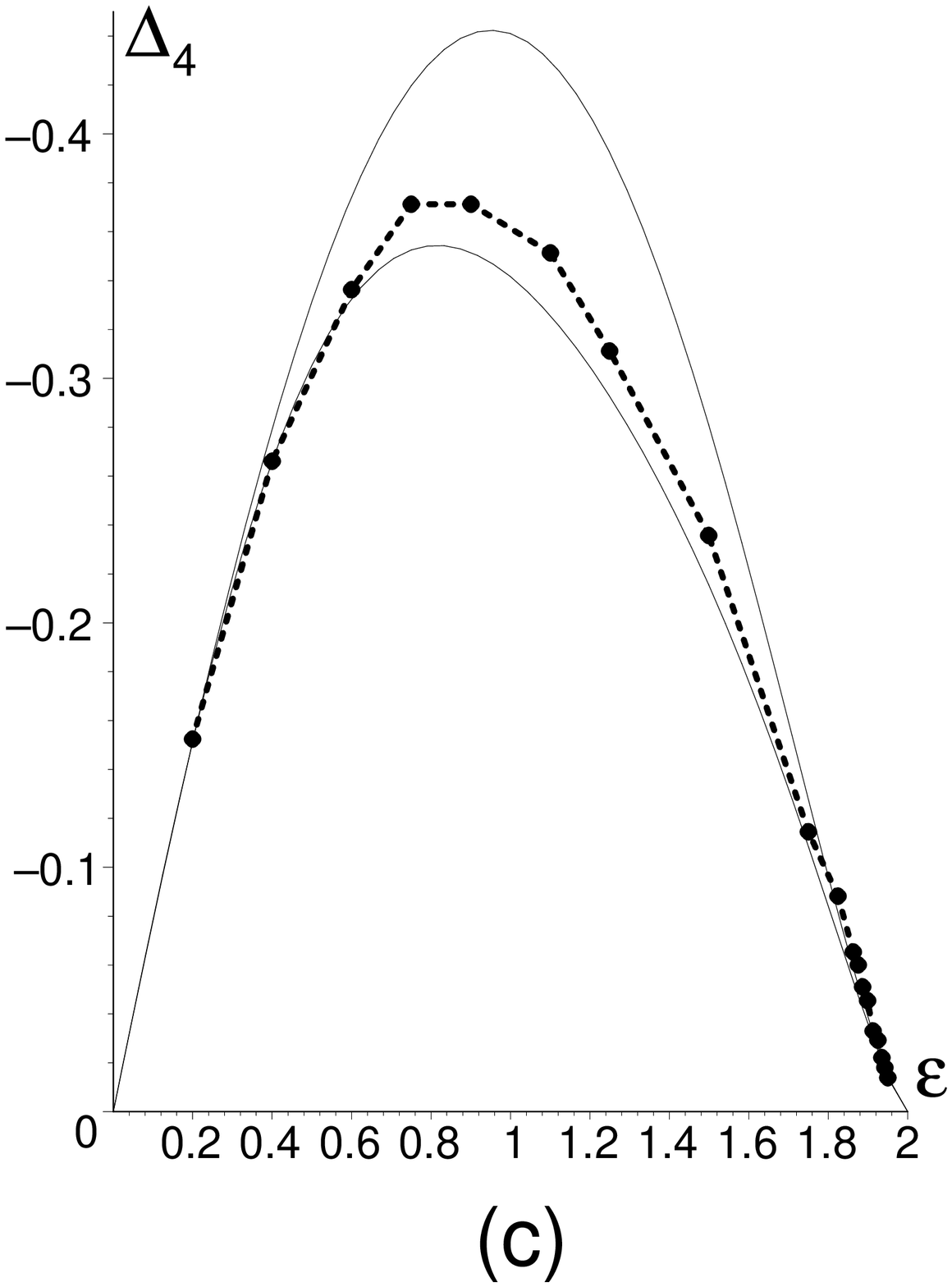,height=4cm,width=4cm}}
\bigskip
\caption{
Critical dimensions $\Delta_{n}$ for $d=3$: $n=6$ ({\it a}) and $n=4$
({\it b,c}). Dots connected by dashed lines: numerical simulations by
Refs.~\protect\cite{MM} ($n=6$) and~\protect\cite{VMF1,VMF2} ($n=4$). Solid
lines in Figs.~5{\it a,b}: the $O(\eps)$ slope, third-order approximation
of the $\gamma$ expansion, and third-order approximation of the plain $\eps$
expansion (from above to below). Solid lines in Fig.~5{\it c}: approximations
for $n=4$, obtained using the interpolation formula (\protect\ref{interpol})
with $p=1$ (upper curve) and $p=3$ (lower curve).}
\end{figure}

\section{Conclusion} \label {sec:Con}

To conclude with, we have studied the inertial-range anomalous scaling of a
passive scalar quantity advected by the Gaussian velocity field, white in
time and self-similar in space. The corresponding stochastic problem
(\ref{1})--(\ref{3}) can be reformulated as a field theoretic model
(\ref{action}), which allows one to identify the anomalous exponents with
the critical dimensions of certain scalar and tensor composite operators
built of the scalar gradients and to calculate them within the RG and OPE
approach in the form of a regular perturbation expansion, similar to the
well-known $\eps$ expansion in the RG theory of critical behavior.

Earlier, the anomalous exponents were presented to orders $\eps^2$
\cite{RG,RG1,RG2,Juha} and $\eps^3$ \cite{cube}; the main goal of the present
paper has been the detailed explanation of the corresponding calculational
techniques and derivation of the three-loop result, including the anisotropic
sectors. Owing to the comparative universality of the RG and OPE formalism,
these techniques can be applied to other models of dynamical critical
phenomena and systems far from equilibrium: passive advection by the
non-Gaussian velocities with finite correlation time, stochastic
Navier--Stokes equation and so on; see Refs.~\cite{RG3,UFN,turbo}.

Another scope of the paper has been the discussion of the convergence
properties of the $\eps$ expansion and the possibility of its extrapolation
to finite values of $\eps$. It was shown that the knowledge of three
terms allows one to obtain reasonable predictions for finite $\eps\sim1$;
even the plain $\eps$ expansion captures some subtle qualitative features
of the anomalous exponents established in numerical experiments.

We believe that the framework of the renormalization group and operator
product expansion, the concept of dangerous composite operators and the
$\eps$ expansion will become the necessary elements of the appearing
theory of the anomalous scaling in fully developed turbulence.

\acknowledgments
The authors are thankful to Michal Hnatich, Juha Honkonen, Antti Kupiainen,
Andrea Mazzino, Paolo Muratore Ginanneschi and Anton Runov for discussions.
The work was supported in part by the Nordic Grant for Network Cooperation
with the Baltic Countries and Northwest Russia (Grant No.~FIN-18/2001), the
Russian Foundation for Fundamental Research (Grant No.~99-02-16783) and the
Grant Center for Natural Sciences (Grant No.~E00-3-24).

\appendix
\section{Coefficients $A_{\lowercase{i}}$ for the three-loop diagrams}

Below we give the pole parts of the coefficients $A_{i}$ from Eq. (\ref{V6})
for all normal (not factorizable) three-loop diagrams.
We use the following notation:
\[ A_{i} = u^{3}(\mu/m)^{3\eps} \left\{ A^{(3)}_{i}  \eps^{-3} +
A^{(2)}_{i}  \eps^{-2} + A^{(1)}_{i}  \eps^{-1} \right\} ,
\qquad A^{(3)}_{i} = (d-1)\, a_{i}(d) / 432 d^{2}(d+2)^{2}. \]
Coefficients $A^{(1,2)}_{i}$ are given in Tables~1 and 2 for $d=2$
and $d=3$, respectively; coefficients $a_{i}(d)$ are given in Table~3
for general $d$.

\begin{table}[ht]
\caption{Coefficients $A^{(1,2)}_{i}$ for $d=2$.}
\label{table1}
\begin{tabular}{ccccccc}
No & $A^{(2)}_{1}\cdot 10^{3}$ & $A^{(1)}_{1}\cdot 10^{3}$  &
$A^{(2)}_{2}\cdot 10^{3}$ &  $A^{(1)}_{2}\cdot 10^{3}$ &
$A^{(2)}_{3}\cdot 10^{3}$ &  $A^{(1)}_{3}\cdot 10^{3}$  \\
\tableline
23 &\ \ 0.23875 & $-0.06103$ & \ \ 0 &\  \ 0 & --- & --- \\
32 & $-0.56396$ &\ \  0.32532 &\ \  0.49056 & $-0.56645$ & --- & --- \\
33 &\ \  1.11953 &  $-0.49834$ &\ \  0.93382 & $-1.04911$ & --- & --- \\
34 & $-1.05470$ & $-0.12495$ &\ \  0.2434\ \ & $-0.5770$\ \ & --- & --- \\
35 & $-0.41468$ &\ \  0.30671 & $-0.63626 $ &\ \  0.22330 & --- & --- \\
42 & $-0.00053$ & $-0.11026$ &\ \  0.06564 & $-0.16222$ & $-0.54061$ &
\ \ 0.32914 \\
43 &\ \  0.02902 &\ \  0.00249 &\ \  0.03161 & $-0.00715$ & $-0.33590$ &
\ \ 0.07208 \\
44 & $-0.03113$ & $-0.01477$ &\ \  0.03350 &$- 0.00286$ & $-0.19626$ &
\ \ 0.21410 \\
45 & $-0.26802$ &\ \  0.26530 & $-0.02352$ &\ \  0.03554 &\ \ 0.88599 &
$-0.88934$ \\
\end{tabular}
\end{table}

\begin{table}[ht]
\caption{Coefficients $A^{(1,2)}_{i}$ for $d=3$.}
\label{table2}
\begin{tabular}{ccccccc}
No & $A^{(2)}_{1}\cdot 10^{3}$ & $A^{(1)}_{1}\cdot 10^{3}$  &
$A^{(2)}_{2}\cdot 10^{3}$ &  $A^{(1)}_{2}\cdot 10^{3}$ &
$A^{(2)}_{3}\cdot 10^{3}$ &  $A^{(1)}_{3}\cdot 10^{3}$  \\
\tableline
23 &\ \ 1.03775 & $-0.14683 $ & \ \ 0 &\  \ 0 & --- & --- \\
32 & $-0.52289$ &\ \  0.39815 &\ \  0.21884 & $-0.50345$ & --- & --- \\
33 &\ \  1.57028 &\ \  0.00156 &\ \  1.56822 & $-0.76955$ & --- & --- \\
34 & $-1.64375$ & $-0.82315$ & $-0.43185$ & $-1.46450$ & --- & --- \\
35 & $-0.43289$ &\ \  0.51544 & $-1.07015$ &\ \  0.37916 & --- & --- \\
42 & $-0.07198$ & $-0.08409$ &\ \  0.02952 & $-0.22968$ & $-0.83453$ &
\ \ 0.37471 \\
43 &\ \  0.02808 &\ \  0.00152 &\ \  0.03587 & $-0.01186$ & $-0.62532$ &
\ \ 0.07298 \\
44 & $-0.07726$ &\ \  0.01392 &\ \  0.00504 &\ \  0.02341 & $-0.36121$ &
\ \ 0.37991 \\
45 & $-0.33881$ &\ \  0.33509 & $-0.10389$ &\ \  0.11756 &\ \  0.94431 &
$-0.95923$ \\
\end{tabular}
\end{table}

\begin{table}[ht]
\caption{Coefficients $a_{i}(d)$ for general $d$.}
\label{table3}
\begin{tabular}{cccc}
No & $a_{1}(d)$ & $a_{2}(d)$  & $a_{3}(d)$  \\
\tableline
23 & $9(d+1)(2d^{2}+2d-5)$ & $27(d-1)^{2}(d+2)^{2}$ &  --- \\
32 & $6(d+1)(d^{2}+d-3)$ & $2(d^{4}+3d^{3}-3d^{2}-7d+10) $ &  --- \\
33 & $-12(d+1)$ & $-4(d^{2}+2d-2) $ &  --- \\
34 & $3(d+1)(3d^{2}+3d-8)$ & $3d^{4}+8d^{3}-13d^{2}-22d+32 $ &  --- \\
35 & $-12(d+1)$ & $-4(d^{2}+2d-2) $ &  --- \\
42 & $3(d+1)(d^{2}+d-8)/4$ & $(d^{4}+6d^{3}+5d^{2}-16d+32)/8 $  &
      $-4(d^{2}+2d-2) $\\
43 & $ -6(d+1)$ & $(d^{3}+d^{2}-4d+8)/2$ & $d^{4}+4d^{3}+d^{2}-6d+8$\\
44 & $ -6(d+1)$ & $(d^{3}+d^{2}-4d+8)/2$ & $d^{4}+4d^{3}+d^{2}-6d+8$\\
45 & $3(d+1)(d^{2}+d-8)/4$ & $(d^{4}+6d^{3}+5d^{2}-16d+32)/8 $  &
      $-4(d^{2}+2d-2) $\\
\end{tabular}
\end{table}

\end{document}